\documentclass[aps,superscriptaddress,nofootinbib,eqsecnum,tightenlines,floatfix]{revtex4}
\usepackage{epsfig,rotating}
\usepackage{amsmath,amssymb}
\usepackage{dsfont}
\usepackage{graphics}
\usepackage{rotate}
\usepackage{psfrag}
\begin{document}
\newcommand{\be}{\begin{equation}}
\newcommand{\ee}{\end{equation}}
\newcommand{\avg}[1]{\langle #1 \rangle}
\newcommand{\vx}{{\boldsymbol{x}}}
\newcommand{\rv}{{\boldsymbol{r}}}
\newcommand{\vq}{\ensuremath{\vec{q}}}
\newcommand{\pv}{\ensuremath{\vec{p}}}
\def\AJ{{\it Astron. J.} }
\def\ARAA{{\it Annual Rev. of Astron. \& Astrophys.} }
\def\ApJ{{\it Astrophys. J.} }
\def\ApJL{{\it Astrophys. J. Letters} }
\def\ApJS{{\it Astrophys. J. Suppl.} }
\def\ApP{{\it Astropart. Phys.} }
\def\AA{{\it Astron. \& Astroph.} }
\def\AAR{{\it Astron. \& Astroph. Rev.} }
\def\AAL{{\it Astron. \& Astroph. Letters} }
\def\AASu{{\it Astron. \& Astroph. Suppl.} }
\def\AN{{\it Astron. Nachr.} }
\def\IJMP{{\it Int. J. of Mod. Phys.} }
\def\JGR{{\it Journ. of Geophys. Res.}}
\def\JHEP{{\it Journ. of High En. Phys.} }
\def\JPhG{{\it Journ. of Physics} {\bf G} }
\def\MNRAS{{\it Month. Not. Roy. Astr. Soc.} }
\def\Nature{{\it Nature} }
\def\NewAR{{\it New Astron. Rev.} }
\def\NJPh{{\it New Journ. of Phys.} }
\def\PASP{{\it Publ. Astron. Soc. Pac.} }
\def\PhFl{{\it Phys. of Fluids} }
\def\PLB{{\it Phys. Lett.}{\bf B} }
\def\PhysRep{{\it Phys. Rep.} }
\def\PR{{\it Phys. Rev.} }
\def\PRD{{\it Phys. Rev.} {\bf D} }
\def\PRL{{\it Phys. Rev. Letters} }
\def\RMP{{\it Rev. Mod. Phys.} }
\def\Science{{\it Science} }
\def\ZfA{{\it Zeitschr. f{\"u}r Astrophys.} }
\def\ZfN{{\it Zeitschr. f{\"u}r Naturforsch.} }
\def\etal{{\it et al.}}
\hyphenation{mono-chro-matic sour-ces Wein-berg
chang-es Strah-lung dis-tri-bu-tion com-po-si-tion elec-tro-mag-ne-tic
ex-tra-galactic ap-prox-i-ma-tion nu-cle-o-syn-the-sis re-spec-tive-ly
su-per-nova su-per-novae su-per-nova-shocks con-vec-tive down-wards
es-ti-ma-ted frag-ments grav-i-ta-tion-al-ly el-e-ments me-di-um
ob-ser-va-tions tur-bul-ence sec-ond-ary in-ter-action
in-ter-stellar spall-ation ar-gu-ment de-pen-dence sig-nif-i-cant-ly
in-flu-enc-ed par-ti-cle sim-plic-i-ty nu-cle-ar smash-es iso-topes
in-ject-ed in-di-vid-u-al nor-mal-iza-tion lon-ger con-stant
sta-tion-ary sta-tion-ar-i-ty spec-trum pro-por-tion-al cos-mic
re-turn ob-ser-va-tion-al es-ti-mate switch-over grav-i-ta-tion-al
super-galactic com-po-nent com-po-nents prob-a-bly cos-mo-log-ical-ly
Kron-berg Berk-huij-sen}

\def\refalt@jnl#1{{\rm #1}}

\def\aj{\refalt@jnl{AJ}}                   
\def\araa{\refalt@jnl{ARA\&A}}             
\def\apj{\refalt@jnl{ApJ}}                 
\def\apjl{\refalt@jnl{ApJ}}                
\def\apjs{\refalt@jnl{ApJS}}               
\def\ao{\refalt@jnl{Appl.~Opt.}}           
\def\apss{\refalt@jnl{Ap\&SS}}             
\def\aap{\refalt@jnl{A\&A}}                
\def\aapr{\refalt@jnl{A\&A~Rev.}}          
\def\aaps{\refalt@jnl{A\&AS}}              
\def\azh{\refalt@jnl{AZh}}                 
\def\baas{\refalt@jnl{BAAS}}               
\def\jrasc{\refalt@jnl{JRASC}}             
\def\memras{\refalt@jnl{MmRAS}}            
\def\mnras{\refalt@jnl{MNRAS}}             
\def\pra{\refalt@jnl{Phys.~Rev.~A}}        
\def\prb{\refalt@jnl{Phys.~Rev.~B}}        
\def\prc{\refalt@jnl{Phys.~Rev.~C}}        
\def\prd{\refalt@jnl{Phys.~Rev.~D}}        
\def\pre{\refalt@jnl{Phys.~Rev.~E}}        
\def\prl{\refalt@jnl{Phys.~Rev.~Lett.}}    
\def\pasp{\refalt@jnl{PASP}}               
\def\pasj{\refalt@jnl{PASJ}}               
\def\qjras{\refalt@jnl{QJRAS}}             
\def\skytel{\refalt@jnl{S\&T}}             
\def\solphys{\refalt@jnl{Sol.~Phys.}}      
\def\sovast{\refalt@jnl{Soviet~Ast.}}      
\def\ssr{\refalt@jnl{Space~Sci.~Rev.}}     
\def\zap{\refalt@jnl{ZAp}}                 
\def\nat{\refalt@jnl{Nature}}              
\def\iaucirc{\refalt@jnl{IAU~Circ.}}       
\def\aplett{\refalt@jnl{Astrophys.~Lett.}} 
\def\apspr{\refalt@jnl{Astrophys.~Space~Phys.~Res.}}
\def\bain{\refalt@jnl{Bull.~Astron.~Inst.~Netherlands}}
\def\fcp{\refalt@jnl{Fund.~Cosmic~Phys.}}  
\def\gca{\refalt@jnl{Geochim.~Cosmochim.~Acta}}   
\def\grl{\refalt@jnl{Geophys.~Res.~Lett.}} 
\def\jcp{\refalt@jnl{J.~Chem.~Phys.}}      
\def\jgr{\refalt@jnl{J.~Geophys.~Res.}}    
\def\jqsrt{\refalt@jnl{J.~Quant.~Spec.~Radiat.~Transf.}}
\def\memsai{\refalt@jnl{Mem.~Soc.~Astron.~Italiana}}
\def\nphysa{\refalt@jnl{Nucl.~Phys.~A}}   
\def\physrep{\refalt@jnl{Phys.~Rep.}}   
\def\physscr{\refalt@jnl{Phys.~Scr}}   
\def\planss{\refalt@jnl{Planet.~Space~Sci.}}   
\def\procspie{\refalt@jnl{Proc.~SPIE}}   

\title{\Large HIGHLIGHTS and CONCLUSIONS 

\medskip

of the Chalonge 14th Paris Cosmology Colloquium 2010:

\medskip

`The Standard Model of the Universe: Theory and Observations',    

\medskip

Ecole Internationale d'Astrophysique Daniel Chalonge

Observatoire de Paris

in the historic Perrault building, 22-24 July 2010.}

\author{\Large \bf   H. J. de Vega$^{(a,b)}$,  M.C. Falvella$^{(c)}$, N. G. Sanchez$^{(b)}$}

\date{\today}

\affiliation{$^{(a)}$ LPTHE, Universit\'e
Pierre et Marie Curie (Paris VI) et Denis Diderot (Paris VII),
Laboratoire Associ\'e au CNRS UMR 7589, Tour 24, 5\`eme. \'etage, 
Boite 126, 4, Place Jussieu, 75252 Paris, Cedex 05, France. \\
$^{(b)}$ Observatoire de Paris,
LERMA. Laboratoire Associ\'e au CNRS UMR 8112.
 \\61, Avenue de l'Observatoire, 75014 Paris, France.\\
$^{(c)}$ Italian Space Agency and MIUR, Viale Liegi n.26,
00198 Rome, Italy.}

\begin{abstract}

The Chalonge 14th Paris Cosmology Colloquium 2010 was held on 22-24 July in the historic 
Paris Observatory's Perrault building, in the
Chalonge School spirit combining real cosmological/astrophysical data and hard theory
predictive approach connected to them in the Standard Model of the Universe:
News and reviews from WMAP7, BICEP, QUAD, SPT, AMI, ACT, Planck, QUIJOTE, Herschel, SPIRE,
ATLAS and HerMES surveys; astrophysics 
and particle physics dark matter (DM) searches and galactic observations; related theory and simulations,
with the aim of synthesis, progress and clarification.  Peter Biermann,  Daniel Boyanovsky, 
Asantha Cooray, Claudio Destri, 
Hector de Vega, Gerry Gilmore, Stefan Gottl\"ober, Eiichiro Komatsu,  Stacy McGaugh,
Anthony Lasenby, Rafael Rebolo, Paolo Salucci, Norma  Sanchez and Anton Tikhonov
present here their highlights of the Colloquium. Inflection points in several current 
research lines emerged, particularly on dark matter (DM) where $\Lambda$WDM 
(Warm Dark Matter) emerges impressively
over $\Lambda$CDM whose ever-increasing galactic scale problems are staggering.
The summary and conclusions by H. J. de Vega, M. C. Falvella and N. G. Sanchez 
stress among other points: 
(i) Data confirm primordial CMB gaussianity. Inflation effective theory predicts
negligible primordial non-gaussianity, negligible scalar index running  
and a tensor to scalar ratio $ \sim 0.05-0.04 $ at reach/border line of next CMB observations;
the present data with this theory  clearly prefer new inflation; 
early fast-roll inflation is generic and provides lowest multipoles depression.
CMB secondary anisotropies progress rapidly with new CMB high-$l$ constraints and 
Sunyaev-Zeldovich (SZ) amplitudes smaller than expected: CMB and X-ray data agree but 
intracluster medium models 
need revision (they overestimate SZ), relaxed and non-relaxed clusters need distinction as 
WMAP7 shows.(ii) The Milky Way is not formed from dSph-like systems. 
Feedback does not operate as already suggested: it was extremely mild in the lowest 
luminosity galaxies, it does not substantially modify initial conditions.
(iii) The cosmic ray positron and electron excess recently observed is 
explained by natural astrophysical processes, while annihilating/decaying 
dark matter models face growing tailoring to explain observations. 
(iv) Cored (non cusped) DM halos and warm
(keV scale mass) DM are increasingly favored from theory and 
astrophysical observations, they naturally produce the observed
small scale structures; sterile neutrinos are suitable candidates. Wimps (heavier 
than 1 GeV) are strongly disfavoured combining theory with galaxy 
observations. Putting all together, evidence that 
$\Lambda$CDM does not work at small scales is staggering. 
$\Lambda$WDM simulations with 1 keV particles 
reproduce observations, sizes of local 
minivoids and velocity functions. Overall, keV scale DM particles deserve dedicated
searchs and simulations. Peter Biermann presents his live minutes of the 
Colloquium and concludes that a right-handed sterile neutrino of mass of a 
few keV is the most interesting DM candidate. Photos of the Colloquium are included.

\end{abstract}

\maketitle

\tableofcontents

\newpage

\section{Purpose of the Colloquium and Introduction}

The main aim of the series `Paris Cosmology Colloquia', in the framework of the International 
School of Astrophysics  {\bf `Daniel Chalonge'},  is to put together real cosmological and 
astrophysical  data and hard theory approach connected to them. The Chalonge Paris Cosmology Colloquia 
bring together physicists, astrophysicists and astronomers from the world over. Each year these 
Colloquia are more attended and appreciated both by PhD students, post-docs and lecturers. 
The format of the Colloquia is intended to allow easy and fruitful mutual contacts and 
communication.

\bigskip

The subject of the 14th Paris Cosmology Colloquium 2010  was `THE STANDARD MODEL OF 
THE UNIVERSE: THEORY AND OBSERVATIONS'. 

\bigskip

The  Colloquium took  place during three full days  (Thursday July 22, Friday 23 and 
Saturday July 24) at the parisian campus of  Paris Observatory (HQ), in the historic 
Perrault building.

\bigskip

The {\bf 14th Paris Cosmology Colloquium 2010} was within the astrofundamental physics 
spirit of the Chalonge School, focalized on recent observational and theoretical progress 
on the CMB and inflation with predictive power, dark matter, dark energy, dark ages and LSS 
in the context of the Standard Model of the Universe. Never as in this period, the Golden 
Age of Cosmology, the major subjects of the Daniel Chalonge School were so timely and in 
full development: the WMAP mission released in 2010 the new survey (7 years of 
observations) and the PLANCK mission launched in May 2009 is performing its 
First Survey.

\bigskip

The {\bf main topics} were: Observational and theoretical progress in deciphering the 
nature of dark matter, large and small scale structure, Warm (keV) dark matter and 
sterile neutrinos. Inflation after WMAP (in connection with the CMB and LSS data), slow roll and 
fast roll inflation, quadrupole suppression and initial conditions. CMB polarization. 
CMB measurements by the Planck mission and its science perspectives. 

\bigskip

All Lectures  were  plenary and followed by a discussion. 
Enough time was provided to the discussions.  

\begin{center}

Informations of the Colloquium are available on

 {\bf http://www.chalonge.obspm.fr/colloque2010.html}

\end {center}

\bigskip

Informations on the previous Paris Cosmology Colloquia and  
on the Chalonge school events are available at  

\begin{center}

 {\bf http://chalonge.obspm.fr}

(lecturers, lists of participants, lecture files and photos during the Colloquia).

\end {center}

\bigskip

This Paris Colloquia series started in 1994 at the Observatoire de Paris. 
The series cover selected topics of high current interest in the interplay between 
cosmology and fundamental physics. The PARIS COSMOLOGY COLLOQUIA are informal meetings. 
Their purpose is an updated understanding, from a fundamental point of view, of the progress 
and current problems in the early universe, cosmic microwave background radiation, large scale 
structure and neutrinos in astrophysics and the interplay between them. Emphasis is given to the 
mutual impact of fundamental physics and cosmology, both at theoretical and experimental 
-or observational- levels. 

\bigskip

Deep understanding, clarification, synthesis, a careful interdisciplinarity within a 
fundamental physics approach, are goals of this series of Colloquia.

\bigskip

Sessions last for three full days and leave enough time for private discussions and to enjoy 
the beautiful parisian campus of Observatoire de Paris (built on orders from Colbert and to 
plans by Claude Perrault from 1667 to 1672).

\bigskip

Sessions took  place in the Cassini Hall, on the meridean of Paris, in 'Salle du Conseil'
(Council Room) in the historic Perrault building ('B\^atiment Perrault") of 
Observatoire de Paris HQ, 
under the portraits of Laplace, Le Verrier, Lalande, Arago, Delambre and Louis XIV and in the 
'Grande Galerie' (the Great Gallery).

\bigskip

An {\bf Exhibition} retraced the 19 years of activity of the Chalonge School and of George Smoot 
participation to the School.
The books and proceedings of the School since its creation, as well as historic 
Daniel Chalonge material, instruments and the Daniel Chalonge Medal were on exhibition at the 
Grande Galerie.

\bigskip

After the Colloquium, a visit of the Perrault building took place guided by Professor Suzanne Debarbat 

\bigskip

More information on the Colloquia of this series can be found in the Proceedings
(H.J. de Vega and N. Sanchez, Editors) published by World Scientific Co. since 1994 and by 
Observatoire de Paris, and the Chalonge School Courses published by World Scientific Co 
and by Kluwer Publ Co. since 1991.

\bigskip

We want to express our grateful thanks to all the sponsors of the Colloquium,  
to all the lecturers for their excellent and polished presentations, to all the 
lecturers and participants for their active participation and their contribution to 
the outstanding discussions and lively atmosphere, to the assistants, secretaries and 
all collaborators of the Chalonge School,  who made this event so harmonious, wonderful and successful . 

\bigskip

\begin{center}
                                   
With Compliments and kind regards,

\bigskip  
                                            
{\bf Hector J de Vega, Maria Cristina Falvella,  Norma G Sanchez}

\end{center}

\bigskip

\bigskip

\begin{figure}[ht]
\includegraphics[scale=.9]{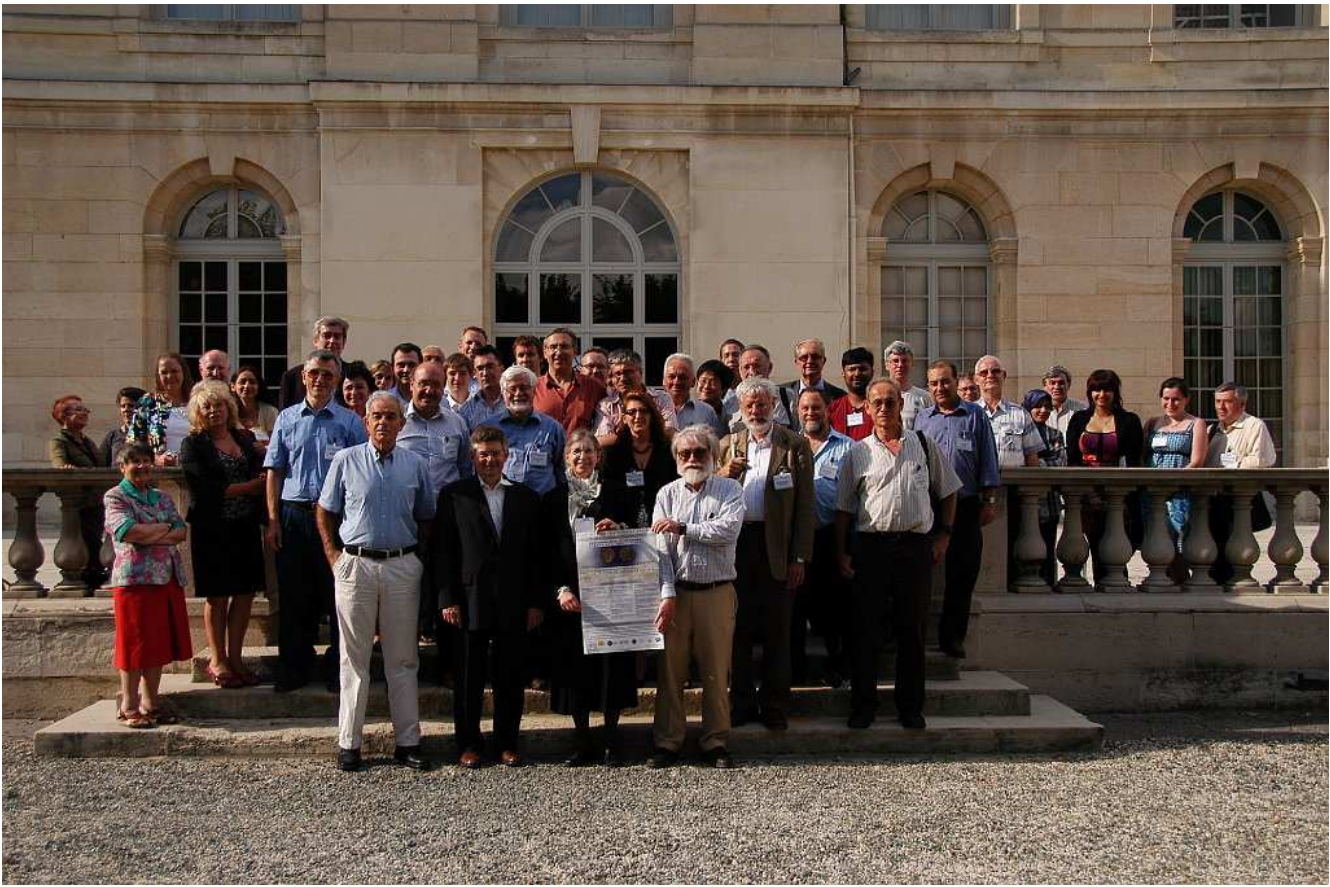}
\caption{Photo of the Group}
\end{figure}

\newpage

\begin{figure}[ht]
\includegraphics[height=20cm,width=15cm]{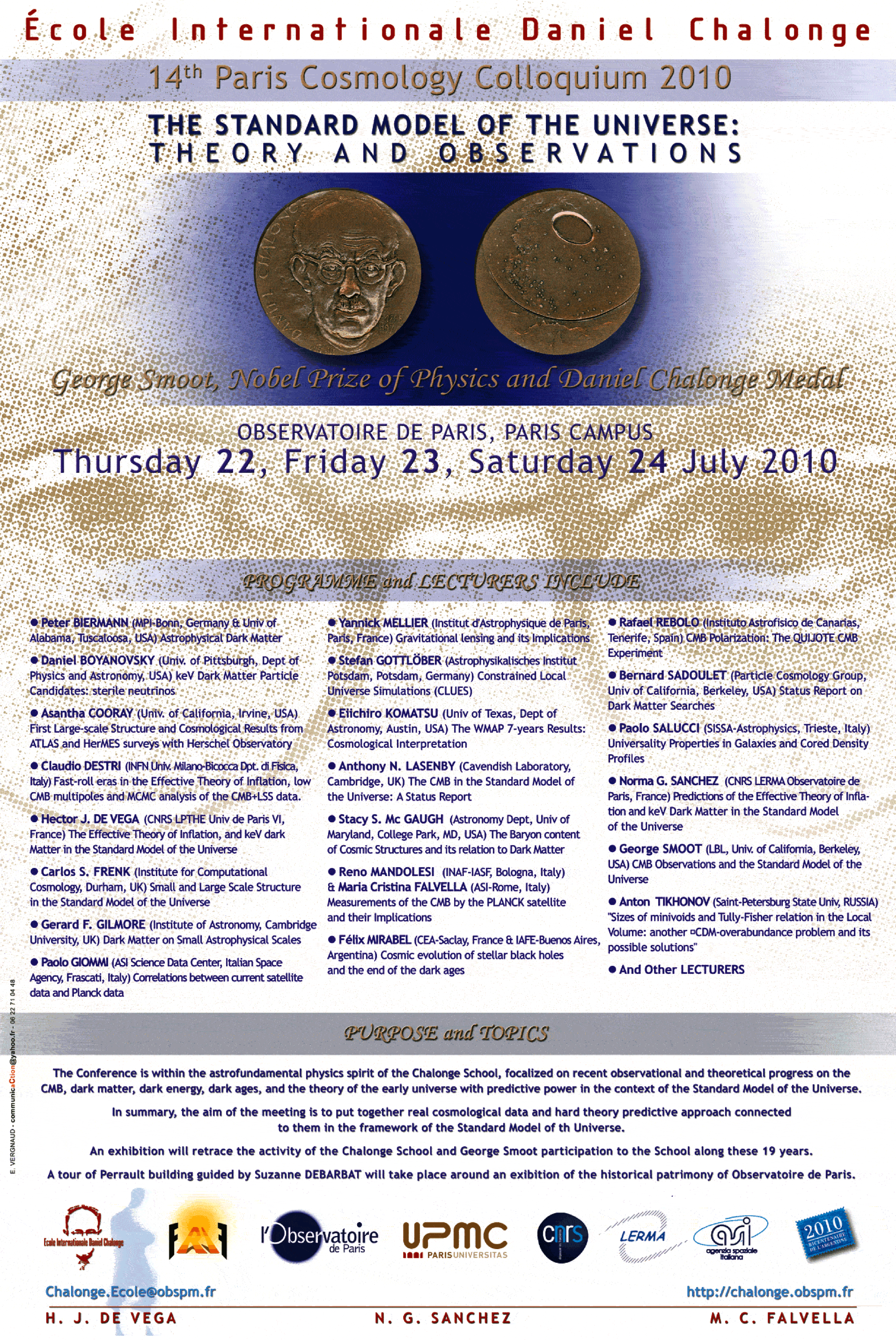}
\caption{Poster of the 14th Paris Cosmology Colloquium 2010}
\end{figure}

\newpage

\section{Programme and Lecturers}


\begin{itemize}

\item{{\bf Peter BIERMANN} (MPI-Bonn, Germany \& Univ of Alabama, Tuscaloosa, USA)\\
Astrophysical Dark Matter}

\item{{\bf Daniel BOYANOVSKY} (Univ. of Pittsburgh, Dept of Physics and Astronomy, USA)\\
keV Dark Matter Particle Candidates: sterile neutrinos}

\item{{\bf Asantha COORAY} (University of California, Irvine, USA)\\
First Large-scale Structure and Cosmological Results from ATLAS and HerMES surveys with 
Herschel Observatory}

\item{{\bf Claudio DESTRI} (INFN Univ. Milano-Bicocca Dpt. di Fisica G. Occhialini, Italy)\\
Fast-roll eras in the Effective Theory of Inflation, low CMB multipoles and MCMC analysis of 
the CMB+LSS data}.

\item{{\bf Hector J. DE VEGA} (CNRS LPTHE Univ de Paris VI, P \& M Curie, Paris, France)\\
The Effective Theory of Inflation, and keV dark Matter in the Standard Model of the Universe}

\item{{\bf Carlos S. FRENK} (Institute for Computational Cosmology, Durham, UK)\\
Small and Large Scale Structure in the Standard Model of the Universe}

\item{{\bf Gerard F. GILMORE} (nstitute of Astronomy, Cambridge University, UK)\\
Dark Matter on Small Astrophysical Scales}

\item{{\bf Stefan GOTTL\"OBER} (Astrophysikalisches Institut Potsdam, Potsdam, Germany)\\
Constrained Local UniversE Simulations (CLUES)}

\item{{\bf Eiichiro KOMATSU} (Univ of Texas, Dept of Astronomy, Austin, USA)\\
The WMAP 7-years Results: Cosmological Interpretation}

\item{{\bf Anthony N. LASENBY} (Cavendish Laboratory, Cambridge University, UK)\\
The CMB in the Standard Model of the Universe: A Status Report}

\item{{\bf Stacy S. McGAUGH} (Astronomy Dept, Univ of Maryland, College Park, MD, USA)\\
The Baryon content of Cosmic Structures and its relation to Dark Matter}

\item{{\bf Reno MANDOLESI (INAF-IASF, Bologna, Italy) \& Paolo NATOLI} (Univ. Roma
Due Tor Vergata and ASI Science Data Center, Frascati, Italy)\\
Measurements of the CMB by the PLANCK satellite and their Implications}

\item{{\bf F\'elix MIRABEL} (CEA-Saclay, France \& IAFE-Buenos Aires, Argentina)\\
Cosmic evolution of stellar black holes and the end of the dark ages}

\item{{\bf Rafael REBOLO} (Instituto Astrofisico de Canarias, Tenerife, Spain)\\
CMB Polarization: The QUIJOTE CMB Experiment}

\item{{\bf Bernard SADOULET} (Particle Cosmology Group, Univ of California, Berkeley, USA)\\
Status Report on Dark Matter Searches}

\item{{\bf Paolo SALUCCI} (SISSA-Astrophysics, Trieste, Italy)\\
Universality Properties in Galaxies and Cored Density Profiles}

\item{{\bf Norma G. SANCHEZ} (CNRS LERMA Observatoire de Paris, Paris, France)\\
Predictions of the Effective Theory of Inflation and keV 
Dark Matter in the Standard Model of the Universe}
 
\item{{\bf Anton TIKHONOV} (Saint-Petersburg State Univ, RUSSIA)\\
 Sizes of minivoids and Tully-Fisher relation in the Local Volume: another 
$\Lambda$CDM-overabundance problem and its possible solutions} 

\end{itemize}

\newpage

\section{Highlights by the Lecturers}

\begin{center}
  
More  informations on the Colloquium Lectures  are at: 

\bigskip

{\bf http://www.chalonge.obspm.fr/colloque2010.html}

\end{center}

\subsection{Peter Biermann 1,2,3,4,5 }

\begin{center}
{{\bf The nature of dark matter}}\\

\bigskip

P.L. Biermann 1,2,3,4,5 with help from J.K. Becker$^{6}$, L. Caramete$^{1,7}$, L. Clavelli$^{3}$, J. Dreyer$^{6}$, B. Harms$^{3}$, A. Meli$^{8}$, E.-S. Seo$^{9}$, \& T. Stanev$^{10}$\\

\vskip0.5cm

$^{1}$ MPI for Radioastronomy, Bonn, Germany; 
$^{2}$ Dept. of Phys., Karlsruher Institut f{\"u}r Technologie KIT, Germany, 
$^{3}$ Dept. of Phys. \& Astr., Univ. of Alabama, Tuscaloosa, AL, USA; 
$^{4}$ Dept. of Phys., Univ. of Alabama at Huntsville, AL, USA; 
$^{5}$ Dept. of Phys. \& Astron., Univ. of Bonn, Germany ; 
$^{6}$ Dept. of Phys., Univ. Bochum, Bochum, Germany; 
$^{7}$ Institute for Space Sciences, Bucharest, Romania; 
$^{8}$ ECAP, Physik. Inst. Univ. Erlangen-N{\"u}rnberg, Germany; 
$^{9}$ Dept. of Physics, Univ. of Maryland, College Park, MD, USA; 
$^{10}$ Bartol Research Inst., Univ. of Delaware, Newark, DE, USA\\
\end{center}

\bigskip

Dark matter has been detected since 1933 (Zwicky) and basically behaves like a 
non-EM-interacting gravitational gas of particles.  From particle physics Supersymmetry 
suggests with an elegant argument that there should be a lightest supersymmetric particle, 
which is a dark matter candidate, possibly visible via decay in odd properties of energetic 
particles and photons:  

\bigskip

Observations have discovered: (i) an upturn in the CR-positron fraction 
(Pamela: Adriani et al. 2009 Nature), (ii) an upturn in the CR-electron spectrum 
(ATIC: Chang et al. 2008 Nature; Fermi: Aharonian et al. 2009 AA), (iii) a flat radio 
emission component near the Galactic Center (WMAP haze: Dobler \& Finkbeiner 2008 ApJ), 
(iv) a corresponding IC component in gamma rays (Fermi haze: Dobler et al. 2010 ApJ, 
Su et al. 2010 arXiv), (v) the 511 keV annihilation line also near the Galactic Center 
(Integral: Weidenspointner et al. 2008 NewAR), and most recently, (vi) an upturn in the
 CR-spectra of all elements from Helium (CREAM: Ahn et al. 2009 ApJ, 2010 ApJL; 
for H and He the upturn has been confirmed by Pamela, shown at the COSPAR meeting July 2010).  

\bigskip

All these features can be quantitatively explained with the action of cosmic rays 
accelerated in the magnetic winds of very massive stars, when they explode 
(Biermann et al. 2009 PRL, 2010 ApJL), based on well-defined predictions from 1993 
(Biermann 1993 AA, Biermann \& Cassinelli 1993 AA, Biermann \& Strom 1993 AA, Stanev 
et al 1993 AA).  

\bigskip

While the leptonic part of these observations may be explainable with pulsars and their winds, 
the hadronic part clearly needs very massive stars, such as Wolf-Rayet stars, their winds and
 their explosions. What the cosmic ray work (Biermann et al., from 1993 through 2010) shows, 
that allowing for the magnetic field topology of Wolf Rayet star winds 
(see, e.g. Parker 1958 ApJ), both the leptonic and the hadronic part get readily and 
quantitatively explained, close to what had been predicted, without any significant
 free parameter, so by Occam's razor the Wolf-Rayet star wind proposal is much simpler.    

\bigskip

This allows to go back to galaxy data to derive the key properties of the dark matter 
particle: Work by Hogan \& Dalcanton (2000 PRD, 2001 ApJ), Gilmore et al. 
(from 2006 MNRAS, 2007 ApJ, etc.), Strigari et al. (2008 Nature), Gentile et al. 
(2009 Nature); work by Boyanovsky et al. (2008 PRD), de Vega \& Sanchez (2010 MNRAS) 
clearly points to a keV particle. 

A right-handed neutrino is a candidate to be this particle (e.g. Kusenko \& Segre 1997 PLB; 
Fuller et al. 2003 PRD; Kusenko 2004 IJMP; for a review see Kusenko 2009 PhysRep; 
Biermann \& Kusenko 2006 PRL; Stasielak et al. 2007 ApJ; Loewenstein et al. 2009 ApJ; 
Loewenstein \& Kusenko 2010 ApJL): 
 
\bigskip
 
This particle has the advantage to allow star formation very early, near redshift 80, 
and so also allows the formation of supermassive black holes, possibly formed out of 
agglomerating massive stars in the gravitational potential of a dark matter clump; the 
stellar wind limit derived by Yungelson et al. 2008 AA does not apply for stars at near 
zero heavy elements, since such stars have weak winds.  Black holes in turn also merge, 
but in this manner start their mergers at masses of a few million solar masses; the mass 
is given by the instability of stars at such a mass due to General Relativity and radiation 
effects.  This readily explains the supermassive black hole mass function as the result of 
mergers between black holes.  The corresponding gravitational waves are not constrained 
by any existing limit, and could have given a substantial energy contribution at high redshift.  

\bigskip

Our conclusion is that a right-handed neutrino of a mass of a few keV is the most 
interesting candidate to constitute dark matter.  

\bigskip

A consequence should be Lyman alpha emission and absorption at around a few microns; 
corresponding emission and absorption lines might be visible from molecular Hydrogen H$_2$ 
(Tegmark et al. 1997 ApJ) and H$_3$ (Goto et al. 2008 ApJ) and their ions, in the far 
infrared and sub-mm wavelength range.  

\bigskip

The detection at very high redshift of massive star formation, stellar evolution and the 
formation of the first super-massive black holes would constitute the most striking and 
testable prediction of this specific dark matter particle proposal.

\newpage

\subsection{Daniel Boyanovsky}

\vskip -0.4cm

\begin{center}

Physics \& Astronomy, University of Pittsburgh, Pittsburgh, PA 15260.

\bigskip

{\bf The case for sterile neutrinos as warm dark matter candidates.} 

\end{center}


\vspace{1mm}


Sterile neutrinos with mass in the $ \sim $ keV range are suitable warm dark matter 
candidates that may help solve some possible small scale problems of the $\Lambda$ CDM
concordance model. I review some of the non-resonant production mechanisms and analyze their 
transfer function and power spectra at small scales.

\vspace{1mm}

.....................................................................................................................................................................

\vspace{1mm}

 In the \emph{concordance} $\Lambda$CDM standard cosmological
model  dark matter (DM) is composed of primordial particles which are
cold and collisionless.  In this  cold dark matter (CDM) scenario  particles
feature negligible small velocity dispersion  leading to a power spectrum that favors small
scales. Structure formation proceeds in a hierarchical ``bottom up''
approach: small scales become non-linear and collapse first and
their merger and accretion leads to structure on larger scales, dense clumps that survive the
merger process form satellite galaxies.

\medskip

Large scale simulations seemingly yield  an over-prediction of
satellite galaxies [1]. Simulations
within the $\Lambda$CDM paradigm also yield a density profile in
virialized (DM) halos that increases monotonically towards the
center and features a cusp, such as the
Navarro-Frenk-White (NFW) profile [2]. These density
profiles accurately describe clusters of galaxies but there is an
accumulating body of observational evidence [3,4] that suggest that the 
central regions of (DM)-dominated
dwarf spheroidal satellite (dSphs) galaxies
 feature smooth cores instead of cusps as predicted by (CDM).

\medskip

 Salucci et. al. [5] reported that the mass distribution of spiral disk galaxies 
can be best fit by a cored Burkert-type  profile. Warm dark matter (WDM) particles were
invoked [6] as possible solutions to these
discrepancies. A model independent analysis suggests that dark matter
particles with a mass in the keV range is a suitable (WDM)
candidate [7,8], and sterile   neutrinos with masses in the 
$ \sim $ keV range are compelling
(WDM) candidates [9,10].
These neutrinos can decay into an active-like neutrino and an X-ray
photon , and recent  astrophysical evidence in favor of   a
$ 5~ $ keV line has been presented in ref. [11].
Abundance and phase space density of dwarf spheroidal galaxies constrain the
mass to be in the $ \sim $ keV  range [8].

\medskip

In ref. [13] we analyze the small scale aspects of sterile neutrinos with mass 
in the keV range produced by two different non-resonant mechanisms: sterile-active
 mixing or Dodelson-Widrow [9] (DW) and by the decay of vector or scalar bosons at 
the EW scale (BD) [12]
The transfer function and power spectra are obtained by solving the collisionless 
Boltzmann equation during radiation and matter domination. There are three stages 
in the evolution: (i) (RD) when the WDM particle is relativistic, (ii) (RD) 
when the particle is non-relativistic and (iii) (MD). 
During stages (i) and (ii) the gravitational potential is dominated by the
acoustic oscillations in the radiation fluid. The evolution throughout these 
two stages determine the initial condition to stage (iii) during which WDM 
density perturbations dominate the gravitational potential and the Boltzmann-Poisson 
equation can be related to a fluid like equation. 
The power spectra features WDM acoustic
oscillations on mass scales $ \sim 10^8-10^9 \, M_{\odot} $. Details are available in ref. [13].

\medskip

There are two relevant scales: $ k_{eq} \sim 0.01\,(\mathrm{Mpc})^{-1} $ 
which is the wavevector of modes that enter the Hubble radius at matter-radiation 
equality, and the free streaming scale 
$ k_{fs} = \sqrt3 \,k_{eq}/[2\,\langle V^2_{eq}\rangle^{\frac12}]  $ 
where $ \langle V^2_{eq}\rangle^\frac12 $ is the mean square root velocity dispersion 
of the WDM particle at \emph{matter-radiation equality}. For a WDM candidate with 
$ m \sim $ keV  produced non-resonantly and    decoupling either at the 
electroweak or QCD scale $ k_{fs} \gtrsim 10^3 \,k_{eq} $. The free streaming length 
scale $ 1/k_{fs} $ is proportional to the distance traveled by a  non-relativistic 
particle with average velocity $ \langle V^2_{eq}\rangle^\frac12 $ from matter-radiation 
equality until today, \emph{and} it also determines the size of the (comoving) horizon 
(conformal time) when the WDM particle transitions from relativistic to non-relativistic: 
$ \eta_{NR} = \sqrt3/[\sqrt2 \, k_{fs} ] $.
This means that perturbations with $ k > k_{fs} $ enter the horizon when the WDM 
particle is still relativistic and undergo suppression by relativistic free 
streaming between the time of horizon entry until $ \eta_{NR} $.  

\begin{figure}[ht]
\begin{center}
\includegraphics[height=7cm,width=13cm,keepaspectratio=true]{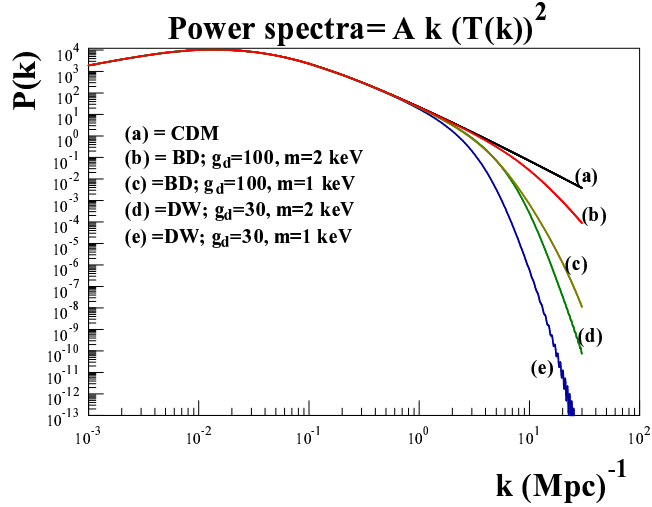}
\caption{The matter power spectra: $P(k)= A\,k \, (T(k))^2$ for
$n_s=1$ (A is the normalization amplitude) for CDM, DW and BD for
$m=1,2 \mathrm{keV}$. Note the quasi degeneracy for DW with
$m=2\mathrm{keV}$ (d) and BD with $m=1\mathrm{keV}$ (c) in a large range of
$k\lesssim 12\,(\mathrm{Mpc})^{-1}$. } \label{fig:powerspectra}
\end{center}
\end{figure}

During the RD era  acoustic oscillations in the radiation fluid 
determine the gravitational potential $ \phi $. The time dependence of $\phi$ 
induces an early ISW that results in an \emph{enhancement} of the amplitude of 
WDM density perturbations for wavelengths larger than the sound horizon of the 
radiation fluid at $ \eta_{NR} $, namely $ \eta_{NR}/\sqrt{3}$,  but  those with 
$ k\eta_{NR}/\sqrt{3} \gg 1 $ are suppressed by relativistic free streaming. 

In stage (iii) the Boltzmann-Poisson equation is related to a fluid-like 
equation, whose solutions yield WDM acoustic oscillations.

We obtain a semi-analytic expression for the transfer function and power 
spectrum valid for arbitrary distribution functions , mass and decoupling temperature. 
These are computed for two different scenarios of sterile neutrinos
         produced non-resonantly (DW) and (BD). The transfer functions are very different 
\emph{even for the same mass}. 

\bigskip

[1] B. Moore \emph{et. al.}, Astrophys. J. Lett.
\textbf{524}, L19 (1999); S. Ghigna \emph{et.al.} Astrophys.J. \textbf{544},616
(2000); A. Klypin \emph{et. al.} Astrophys. J.
\textbf{523}, 32 (1999); Astrophys. J. \textbf{522}, 82 (1999).

\medskip

[2] J. F. Navarro, C. S. Frenk, S. White, Mon. Not. R.
Astron. Soc. \textbf{462}, 563 (1996); J. Diemand \emph{et.al.} 
Mon. Not. Roy. Astron. Soc. \textbf{364}, 665
(2005). 

\medskip

[3] R. F.G. Wyse and  G. Gilmore, arXiv:0708.1492; G.
Gilmore \emph{et. al.} arXiv:astro-ph/0703308; G. Gilmore \emph{et.al.}
arXiv:0804.1919 (astro-ph); G. Gilmore, arXiv:astro-ph/0703370.

\medskip

[4] G.Gentile \emph{et.al} Astrophys. J. Lett. \textbf{634},
L145 (2005); G. Gentile \emph{et.al.}, Mon. Not. Roy. Astron. Soc.
\textbf{351}, 903 (2004); V.G. J.  De Blok \emph{et.al.} Mon. Not.
Roy. Astron. Soc. \textbf{340}, 657 (2003), G. Gentile
\emph{et.al.},arXiv:astro-ph/0701550;  P. Salucci, A. Sinibaldi,
Astron. Astrophys. \textbf{323}, 1 (1997).

\medskip

[5] P. Salucci et al. MNRAS \textbf{378}, 41 (2007); P. Salucci, arXiv:0707.4370.

\medskip

[6] B. Moore, \emph{et.al.} Mon. Not. Roy.
Astron. Soc. \textbf{310}, 1147 (1999);  
P. Bode, J. P. Ostriker, N. Turok, Astrophys. J \textbf{556}, 93
(2001); V. Avila-Reese \emph{et.al.} Astrophys. J. \textbf{559}, 516
(2001).

\medskip

[7] H. J. de Vega, N. Sanchez, 
Mon. Not. R. Astron. Soc. \textbf{404}, 885 (2010); arXiv:0907.0006; 
H. J. de Vega, P. Salucci, N. G. Sanchez,      arXiv:1004.1908 .

\medskip

[8] D. Boyanovsky, H. J. de Vega, N. G. Sanchez,
Phys. Rev. D 78, 063546 (2008); D. Boyanovsky, H. J. de Vega, N. G. Sanchez,
arXiv:0710.5180, Phys. Rev.  D 77, 043518 (2008).

\medskip

[9] S. Dodelson, L. M. Widrow, Phys. Rev. Lett. \textbf{72}, 17 (1994).

\medskip

[10] A. Kusenko, arXiv:hep-ph/0703116;     Int.J.Mod.Phys.\textbf{D16},2325
(2008); T. Asaka, M. Shaposhnikov, A. Kusenko; Phys.Lett.
\textbf{B638}, 401 (2006);  P. L. Biermann, A. Kusenko, Phys. Rev.
Lett. \textbf{96}, 091301 (2006); A. Kusenko, Phys. Rev. Lett. \textbf{97}, 241301 (2006);  
K. Petraki, A. Kusenko, Phys. Rev. \textbf{D 77}, 065014
(2008); K. Petraki, Phys. Rev. \textbf{D 77}, 105004 (2008); 
A. Kusenko, Phys.Rept.\textbf{481}, 1 (2009); 
A. Boyarsky, O. Ruchayskiy, M. Shaposhnikov, Ann.Rev.Nucl.Part.Sci.\textbf{59}, 191 (2009).

\medskip

[11] M. Lowenstein, A. Kusenko, Astrophys.J.\textbf{714},652 (2010);  Astrophys.J \textbf{714 },
652 (2010); M. Loewenstein, A. Kusenko, P. L. Biermann;
 Astrophys.J.\textbf{700}, 426 (2009).

\medskip

[12] D. Boyanovsky, Phys.Rev.\textbf{D78}, 103505 (2008); J. Wu, C.-M.Ho, 
D. Boyanovsky, Phys. Rev. \textbf{D80}, 103511 (2009).

\medskip

[13] D. Boyanovsky, J. Wu, arXiv: 1008.0992

\newpage

\subsection{Asantha Cooray}

\vskip -0.3cm

\begin{center}

University of California, Department of Physics and Astronomy, Irvine, CA USA

\bigskip

{\bf  First Results from Herschel Extragalactic Surveys: HerMES and H-ATLAS}

\end{center}

\bigskip

I presented a summary talk focusing on several key results obtained by two consortia that are
primarily using SPIRE instrument for extragalactic observations with the Herschel Space Observatory.

\bigskip

I focused on the results that appear in the Special Issue of Astronomy and Astrophysics journal 
(volume 518)
that contains first results from the Science Demonstration Phase (SDP) data from Herschel.

\medskip

The results I presented include: 

\bigskip

(1) number counts of sub-mm galaxies at 250, 350 and 500 microns
from HerMES (Oliver et al. 2010) and ATLAS \\
\noindent (Clements et al. 2010).

\bigskip

(2) the clustering of the
resolved sources (Cooray et al. 2010 for HerMES and Maddox et al. 2010 for ATLAS), 

\bigskip

(3) the surface density of lensed sub-mm galaxies (from Negrello et al. submitted for ATLAS).

\bigskip

In terms of clustering I showed that we have a marginal detection of the 2-halo to 1-halo transition in
the correlation function of bright SMGs; this sets a mass scale of few times 
10$^{12}$ M$-{\odot}$ for sub-mm sources
assuming that the 350 micron bright sources are mostly at z of 2. These bright sources, 
however, only
account for 15\% of the extragalatic background light. 

\medskip

To understand the nature of faint sources 
(those responsible for the
confusion noise and fainter than the confusion noise), I showed that one can treat the map as a 
map of fluctuating
field similar to maps of the CMB anisotropies and one can directly study the angular power 
spectrum of the map.
A measurement of the unresolved fluctuation power spectrum is described in 
Amblard et al. (2010; submitted)
and I summarized some of the results related to this work.

\medskip

Finally, I outline the science goals of the Herschel-SPIRE Legacy Survey (HSLS), a 
recently proposed open-time program to ESA
with SPIRE on Herschel to cover 4000 square degrees on the sky with SPIRE scanning the sky in 
fast mode with maps made of single scans. The survey is aimed at finding 2.5 to 3 million 
individual sources down to the 50\% completeness level of the
catalogs at 25, 26, and 30 mJy at 250, 350 and 500 microns respectively. 

\medskip

The HSLS maps will be used for a wide variety of
cosmological studies including cross-correlation with the CMB maps to look for 
ISW and CMB lensing signal traced by dusty, starbursts
at redshifts of 1 to 3. Among the 2.5 to 3 million detected sources will be at least 
2000 bright lensed galaxies and more than 10,000
dusty galaxies at redshifts greater than 5.

\bigskip

{\bf References}

\bigskip

A\&A Special Issue on Herschel First Science Results vol 518 (July 2010).

\medskip

S.J. Oliver et al, arXiv:1005.2184,
HerMES: SPIRE galaxy number counts at 250, 350 and 500 microns,
A\&A Special Issue on Herschel First Science Results,
vol 518 (July 2010).

\medskip
 
L. Clements et al, arXiv:1005.2184, 
The Herschel-ATLAS: Extragalactic Number Counts from 250 to 500 Microns,
A\&A Special Issue on Herschel First Science Results,
 vol 518 (July 2010).

\medskip

A. Cooray et al, arXiv:1005.3303, 
HerMES: Halo Occupation Number and Bias Properties of Dusty Galaxies from Angular 
Clustering Measurements, 
A\&A Special Issue on Herschel First Science Results,
 vol 518 (July 2010).

\medskip

S. J. Maddox,arXiv:1005.2406,
Herschel ATLAS: The angular correlation function of submillimetre galaxies at high and low redshift
A\&A Special Issue on Herschel First Science Results,
 vol 518 (July 2010).

\medskip

A. Cooray et al.arXiv:1007.3519, 
The Herschel-SPIRE Legacy Survey (HSLS): the scientific goals of a shallow 
and wide submillimeter imaging survey with SPIRE.

\newpage

\subsection{Claudio Destri}

\vskip -0.3cm

\begin{center}

Dipartimento di Fisica G. Occhialini, Universit\`a
Milano-Bicocca and INFN, sezione di Milano-Bicocca, 

Piazza della Scienza 3,
20126 Milano, Italia. Claudio.Destri@mib.infn.it

\bigskip

{\bf Fast--roll eras, primordial fluctuations \\
  and the lowest CMB multipoles: theory and observations} \\

\end {center}

\bigskip

Cosmic inflation is by now the standard paradigm for the description of the
evolution of the Universe before the Hot Big Bang. Inflation solves the horizon
and flatness problems, provides an explanation for the large entropy of the
observed universe and naturally generates the scalar fluctuations that seed
CMB anisotropies. 

\medskip

Single-field inflation is based on a scalar field (the inflaton) whose fairly
flat potential determines a slow-roll evolution with a sufficiently large number
$ N\gtrsim 60 $ of efolds (to solve the horizon and flatness problems) as well as
a nearly scale-invariant power spectrum of scalar fluctuations. In particular,
the simplest double-well inflaton potential, the so-called Binomial New
Inflation (BNI), provides a stable (in the Ginsburg-Landau sense) realization of the
inflaton paradigm in very good agreement with present CMB observations
of the spectral index $ n_s $ and upper limits on the tensor to scalar ratio $ r $.
[D. Boyanovsky, C. Destri, H.J. de Vega, N. Sanchez, IJMP 24, no.  20-21, 3669
(2009)]. It must be stressed that the model also provides a lower bound on 
$ r $, namely  $r> 0.023 $ to 95\% C.L. quite close to the detetectability 
limits of Planck.
 
\medskip

In this talk the main focus is on the inflaton evolution before the slow-roll
stage. Indeed this attractor stage is generically preceeded by a fast-roll stage
which in turn follows a period of decelerating expansion. The corresponding
evolution of the scale factor, the inflaton and the scalar quantum fluctuations
are studied in detail both semi-analytically and numerically with the BNI
potential. Particular attention is placed on the effect that 
Bunch-Davies initial conditions at finite time
have on the observable CMB spectrum. This effect
is embodied in the transfer function $ D(k) $ which relates the power spectrum
$ P(k) $ as
\begin{equation*}
  P(k) = P_\infty(k)\Big[1+D(k) \Big]  
\end{equation*}
to the standard pure slow-roll spectrum $P_\infty(k)$ obtained when the
Bunch-Davies conditions are imposed in the infinite past. 

Several results for
$ D(k) $ are presented from accurate numerical calculations as well as from
semi-analytic approximations. In particular, it is shown quite generally how
$ 1+D(k) $ vanishes as $ k^2 $ when $ k\to0 $ while $ D(k) $ vanishes as $ 1/k^2 $ times
oscillating terms with zero average when $ k\to\infty $. The latter property is crucial
for a negligible back-reaction on the metric, ensuring the applicability of
semi-classical gravity.

\medskip

Also the important, in principle observable effect of a nonzero  $ D(k) $ on the
angular multipoles $ C_\ell $ is exhibited. When finite-time Bunch-Davies
conditions are imposed at the transition from fast-roll to slow-roll the lowest
CMB multipoles are depressed. Comparison with current data through standard
Monte Carlo Markov chains analysis shows a small likelihood improvement over
the assumption that the low quadrupole is simply a manifestation of cosmic
variance. 

\medskip

The core idea put forward is that the large scale CMB anisotropies may provide
information on the beginning of inflation. Indeed the early fast-roll inflation
is generic and provides a simple mechanism for quadrupole depression, setting
around $64$ the total number of inflation efolds. The near saturation of the
entropy bound may suggest a deep connection between the duration of inflation
and the entropy production at reheating.

\bigskip

\noindent
References:\\

\noindent
D. Boyanovsky, C. Destri, H. J. de Vega, N. Sanchez, ``The Effective Theory of
Inflation in the Standard Model of the Universe and the CMP + LSS Data
Analysis'', Int. J. Mod. Phys. {\bf A 24}, 3669-3864 (2009)
and author's references therein.\\

\noindent
C. Destri, H. J. de Vega and N. G. Sanchez, ``Preinflationary and
  inflationary fast-roll eras and their signatures in the low CMB
  multipoles'', Phys. Rev. D 81, 063520 (2010).

\newpage

\subsection{Hector J. de Vega and Norma G. Sanchez}

\vskip -0.3cm

\begin{center}

H.J.dV: LPTHE, CNRS/Universit\'e Paris VI-P. \& M. Curie \& Observatoire de Paris, Paris, France\\
N.G.S: LERMA, CNRS/Observatoire de Paris, Paris, France

\medskip

{\bf Predictions of the Effective Theory of Inflation in the Standard Model of the 
Universe and the CMB+LSS data analysis}

\end {center}

\medskip

Inflation is today a part of the Standard Model of the Universe supported by
the cosmic microwave background (CMB) and large scale structure (LSS)
datasets. Inflation solves the horizon and flatness problems and
naturally generates  density fluctuations that seed LSS and CMB anisotropies,
and tensor perturbations (primordial gravitational waves). 
Inflation theory is
based on a scalar field  $ \varphi $ (the inflaton) whose potential
is fairly flat leading to a slow-roll evolution. 

\medskip

We focuse here on the following new aspects of inflation. We present the
effective theory of inflation \`a la {\bf Ginsburg-Landau} in which
the inflaton potential is a polynomial in the field $ \varphi $ and has
the universal form $ V(\varphi) = N \; M^4 \;
w(\varphi/[\sqrt{N}\; M_{Pl}]) $, where $ w = {\cal O}(1) , \;
M \ll M_{Pl} $ is the scale of inflation and  $ N \sim 60 $ is the number
of efolds since the cosmologically relevant modes exit the horizon till
inflation ends. The slow-roll expansion becomes a systematic $ 1/N $ expansion and
the inflaton couplings become {\bf naturally small} as powers of the ratio
$ (M / M_{Pl})^2 $. The spectral index and the ratio of tensor/scalar
fluctuations are $ n_s - 1 = {\cal O}(1/N), \; r = {\cal O}(1/N) $ while
the running index turns to be $ d n_s/d \ln k =  {\cal O}(1/N^2) $
and therefore can be neglected. The {\bf energy scale of inflation }$ M \sim 0.7
\times 10^{16}$ GeV turns to be completely determined by the amplitude of the scalar 
adiabatic fluctuations [1-2]. 

\bigskip

\begin{figure}[ht]
\includegraphics[height=6cm,width=10cm]{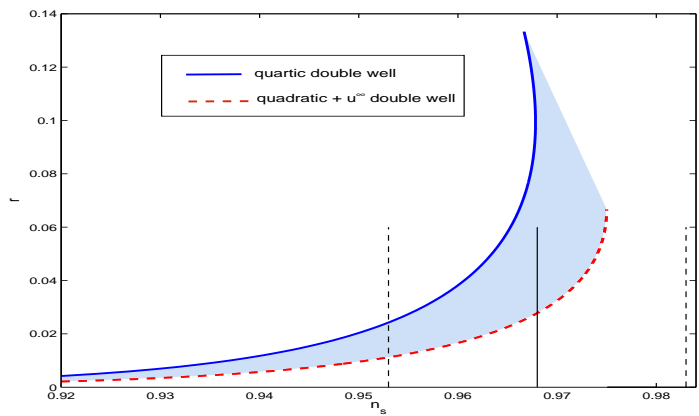}
\caption{The universal banana region $ \cal B $ in the $ (n_s, r) $-plane
  setting $ N = 60 $. The upper border of the region $ \cal B $ corresponds to
  the fourth order double--well potential (new inflation). The lower border is
  described by the potential $ V(\varphi) = \frac12{ m^2} \,
  \left(\frac{m^2}{\lambda} - \varphi^2\right) $ for $ \varphi^2 < m^2/\lambda $
  and $ V(\varphi) = \infty $ for $ \varphi^2 > m^2/\lambda $ [4].  We
  display in the vertical full line the observed value $ n_s = 0.968 \pm
  0.015 $ using the WMAP+BAO+SN data set.  The broken vertical lines delimit
  the $ \pm 1 \, \sigma$ region.}
\label{banana}
\end{figure}

A complete analytic study plus the
Monte Carlo Markov Chains (MCMC) analysis of the available
CMB+LSS data (including WMAP5) with fourth degree trinomial potentials
showed [1-3]: 

\begin {itemize}

\item {{\bf(a)} the {\bf spontaneous breaking} of the
$ \varphi \to - \varphi $ symmetry of the inflaton potential.} 

\medskip

\item {{\bf(b)} a {\bf lower bound} for $ r $ in new inflation:
$ r > 0.023 \; (95\% \; {\rm CL}) $ and $ r > 0.046 \;  (68\% \;
{\rm CL}) $. }

\medskip

\item {{\bf(c)} The preferred inflation potential is a {\bf double
well}, even function of the field with a moderate quartic coupling
yielding as most probable values: $ n_s \simeq 0.964 ,\; r\simeq
0.051 $. This value for $ r $ is within reach of forthcoming CMB
observations. }

\medskip

\item {{\bf(d)} The present data in the effective theory of
inflation clearly {\bf prefer new inflation}. }

\medskip

\item { {\bf(e)} Study of higher degree
inflaton potentials show that terms of degree higher than four do not
affect the fit in a significant way. In addition, horizon exit happens for
$ \varphi/[\sqrt{N} \; M_{Pl}] \sim 0.9 $ making higher order terms
in the potential $ w $ negligible [4]. }

\item { {\bf(f)} Within the Ginsburg-Landau potentials in new inflation,
$ n_s$ and  $r$ in the $(n_s, r)$ plane are within the universal banana region 
fig. \ref{banana} and $ r $ is in the range $ 0.021 < r < 0.053 $ [4].}

\end {itemize}


We summarize the physical effects of
{\bf generic} initial conditions (different from Bunch-Davies) on the
scalar and tensor perturbations during slow-roll and
introduce the transfer function $ D(k) $ which encodes the observable
initial conditions effects on the power spectra.
These effects are more prominent in the \emph{low}
CMB multipoles: a change in the initial conditions during slow roll can
account for the observed CMB {\bf quadrupole suppression} [1].

\medskip

Slow-roll inflation is generically preceded by a
short {\bf fast-roll} stage. Bunch-Davies initial conditions are the
natural initial conditions for the fast-roll perturbations. During
fast-roll, the potential in the wave equations of curvature and tensor
perturbations is purely attractive and leads to a suppression of the
curvature and tensor CMB quadrupoles [1]. 

\medskip

A MCMC analysis of the WMAP+SDSS data {\bf including fast-roll} shows that the quadrupole
mode exits the horizon about 0.2 efold before fast-roll ends and its
amplitude gets suppressed. In addition, fast-roll fixes the {\bf initial
inflation redshift} to be $ z_{init} = 0.9 \times 10^{56} $ and
the {\bf total number} of efolds of inflation to be $ N_{tot} \simeq 64 $ [1,3].
Fast-roll fits the TT, the TE and the EE
modes well reproducing the quadrupole supression. 

\medskip

A thorough study of the
{\bf quantum loop corrections} reveals that they are very small and controlled by
powers of $(H /M_{Pl})^2 \sim {10}^{-9} $, {\bf a conclusion that validates the
reliability of the effective theory of inflation [1].} 

\bigskip

This work [1-4] shows how powerful is
the Ginsburg-Landau effective theory of inflation in predicting
observables that are being or will soon be contrasted to observations.

\bigskip

The Planck satellite is right now measuring with unprecedented accuracy 
the primary CMB anisotropies.
The Standard Model of the Universe (including inflation) provides the context to analyze the CMB 
and other data. The Planck performance for $ r $ related to the primordial $ B $ 
mode polarization, will depend on the quality of the data analysis.

\medskip

The Ginsburg Landau approach to inflation allows to take high benefit of the CMB data.
We evaluate the Planck precision to the recovery of cosmological
parameters within a reasonable toy model for residuals of systematic
effects of instrumental and astrophysical origin based on publicly
available information.
We use and test two relevant models: the $\Lambda$CDM$r$ model,
i. e. the standard $\Lambda$CDM model augmented by  $ r $, and the
$\Lambda$CDM$r$T model, where the scalar spectral index, $ n_s $, and $ r $
are related through the theoretical `banana-shaped' curve $ r = r(n_s) $
coming from the double-well inflaton potential (upper boundary of the banana region
fig. \ref{banana}. In the latter case,
$ r = r(n_s) $ is imposed as a hard constraint in the MCMC data
analysis. We take into account the white noise sensitivity of Planck in the 70,
100 and 143 GHz channels as well as the residuals from systematics
errors and foregrounds. Foreground residuals turn to affect
only the cosmological parameters sensitive to the B modes [5].

\medskip

In the Ginsburg-Landau inflation approach, better measurements on $ n_s $, 
as well as on TE and EE modes will improve 
the prediction on $ r $ even if a detection of B modes is still
lacking [5].

\medskip

Forecasted B mode detection probability by
the most sensitive HFI-143 channel:
At the level of foreground residual equal to
30\% of our toy model, only a 68\% CL detectiof $ r $ is very likely.
For a 95\% CL detection the level of
foreground residual should be reduced to 10\%
or lower of the adopted toy model. The possibility to detect
$ r $ is borderline [5].

\begin{figure}[ht]
  \includegraphics[height=7cm]{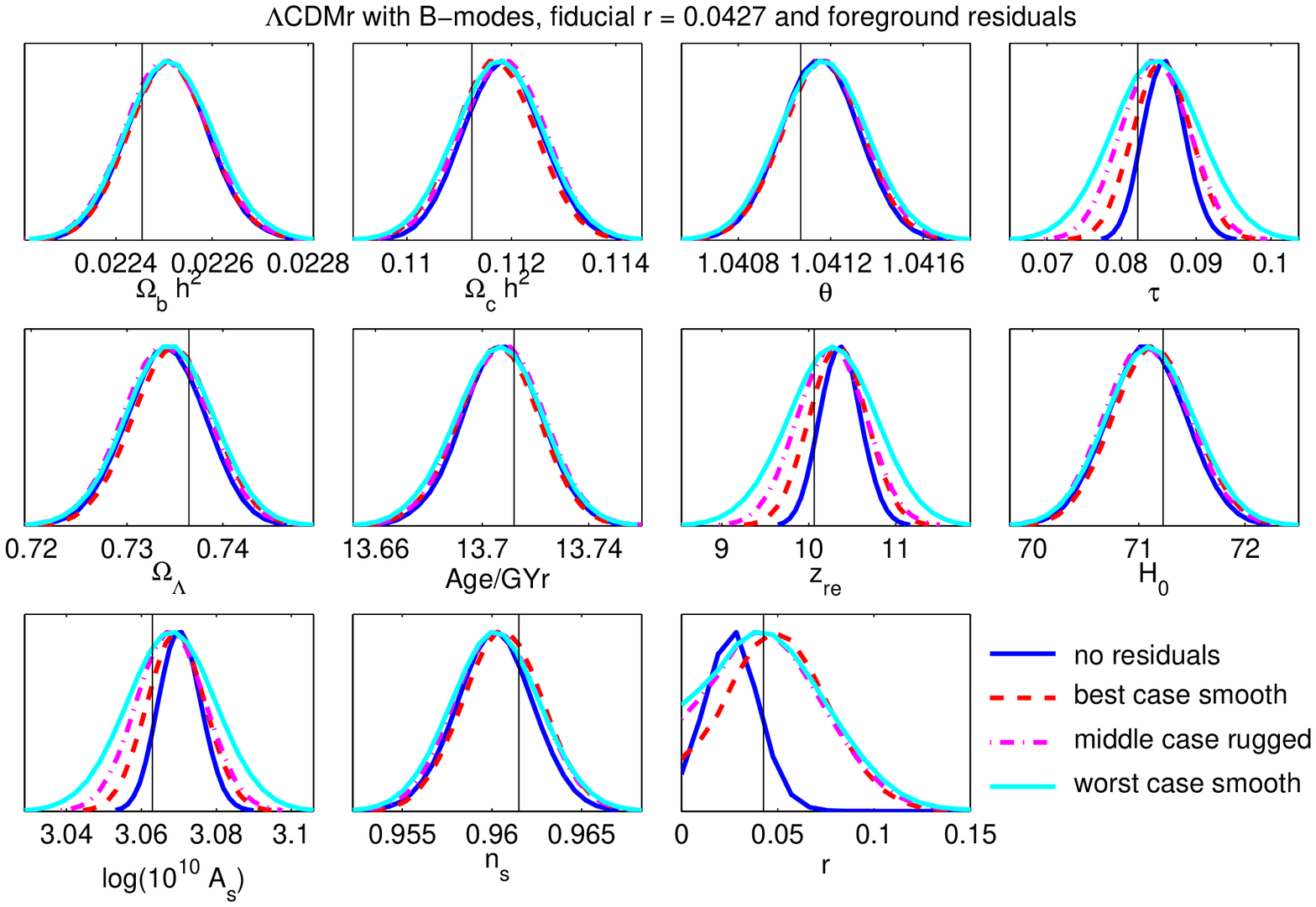}\\
  \includegraphics[height=7cm]{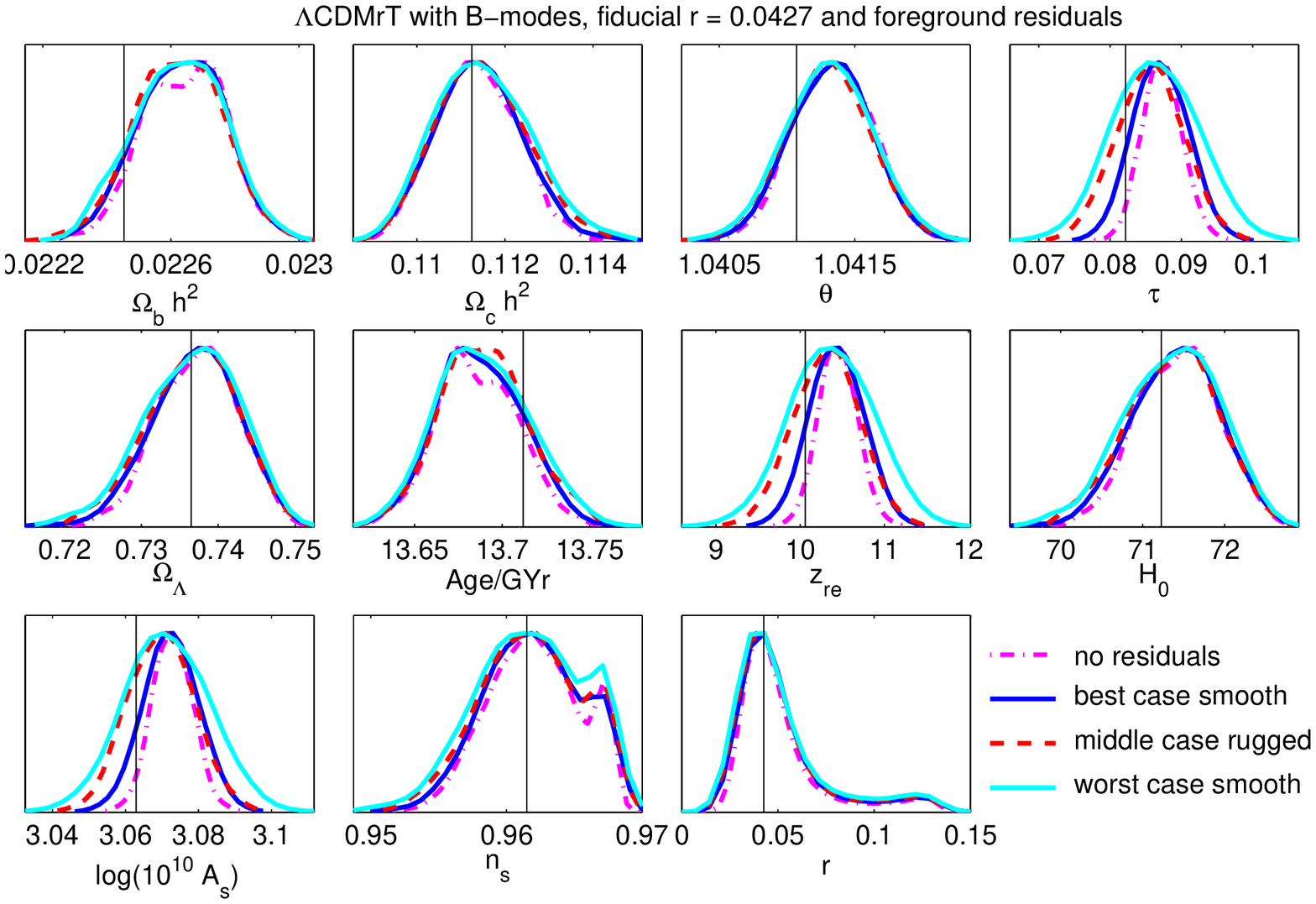}
  \caption{Forecasts for Planck [5]. 
Upper panel: Cumulative $3-$channel marginalized likelihood distributions,
    including $ B $ modes and foreground residuals, of the cosmological
parameters for the $\Lambda$CDM$r$ model.  The fiducial ratio is 
$ r = 0.0427 $. Lower panel:
Cumulative marginalized likelihoods from the three channels for the
    cosmological parameters for the $\Lambda$CDM$r$T model including $ B $ modes
    and fiducial ratio $ r = 0.0427 $ and the foreground residuals. 
We plot the
distributions  in four cases: (a) without residuals, (b) best case smooth:
with 30\% of the toy model residuals in the $TE$ and $E$ modes.
(c) worst case smooth: with the toy model residuals in the $TE$ and $E$ modes.
(d) with 65\% of the toy model 
residuals in the $TE$ and $E$ modes 
and $ 88 \mu K^2 $ in the $T$ modes rugged by 
Gaussian fluctuations of $ 30 \% $ relative strength. }
  \label{ufa7}
\end{figure}

\medskip

{\bf References}

\medskip

[1]  Review article: D. Boyanovsky, C. Destri, H. J. de Vega, N. G. Sanchez

Int. J. Mod. Phys. A24, 3669-3864 (2009) and  author's references therein.

\medskip

[2] C. Destri, H. J. de Vega, N. Sanchez,
Phys. Rev. D77, 043509 (2008), astro-ph/0703417.

\medskip

[3] C. Destri, H. J. de Vega, N. G. Sanchez, arXiv:0804.2387.
Phys. Rev. D 78, 023013 (2008).

\medskip

[4] C. Destri, H. J. de Vega, N. G. Sanchez, arXiv:0906.4102.\\
\noindent D. Boyanovsky, H. J. de Vega, C. M. Ho et N. G. Sanchez, 
Phys. Rev.  D75, 123504 (2007).

\medskip

[5] C. Burigana, C. Destri, H. J. de Vega, A. Gruppuso, N. Mandolesi, P. Natoli, N. G. Sanchez,
arXiv:1003.6108, to apear in ApJ.

\newpage

$ {} {} $

\newpage

\subsection{Gerard Gilmore}

\vskip -0.2cm

\begin{center}

Institute of Astronomy, Cambridge

\bigskip

{\large\bf Dark Matter, Feedback, and the lowest luminosity galaxies}

\end{center}

\medskip

The lowest luminosity dwarf galaxies are extremely valuable probes of
dark matter on small scales, formation of the first bound systems, and
the reality of feedback processes. 
Dwarf galaxies are extremely dark-matter dominated, so that detailed
stellar dynamics analyses are our only viable method to quantify the
actual spatial distribution of matter on small scales. It has long
been of interest that dense cusps are predicted by the simplest
versions of LCDM, while observers find only cores in nature. Is a more
sophisticated LCDM model required, or is astrophysical feedback
confusing everything? In either case current LCDM models need
substantial improvement, so what information can we find to guide
those improvements?

\medskip

How cuspy is CDM? We have been developing for some time the first
full robust distribution function analysis of stellar kinematics in
dSph, in order to determine the true dark matter distribution. The
methodology is a development of that used in the only dynamical
determination of dark matter near the Sun (Kuijken \& Gilmore
1989,1991). It is based on our new studies of several thousand
precision velocities in dSph galaxies, with sophisticated MCMC-based
comparison of models and data. After some years of effort, the method
is in place, and data are being analysed.

\medskip

How massive are the first halos? A popular approach to understanding
the `satellite problem', which is a 3 orders of magnitude discrepancy
between the predicted number of bound satellites and the numbers
observed, is that all galaxies below some threshold fail to form
satrs, while all above it are almost entirely disrupted by
`feedback', and so are invisible. While this concept (due to Joe
Silk long ago) must be true, what is the mass and mass range? Some
recent analyses assert that al dSph have a common mass, over many
magnitudes in luminosity. Is this some `threshold' mass? We are making
progress in two ways here. First, the claimed masses are not based on
observations, but are model extrapolations from central
dispersions. Data show a correlation between velocity dispersion and
radius of measurement, which is the information to explain. Such a
relation applies in any common mass profile. Second, we have developed
improved observational methods which show almost all kinematic data
for the very faint dSph are of inadequate accuracy to resolve teh true
kinematics. New studies are in preparation which are quite
inconsistent with current popular models. Mass ranges are broader than
supposed, feedback does not operate as has been suggested. 

\medskip

How violent is feedback? One of the longest-known challenges in galaxy
formation models is the very considerable difference between the
shape of the galaxy luminosity function and the featureless power-law
mass function predicted by the simplest LCDM models. Among predictions
are that Milky Way-like galay halos are debris from the previously
very common dSph parents. A second critical process is feedback - many
low-mass galaxies are severely affected by astrophysical feedback
early. What that feedback is remains to be discovered.
However, the stellar populations in dSph galaxies are
now well-established to be quite different than stellar populations in
the Milky Way Halo. The Milky Way is not formed from dSph-like systems.
So what is wrong with feedback models? And what is the feedback?
Very substantial progress is being made in quantifying feedback, from
detailed analysis of the chemical distributions and star formation
histories of the dSph. We have established (Norris, Wyse, Gilmore etal
2010 arXiv1008.0137) that the lowest luminosity galaxies host very
broad chemical abundace dispersions, requiring that they continued
star formation from near zero abundance through several enrichment
generations. That is, feedback was extremely mild. In at least these
systems, feedback does not substantially modify initial
conditions. There is room for very considerable improvement in the
models to agree with observations.

\medskip

The next generation of galaxy formation models at low masses will be
able to be based on data. It will be interesting to see what they
predict.

\medskip

{\bf References:}

\medskip

Kuijken \& Gilmore 1989,1991

\medskip

Norris, Wyse, Gilmore et al. 2010 arXiv1008.0137

\newpage

\subsection{Stefan Gottl\"ober}

\vskip -0.2cm

\begin{center}

 Astrophysikalisches Institut Potsdam, Germany

\bigskip

{\large {\bf C}onstrained {\bf L}ocal {\bf U}nivers{\bf E} 
{\bf S}imulations}  

\medskip

The CLUES-project

\end{center}

\medskip

During the last decade cosmological parameters have been determined to
a precision of just a few percent giving rise to the standard model of
cosmology: a flat Friedmann universe whose mass-energy content is
dominated by a cosmological constant, a cold dark matter component and
baryons.  The basic paradigm of structure formation suggests that dark
matter forms halos, within which galaxy formation takes place via
complex baryonic physics.  The dark matter halos grow via the process
of accretion and mergers and galaxy formation proceeds by the combined
action of clumpy and anisotropic gas accretion and mergers with dwarf
galaxies.  On small scales galaxy formation can be observed and
analysed best in the local universe, resulting in the so-called
near-field cosmology. This motivated cosmologists to turn their
attention and study the archaeology of the Local Group in their quest
for understanding galaxy formation. This also motivated us to simulate
the formation of the Local Group in the most realistic possible
way. Within the CLUES project ({\tt http://clues-project.org} ---
Constrained Local UniversE Simulations) we perform numerical
simulations of the evolution of the local universe. The simulations
reproduce the local cosmic web and its key players, such as the Local
Supercluster, the Virgo cluster, the Coma cluster, the Great Attractor
and the Perseus-Pisces supercluster. For these simulations we
construct initial conditions based on observational data of the galaxy
distribution in the local universe. The implementation of the
algorithm of constraining Gaussian random fields
[2] to observational data, the used observational
data and the constrained simulations have been described in [1].

\medskip

In total we have run 5 big collisionless (N-body only) simulations
with 1 billion particles each (i.e. $1024^3$). Two of these
simulations correspond to one realisation with WMAP3 cosmological
parameters, that was done both assuming cold and warm dark matter and
the other three are CDM realisation with WMAP5 cosmological
parameters.  The second set of simulations we performed corresponds to
high resolution re-simulations of the Local Group. These simulations
were performed with dark matter only as well as with dark matter, gas
dynamics, cooling, star formation and supernovae feedback (both in the
CDM and WDM scenario).  All the simulations are described on the web
page of the CLUES project.

\medskip

In order to test the effects of different dark matter candidates on
the formation of the local group, we have performed constrained
simulations assuming that the dark matter is made of a warm candidate
with $ m_{\rm WDM}=1 $ keV.  Due to the large thermal velocities of the
WDM particles power is removed from short scales of the fluctuation
spectrum: the free-streaming length is $ 350 h $ kpc.  We have chosen
this extreme case to study the maximal possible effect of the warm
dark matter on the local structure of the universe. To this end we
have compared our predicted galaxy distribution in the local universe
with the observed one in the ALFALFA survey. We found that our
predictions agree well for the $\Lambda$WDM cosmogony. On the
contrary, the $\Lambda$CDM model predicts a steep rise in the velocity
function towards low velocities and thus forecasts much  more
sources both in Virgo-direction as well as in anti-Virgo-direction
than the ones observed by the ALFALFA survey. These results indicate a
potential problem for the cold dark matter paradigm [4]. 
Also the spectrum of mini-voids points to a
potential problem of the $\Lambda$CDM model. The $\Lambda$WDM model
provides a natural solution to this problem, however, the late
formation of halos in the $\Lambda$WDM model might be a problem for
galaxy formation [3].

\medskip

Constrained simulations are a very useful tool to study the formation
and evolution of our Local Group in the right cosmological environment
and the best possible way to make a direct comparison between
numerical results and observations, minimising the effects of the
cosmic variance.

\bigskip

[1] S. Gottl\"ober,   Y. Hoffman,  G. Yepes, 2010,
Proceedings of "High Performance Computing in Science and Engineering, 
Garching/Munich 2009", Springer-Verlag, (astro-ph/1005.2687)

\medskip

[2] Y. Hoffman,  E. Ribak,  1991, ApJL, 380, L5

\medskip

[3] A.V. Tikhonov, S. Gottl\"ober, G. Yepes, Y. Hoffman, 2009,
MNRAS, 399, 1611

\medskip

[4] J. Zavala, Y. P. Jing, A. Faltenbacher, G. Yepes, Y. Hoffman,
S. Gottl\"ober and  B. Catinella, 2009, ApJ, 700, 1779

\newpage

\subsection{Eiichiro Komatsu}

\vskip -0.2cm

\begin{center}

University of Texas at Austin, Dept of Astronomy, Austin USA

\bigskip

{\large\bf WMAP 7-year Results}

\end{center}

\medskip

We have announced the results from 7 years of observations of the
Wilkinson Microwave Anisotropy Probe (WMAP) on January 26. In this talk,
we presented the cosmological interpretation of the WMAP 7-year data,
including the detection of primordial helium, images of polarization of
microwave background around temperature peaks (see Figure 6), 
and new limits on
inflation and properties of neutrinos. We also reported a significant
detection of the Sunyaev-Zel'dovich (SZ) effect, 
showed that we found, for
the first time in the SZ effect, a significant difference between the
relaxed and non-relaxed clusters.

\medskip

In more detail, our latest determination of the primordial spectral tilt
is $ n_s=0.968\pm 0.012$~(68\%~CL), a measurement which excludes 
$ n_s=1 $
by 99.5\%~CL. The latest 95\% upper limit on the tensor-to-scalar ratio
is $ r<0.24 $ (which is from WMAP+BAO+$H_0$).

\medskip

The 95\% upper limit on the total mass of neutrinos is $ \sum
m_\nu<0.58 $~eV, whereas the effective number of relativistic species is
$ N_{\rm eff}=4.34^{+0.86}_{-0.88} $ (68\%~CL). Whether $N_{\rm eff}\sim
4$ is favored will be tested by the Planck satellite, which is expected
to reduce the error bar by a factor of four. 

\medskip

As for the SZ effect, we find that our SZ measurements agree with the
predictions from the X-ray data very well on a cluster-by-cluster
basis. However, the current {\it theoretical} predictions overestimate
the amount of the gas pressure (hence SZ) in clusters of galaxies. This
will become important when using clusters of galaxies as a cosmological 
probe.

\medskip

Reference:

\medskip

E. Komatsu, et al., `Seven-Year Wilkinson Microwave Anisotropy Probe
(WMAP) Observations: Cosmological Interpretation,' to appear in the
Astrophysical Journal Supplement Series, arXiv:1001.4538.

\bigskip

\begin{figure}[ht]
\includegraphics[width=7cm]{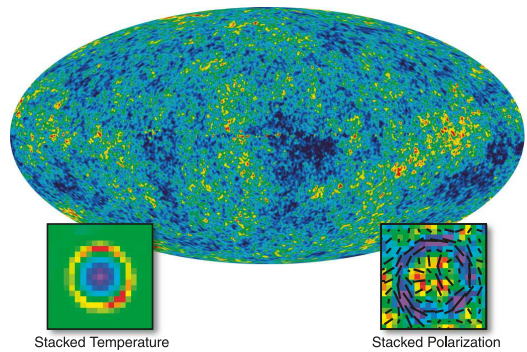}
\caption{The 7-year Internal Linear Combination (ILC) map of the 
cosmic microwave
 background (back). We found and stacked 12,628 cold temperature spots,
 finding the average temperature map around cold spots (the left
 panel). We then averaged the polarization map around cold spots,
 finding the expected tangential polarization pattern at $ 1.2^\circ $
 from the spot center and the radial polarization pattern at 
$ 0.6^\circ $
 from the spot center (the right panel). We also analyzed the hot spots
 in the same way, finding the opposite polarization patterns, consistent
 with the prediction.} 
\label{fig:directions}
\end{figure}

\newpage

\subsection{ Anthony Lasenby}

\vskip -0.2cm

\begin{center}

{\small Addresses: {\it Astrophysics Group, Cavendish Laboratory, J.J. Thomson Avenue,
Cambridge CB3 0HE, U.K. and Kavli Institute for Cosmology, c/o Institute of Astronomy, 
Madingley Road, Cambridge, CB3 0HA, U.K.}}\\

    {\small Email: \verb+a.n.lasenby@mrao.cam.ac.uk+}

\bigskip

{\large\bf CMB Observations: Current Status and Implications for Theory}\\

\end{center}

\medskip

The aim of this talk was to give an overview of the current state of Cosmic Microwave 
Background observations and scientific implications.

\medskip

It is still very much the case that the CMB is one of the most powerful 
tools of modern cosmology, 
and that a significant fraction of all information in cosmology over the past 10 to 15 years 
has come from it.

\medskip

As regards primordial CMB observations and results, we are effectively in an `interregnum' 
period at the moment, although of course some wonderful data is currently being taken by the 
Planck satellite, with the first results for primordial CMB expected during 2012. 
Compared to a year ago, there is no great change as regards results from ground-based
 experiments measuring primordial anisotropies --- BICEP [3] and 
QUAD [2]
are still the leading experiments as regards polarization. 
In space we have do have an increment of knowledge, from WMAP7 [9]
details of which are discussed by Prof. Komatsu. For the first time this has 
given us the opportunity to look at the temperature and polarization patterns around hot 
and cold spots in the CMB. These follow what would be expected for simple adiabatic 
density perturbations at recombination, and are an important test of the standard model. 
This has been carried out in a statistical way by stacking the images of many thousands of 
hot/cold spots. It will interesting with future data to look at individual features, e.g.
the `cold spot' first drawn attention to by Cruz \etal, and which has been advocated 
as due to late time effects from a topological defect remnant known as a `texture' [4].
In the latter case  a quite different pattern of CMB 
polarization around it would be expected, as opposed to that expected for a standard 
primordial CMB perturbation. Although the explanation in terms of a texture is 
speculative, detection of such an object would certainly be very important for cosmology 
and high energy physics theory.

\medskip

A further area in which the CMB can contribute to fundamental physics lies in the 
areas of testing inflation theory, and also assumptions underlying the nature of the 
universe on large scales during inflation itself. Ref.[5] puts forward 
a model in which the early universe is slightly oblate (specifically a Bianchi IX model) 
during the period at which inflationary perturbations are being laid down on the largest 
scales. This has obvious links with questions about whether there is a north/south 
asymmetry in the statistics of the CMB sky distribution, and possible links to issues 
such as a `dark flow' [8]. In a further paper, [8]
have demonstrated the links between their early universe model, which replaces the 
`Big Bang' with a non-singular pancaking event, and the Taub-NUT anisotropic universe 
model. It is shown how the latter, when viewed in a more physically appropriate 
coordinate system, is actually the vacuum limit of the scalar field Bianchi model 
introduced by [5], and instead of having two alternations between 
open, closed and open, it in fact remains always closed, but oscillating indefinitely 
through episodes of (non-singular) pancaking. It is perhaps remarkable that the vacuum 
can go through an infinite succession of exactly repeating oscillations in this way.

\medskip

As regards secondary anisotropies (those imposed e.g. by clusters of galaxies or 
other features of large scale structure at late times) there has been big progress 
over the past year, with the first results from `blank field' Sunyaev-Zeldovich 
samples appearing, and new constraints on the high-$\ell$ CMB power spectrum. 
The telescopes contributing to these developments include the South Pole Telescope 
(SPT), the Arcminute Microkelvin Imager (AMI) in Cambridge, and the Atacama 
Cosmology Telescope (ACT) in Chile. 

\medskip

For AMI, the first results on a blank field candidate cluster detected via its 
SZ effect are about to be published, and results on pointed observations of 
7 known clusters have appeared in [14]. A particularly intriguing 
recent AMI observation is of the `northern hemisphere bullet cluster', A2146, drawn 
attention to by [12]. This demonstrates an offset between the peaks 
of X-ray and SZ emission, consistent with the idea of complex bulk motions.

\medskip

For the SPT, a description of a sample of clusters selected via their SZ effect in 
blank field observations has been given in [13], and an analysis of 
their X-ray properties in [1]. The equivalent in terms of X-ray 
properties for blank field clusters detected using the ACT is contained in 
[11] and both telescopes have produced their first estimates of the 
high-$\ell$ CMB power spectrum, in [10] for the SPT and 
[7] for ACT. Taken together, these results are extremely 
interesting, in that they all show how the amplitude of the SZ effect, whether as 
determined in individual clusters, or in terms of the statistical contribution 
to the high-$\ell$ power spectrum, is {\em significantly smaller\/} than expected
 from current X-ray observations, and models for cluster evolution and scaling. 
This in fact ties in with results from WMAP 7 year data [9],
where the statistical level of SZ cluster effects (got by stacking at the positions 
of known clusters) are systematically smaller than expected from the X-ray observations,
by factors up to 2 in amplitude (similar to the factors involved in the ground-based 
observations). 

\medskip

This hints strongly at a gap of our understanding of processes in clusters, 
and will probably require us to treat clusters on a more individual basis 
(taking account of properties such as whether they are relaxed, and/or contain 
cooling flows) in comparing X-ray and SZ data, as well as improving the models of 
scaling and evolution with cosmic epoch. For both of these areas, observations 
from Planck can be expected to play a pivotal role.

\medskip

{\bf References:}

\medskip

\begin{itemize}
\item{[1] Andersson et~al.(2010) K.~{Andersson} \etal\
X-ray Properties of the First SZE-selected Galaxy Cluster Sample
  from the South Pole Telescope, arXiv 1006.3068.}

\item{[2] Brown et~al.(2009) M.~L. Brown \etal\ 
Improved Measurements of the Temperature and Polarization of the
  Cosmic Microwave Background from QUaD.
 \emph{\apj}, 705:\penalty0 978--999.}

\item{[3] Chiang et~al.(2010)
H.~C. Chiang \etal\ Measurement of Cosmic Microwave Background Polarization Power
  Spectra from Two Years of BICEP Data,
 \emph{\apj}, 711:\penalty0 1123--1140.}

\item{[4] Cruz et~al.(2007)
A Cosmic Microwave Background Feature Consistent with a Cosmic
  Texture.
\newblock \emph{Science}, 318:\penalty0 1612--, December 2007.}

\item{[5] Dechant et~al.(2009)
{P.-P.} {Dechant}, A.~N. {Lasenby}, and M.~P. {Hobson}.
\newblock {Anisotropic, nonsingular early universe model leading to a realistic
  cosmology}.
\newblock \emph{\prd}, 79\penalty0 (4):\penalty0 043524, February 2009.}

\item{[6] Dechant et~al.(2010){Dechant}, {Lasenby}, and
  {Hobson}, Cracking the Taub-NUT,
\newblock \emph{Classical and Quantum Gravity}, 27\penalty0 (18):\penalty0
  185010, September 2010.}

\item{[7] Fowler et~al.(2010) J.~W. {Fowler} \etal\
The Atacama Cosmology Telescope: A Measurement of the
  $600<\ell<8000$ Cosmic Microwave Background Power Spectrum at 148 GHz,
 arXiv:1001.2934.}

\item{[8] Kashlinsky et~al.(2010) 
A.~{Kashlinsky}, F.~{Atrio-Barandela}, H.~{Ebeling}, A.~{Edge}, and
  D.~{Kocevski}.
\newblock A New Measurement of the Bulk Flow of X-Ray Luminous Clusters of
  Galaxies.
\newblock \emph{\apjl}, 712:\penalty0 L81--L85, March 2010.}

\item{[9] Komatsu et~al.(2010) E.~{Komatsu} \etal\
Seven-Year Wilkinson Microwave Anisotropy Probe (WMAP) Observations:
  Cosmological Interpretation,  arXiv:1001.4538.}

\item{[10] Lueker et~al.(2010) M.~{Lueker} \etal\ 
Measurements of Secondary Cosmic Microwave Background Anisotropies
  with the South Pole Telescope,
\newblock \emph{\apj}, 719:\penalty0 1045--1066.}

\item{[11] Menanteau et~al.(2010)
F.~{Menanteau} \etal\ The Atacama Cosmology Telescope: Physical Properties and Purity of a
  Galaxy Cluster Sample Selected via the Sunyaev-Zel'dovich Effect.
 arXiv:1006.5126.}

\item{[12] Russell et~al.(2010) H.~R. {Russell} \etal\ 
\newblock {Chandra observation of two shock fronts in the merging galaxy
  cluster Abell 2146}.
\newblock \emph{\mnras}, 406:\penalty0 1721--1733.}

\item{[13] Vanderlinde et~al.(2010) K.~{Vanderlinde} \etal\
Galaxy Clusters Selected with the Sunyaev-Zel'dovich Effect from
  2008 South Pole Telescope Observations,
arXiv:1003.0003.}

\item{[14] Zwart et~al.(2010) J.~T.~L. {Zwart} \etal\
Sunyaev-Zel'dovich observations of galaxy clusters out to the virial
  radius with the Arcminute Microkelvin Imager,  arXiv:1008.0443.}

\end{itemize}

\newpage

\subsection{Stacy S. McGaugh}

\vskip -0.2cm

\begin{center}

Department of Astronomy, University of Maryland, College Park, MD 20742-2421, USA

\end{center}

\bigskip

\centerline{{\bf The Baryon Content of Cosmic Structures}}

\bigskip

Cosmic parameters like as the mass density ${\Omega_m}$ and expansion 
rate ${H_0}$ are now known with great precision.

\bigskip

The majority of mass in the universe appears to be composed of  
non-baryonic dark matter, whilst the normal baryonic material composing the stars and gas
that we can actually observe constitutes a distinct minority.  
This cosmic baryon fraction is well quantified: ${f_b = 17 \pm 1\%}$.  
An inventory of the detected baryons
in individual cosmic structures like galaxies
and clusters of galaxies falls short of this universal value.  

\bigskip

On the largest scales of clusters,
most but not all of the expected baryons are detected.
The fraction of detected baryons decreases monotonically  
from the cosmic baryon fraction as a function of mass.  
In the smallest dwarf galaxies, fewer
than 1\% of the expected baryons are detected.  
It is an observational challenge to identify the missing baryons, and
a theoretical one to understand the observed variation with mass. \\

\includegraphics[width=6.5in]{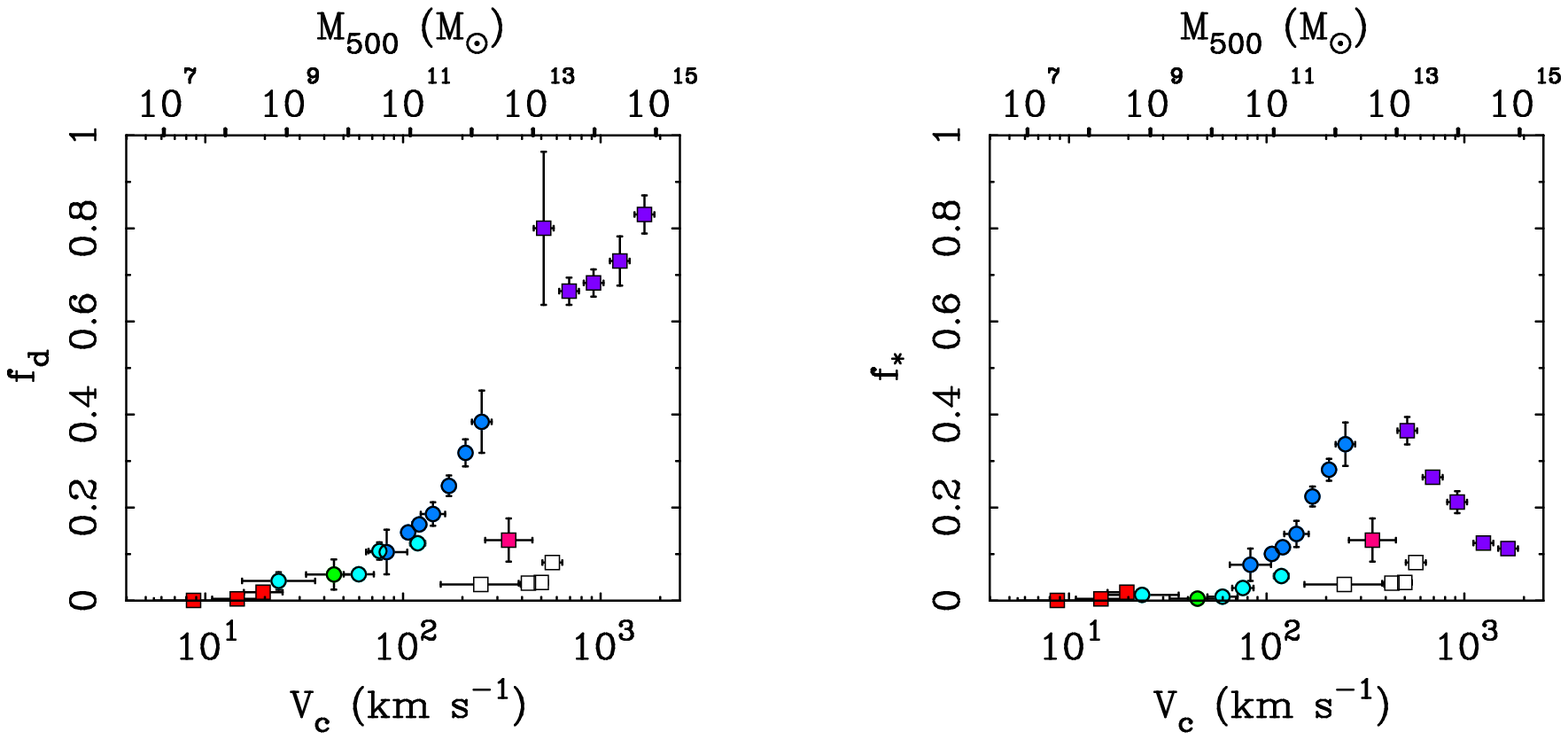}

The fraction of the expected baryons that are detected as a function of mass or
circular velocity (left) and the corresponding fraction that have been
converted to stars (right).
Purple squares are galaxy clusters.
Dark blue circles are star dominated spirals galaxies.  
Light blue circles are gas dominated disks.  
Red squares are Local Group dwarf satellites.  
Giant elliptical galaxies appear to deviate from the trend whether measured
by velocity dispersion (open squares) or gravitational lensing (pink square).
However, large systematic uncertainties for these objects are possible.
\\

\noindent This work is published in \\
McGaugh, S.~S., Schombert, J.~M., de Blok, W.~J.~G., \& Zagursky, M.~J.\ 2010, ApJ, 708, L14 

\newpage





\subsection{Rafael Rebolo}


\begin{figure}[ht]
\includegraphics[height=23.cm,width=16.cm]{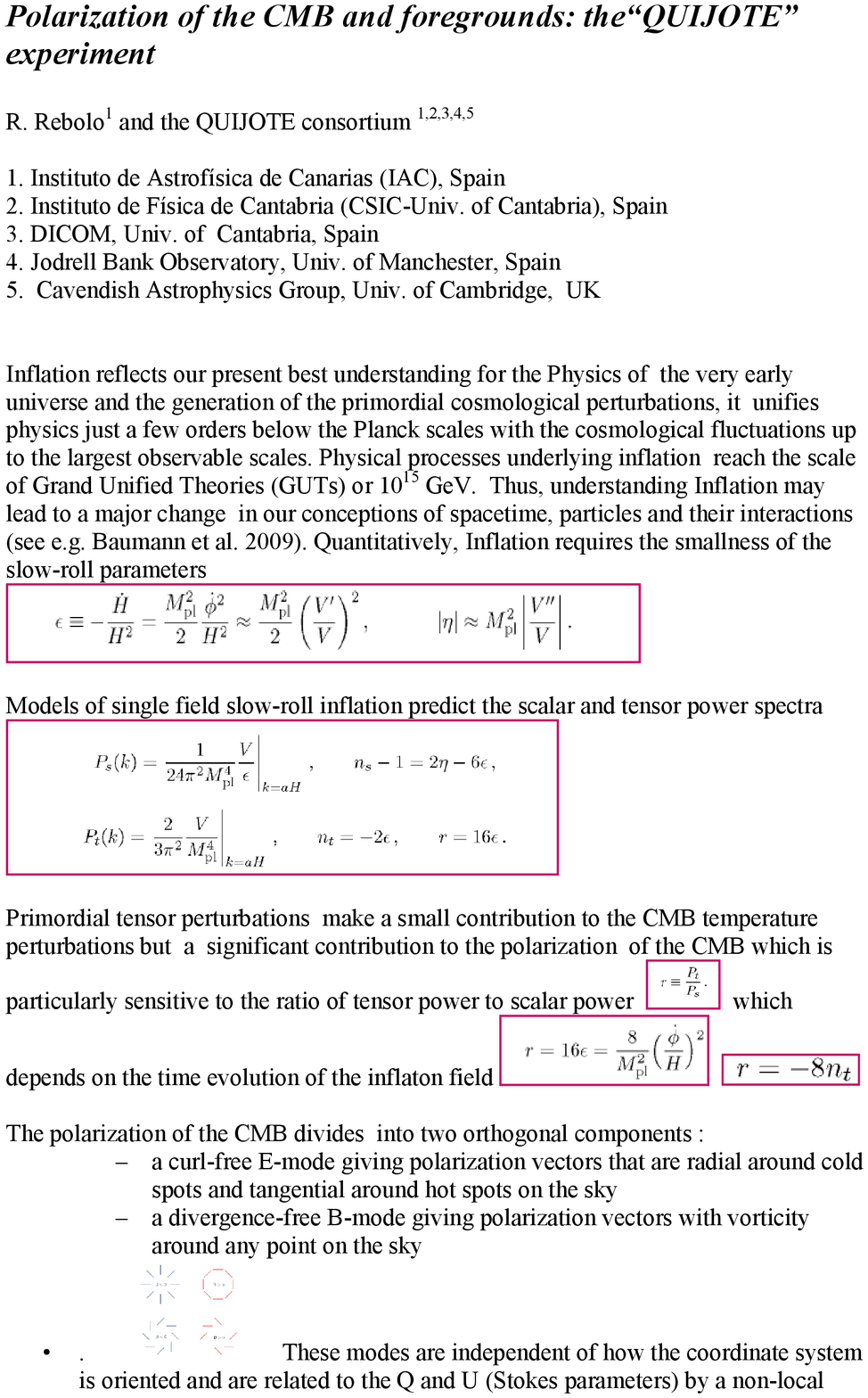}
\end{figure}



\begin{figure}[ht]
\includegraphics[height=23.cm,width=16.cm]{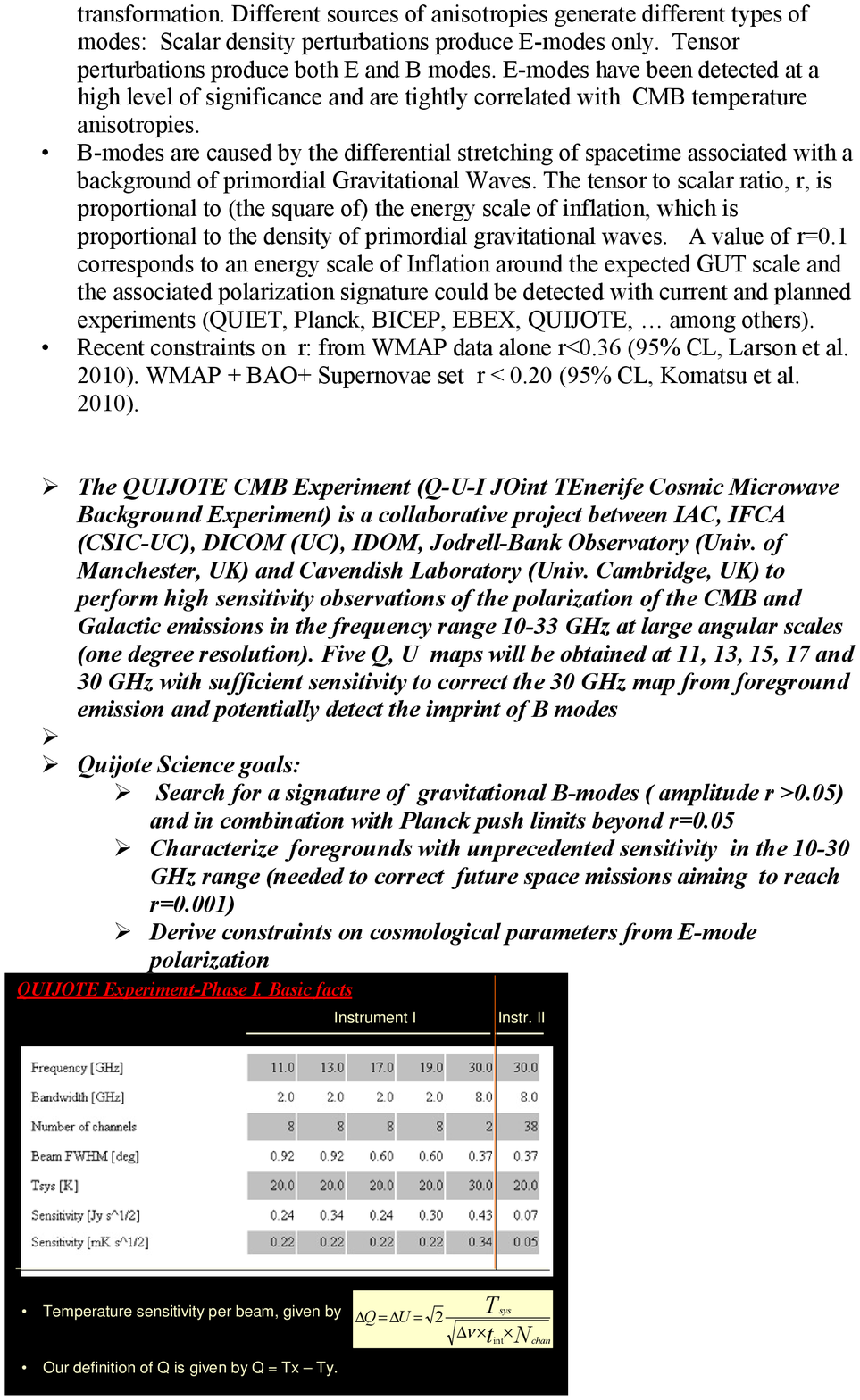}
\end{figure}



\begin{figure}[ht]
\includegraphics[height=23.cm,width=16.cm]{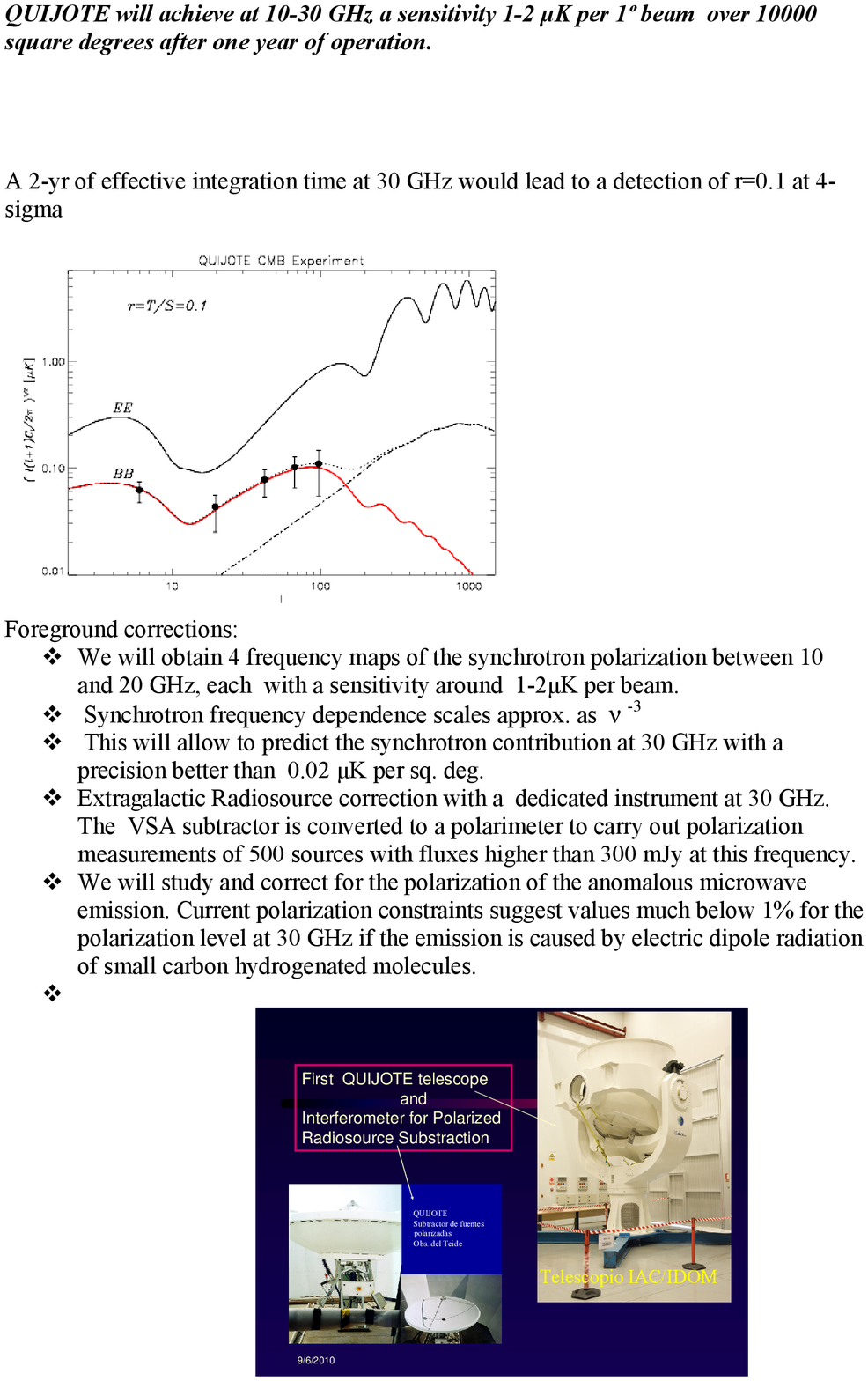}
\end{figure}

\begin{figure}[ht]
\includegraphics[height=23.cm,width=16.cm]{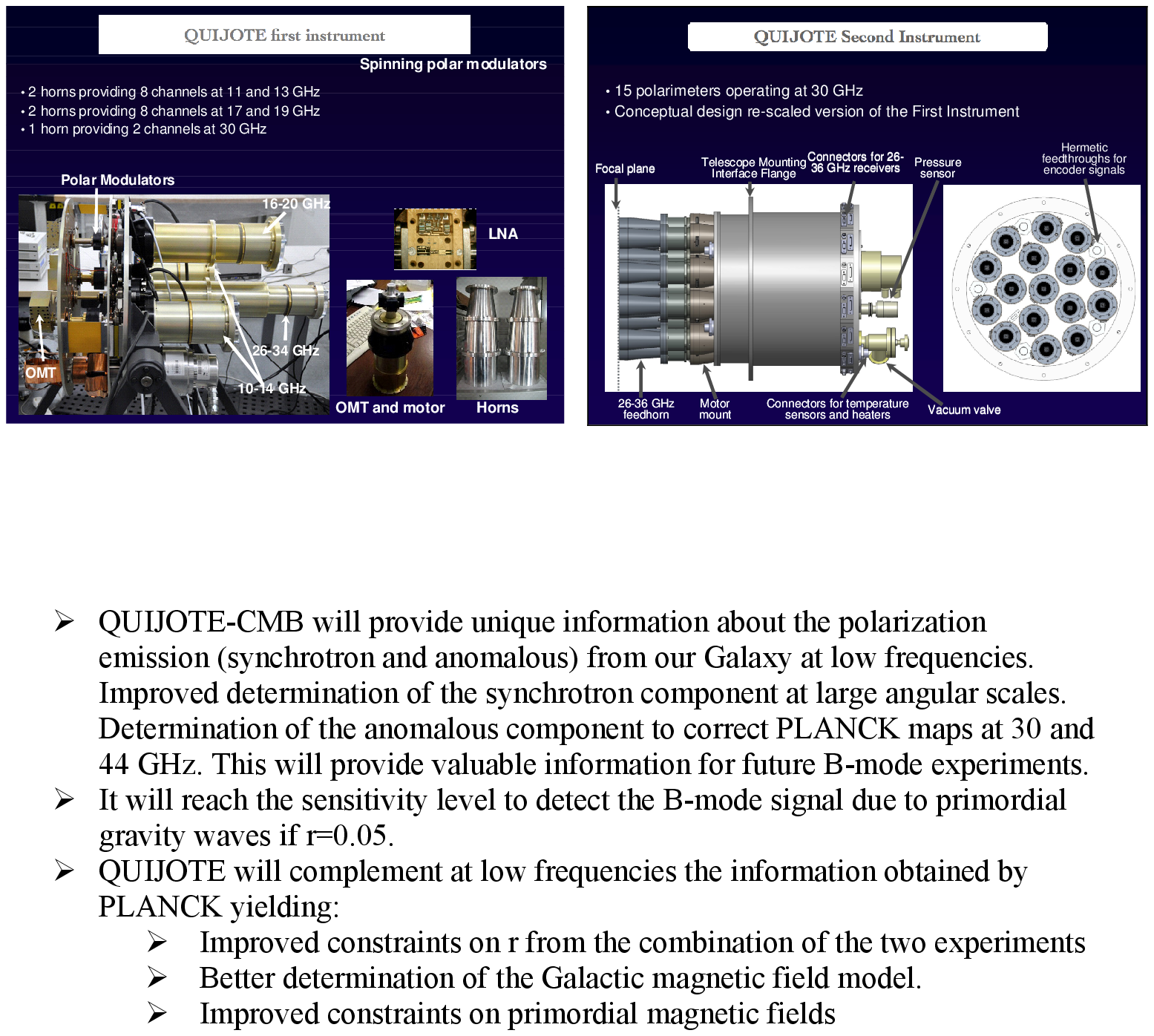}
\end{figure}

\newpage

\subsection{Paolo Salucci}

\vskip -0.2cm

\begin{center}

SISSA, Astrophysics section, Trieste, ITALY

\end{center}

\bigskip

\centerline{{\bf UNIVERSAL PROPERTIES IN GALAXIES AND CORED DM PROFILES}}

\bigskip

The presence of large amounts of unseen matter in  galaxies, distributed  differently from   
stars and gas,  is well established   from rotation curves which do not show the expected 
Keplerian fall-off at large radii,  but  increase, remain flat or start to  gently decrease 
according to a well organized pattern that involves an invisible mass component becoming 
progressively more 
more abundant at outer radii  and in  the less luminous galaxies (Persic, Salucci and Stel,  1996).

\bigskip

In Spirals we have the best opportunity to study the  mass distribution:  the   
gravitational potentials of   a spherical stellar bulge, a dark  halo, a stellar 
disk and  a gaseous disc 
give rise to an observed  equilibrium circular velocity  
$$
V^2(r)=r\frac{d}{dr}\phi_{tot}=V^2_b + V^2_{DM}+V^2_*+V^2_{HI} \; .
$$
 The Poisson equation  relates the surface (spatial)  densities of these components to the 
corresponding  gravitational potentials. 
The investigation is not difficult: e.g.  $\Sigma_*(r)$,  the surface stellar  density is 
proportional (by the  mass-to-light ratio) to the  observed  
surface brightness: 
$$
\Sigma_{*}(r)=\frac{M_{D}}{2 \pi R_{D}^{2}}\: e^{-r/R_{D}}
\quad {\rm and ~ then } \quad
V_{*}^{2}(r)=\frac{G M_{D}}{2R_{D}} x^{2}B\left(\frac{x}{2}\right) \; ,
$$
 where $M_D$ is the disk mass and $R_D$ is the disk scale length and $B(x)$ a 
combination of bessel functions.  

\bigskip

 Dark and luminous matter  in spirals are  coupled:   at any  galactocentric radii $R_n$   
measured in terms of disk length-scale $R_n \equiv (n/5)\ R_{opt}$,  $(R_{opt}=3.2 R_D$ there is a   
{\it Radial}  Tully-Fisher relation (Yegorova and Salucci 2007),  i.e. a relation between  the local 
rotation  velocity  $V(R_n)$ 
and  the  total galaxy luminosity:  
$M_{band} = a_n \log V_n + b_n$.  Spirals present Universal features in their 
kinematics that correlate  
with their  global galactic properties (PSS and by Salucci et al,  2007).

\bigskip

This led to the discovery,  from  3200 individual Rotation Curves (RCs),  of the 
`Universal Rotation Curve' of Spirals   $ V_{URC}(r; L) $ (see PSS, and Fig (1)), 
i.e. of  a function of   galactocentric  radius $r$, that,  tuned by a global galaxy property 
(e.g. the luminosity), well reproduces, out to the virial radius,  the RC of any spiral  
(Salucci \emph{et al.} 2007).  
$V_{URC}$  is the  observational counterpart to which the  circular  
velocity profile emerging in cosmological  simulations must comply.

\bigskip

In the same way of  individual RCs,   it underlies  a mass model  that includes a 
Freeman disk  and a  DM halo  with  a Burkert profile 
$$ 
\rho (r)=\rho_0\, \frac{r_0^3}{(r+r_0)\,(r^2+r_0^2)} \; .
$$
$ r_0 $ is the core radius and $ \rho_0 $  
the central density,  see Salucci and Burkert (2000) for details. 
We obtain the   structural parameters $\rho_0$,  $r_0$,  $M_D$ by $\chi^2$ fitting the   URC  and  
a number of  individual RCs. As result  a set of  scaling laws  among  local and   global galaxy 
quantities    emerges (see Fig 2). 
\begin{figure}[t!]
\centering
\vskip 0.3truecm
\includegraphics[width=6.2truecm]{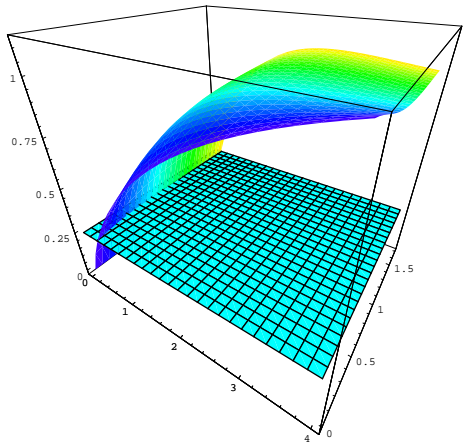}
\vskip -0.1truecm
\caption{The URC. The circular velocity as a function of radius (in units of $R_D$ 
and  out to $4\  R_D$) and luminosity (halo mass). See Salucci et al (2007) for 
details and for the URC out to the virial radius.} 
\label{fig:io_urc1}
\end{figure}

\bigskip

These scaling laws indicate (Salucci et al, 2007)  that spirals have  an Inner Baryon 
Dominance region where the stellar disk  dominates  the total gravitational potential,  
while the   DM halo emerges  farther out.  At any radii, objects  with lower luminosities 
have a larger  dark-to-stellar mass ratio. The baryonic fraction in spirals  is always 
much smaller than the cosmological value $\Omega_b/\Omega_{matter} \simeq 1/6  $, 
and it ranges between $7\times 10^{-3}$   to   $5\times 10^{-2}$,  suggesting that 
processes such as  SN  explosions  must have   removed  a very large fraction of the 
original hydrogen.

\medskip

Smaller spirals are denser, with their  central density spanning 2 order of magnitudes 
over the mass sequence of spirals.
\begin{figure}[t!]
\centering
\vskip -0.3cm
\includegraphics[width=9.5truecm]{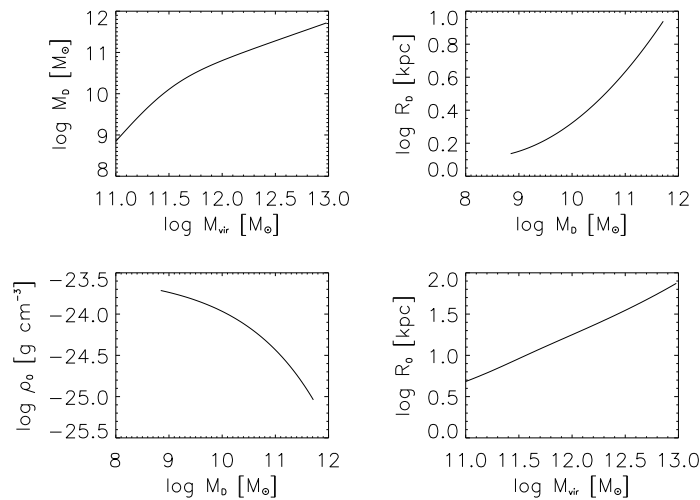}
\vskip -0.6truecm 
\caption{Scaling relations between the structural parameters of the dark and luminous 
mass distribution in spirals.}
\label{fig:scaling_relations}
\vskip -0.5truecm
\end{figure}

\bigskip

To assume a cored halo profile is obligatory. It is well known that  $\Lambda CDM$  
scenario provides a successful picture of the cosmological  structure  formation and 
that   large N-body  numerical simulations performed in  this scenario  lead to the 
commonly used NFW  halo cuspy  spatial density profile. However,   a careful analysis  
of about 100   high quality,  extended and  free from deviations from axial symmetry  
RCs has now ruled out the disk + NFW halo mass model, in favor of cored profiles  
(e.g. Gentile \emph{et al.} 2004, 2005,     Spano \emph{et al.} 2007, de Blok 2008 and  
de Naray \emph{et al.} 2008).

\bigskip

The mass modelling in dSph, LSB and Ellipticals is instead still in its infancy. However, 
data seem to confirm the pattern shown by  in spirals (Gilmore, 2005, Walker, 2010,Nagino 
and Matsushita, 2009).

\bigskip

Regarding the structural properties of the DM distribution a  most important   
finding is  that   the central surface density  $ \propto \mu_{0D}\equiv  \ r_0 \rho_0$,  
where $r_0$ and  $\rho_0$ are  the halo  core radius and central spatial density, is nearly 
constant and  independent of galaxy luminosity.   
Based on the co-added rotation curves of $\sim 1000$
spiral galaxies, mass models of individual dwarf irregular 
and spiral galaxies of late and early types  with
high-quality rotation curves and on  galaxy-galaxy weak lensing signals from a sample of 
spiral and elliptical
galaxies, we find  that  
 $$ \log \mu_{0D} = 2.15 \pm 0.2 \; , $$
 in units of $\log$(M$_{\odot}$
pc$^{-2}$).  This constancy  transpasses the family of disk systems and reaches spherical systems.  
The  internal kinematics of Local Group dwarf spheroidal galaxies  are
consistent with this picture. Our results are obtained for galactic systems spanning over 14  
magnitudes, 
belonging to different Hubble Types, and whose mass profiles have been determined by  
several independent 
methods. Very significantly, in  the same objects, the approximate constancy of $\mu_{0D}$ 
is in sharp contrast to  the
systematical  variations, by several orders of magnitude,  of galaxy properties, 
including $\rho_0$  and
central  stellar surface density see figure (3).

\bigskip
  
The evidence that the DM halo central surface density $\rho_0 r_0$
remains constant to within less than a factor of two over fourteen galaxy magnitudes, and across
several Hubble types,  does indicates  that this quantity  is perhaps  
hiding   the  physical nature of the DM. Considering that DM haloes are (almost)  
spherical systems it is  surprising that their central surface
density plays  a  role in galaxy structure. 

\bigskip
 
Moreover, this evidence, is difficult to  understand   in  an evolutionary scenario as  
the product of the process 
that has turned the primordial cosmological  gas in the stellar galactic structures we  
observe today. 
Such constancy, in fact,   must be achieved in very different  galaxies of different  
morphology and mass, ranging  from dark-matter-dominated  to baryon-dominated objects. In
addition, these galaxies have experienced significantly different
evolutionary histories (e.g. numbers of mergers, significance of
baryon cooling, stellar feedback, etc.). 

\bigskip

The best explanation for our findings stems from the DM nature:
 the DM particle mass and decoupling temperature. Recent theoretical work
 (de Vega, Sanchez 09; de Vega, Salucci, Sanchez 10) points towards
 a DM particle mass in the keV scale leading to the formation of
 cored DM virialized structures.

\bigskip

The results obtained so far indicate  the   distribution of matter galaxies  is  a benchmark for  
understanding dark matter nature   and the    galaxy formation process. In particular, the  
universality of certain structural quantities and the dark-luminous coupling  of the  mass 
distributions,  seem to  bear the direct imprint of the Nature of the  
DM (Donato \emph{et al.} 2009, Gentile \emph{et al.} 2009).

\begin{figure}[t!]
 \vskip -.0truecm
\psfig{file=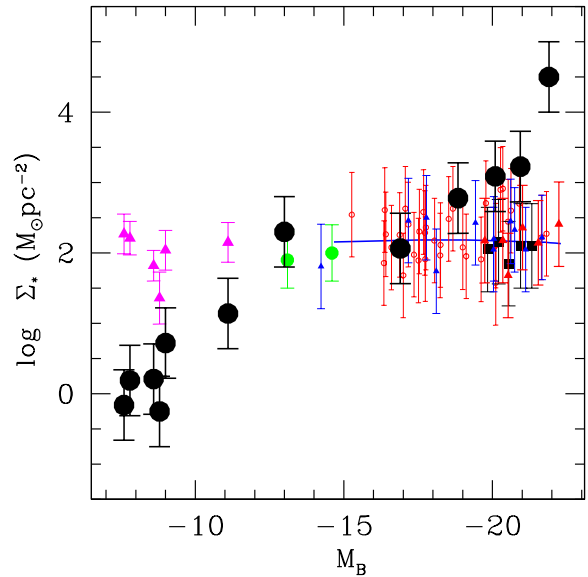,width=0.5\textwidth}   
\vskip -0.8truecm
\caption{Dark matter central surface density in units of $M_\odot$pc$^{-2}$  as a function of
galaxy magnitude,  for different galaxies and Hubble Types. As  a comparison, we also  plot the   
values of the same quantity relative to  the stellar component  (big filled circles). }
\end{figure}

 \begin{itemize}

\item{ de Blok,   W.J.G \emph{et al.}  2008, arXiv:0810.2100. }

\item{ Donato, F.  Gentile, G. ,  Salucci, P.   2004,
  Mon. Not. Roy. Astron. Soc.,  353, L17.}

\item{Donato F., \emph{et al.} 2009,
Mon. Not. Roy. Astron. Soc., 397, 1169. }
 
\item{ Gentile, G. \emph{et al.}  2004,
 Mon. Not. Roy. Astron. Soc.,  351, 903.}
 
\item{Gentile, G.  \emph{et al.} 2005, 
 Astrophys. J.,  634, L145.} 
  
\item{Gentile G., \emph{et al.} 2009,
Nature, 461, 627. }

\item{ Kuzio de Naray, R. \emph{et al.}   2008, Astrophys. J.,  676, 920.}
 
\item{ Navarro,  J.~F., Frenk,   C.~S., White,  S.~D.~M. 1996, Astrophys. J.,  462, 563 (NFW).}
 
\item{Persic, M., Salucci, P. , Stel, F. 1996,
 Mon. Not. R. Astron. Soc.  281, 27 (PSS).}
 
\item{Salucci, P. \emph{et al.} 2007,
 Mon. Not. Roy. Astron. Soc.,  378, 41.}
 
\item{ Shankar, F. \emph{et al.} 2006, Astrophys. J.,   643, 14.}

\item{Spano, M., Marcelin, M., Amram, P. \emph{et al.} 2008, 
Mon. Not. Roy. Astron. Soc.  383, 297.}
 
\item {Nagino, R. and Matsushita K, A \& A , 2009, 501, 157}
  
\item{ Yegorova, I.A.,   Salucci,  P. 2007, Mon. Not. R. Astron. Soc.,  377, 507.}

 \item{Walker, M. et al,  2010, ApJ 717, 87}

\item{ Gilmore, G. et al 2007 ApJ, 663, 948}
\item{de Vega10}
  de Vega, H J,  Salucci P.,  Sanchez N., arXiv:1004.1908 
 
\item{  de Vega, H J,  Sanchez N. G., arXiv0901.0922 }
\end{itemize}

 \newpage

 \subsection{Hector J. de Vega and Norma G. Sanchez}

\vskip -0.3cm

\begin{center}

HJdV: LPTHE, CNRS/Universit\'e Paris VI-P. \& M. Curie \& Observatoire de Paris.\\
NGS: Observatoire de Paris, LERMA \& CNRS

\bigskip

{\bf keV scale dark matter from theory and observations and 
galaxy properties from linear primordial fluctuations} 

\end{center}

In the context of the standard Cosmological model the nature of DM
is unknown. Only the DM gravitational effects are noticed and they are necessary
to explain the present structure of the Universe. DM (dark matter) 
particles must be neutral and so weakly interacting 
such that no effects are  detectable. DM candidates are beyond the standard model
of particle physics.Theoretical analysis combined with astrophysical data from
galaxy observations points towards a DM particle mass in
the {\bf keV scale} (keV = 1/511 electron mass) [1-4].

DM particles can decouple being ultrarelativistic (UR) at 
$ \; T_d \gg m $ or non-relativistic $ \; T_d \ll m $.
They may  decouple at or out of local thermal equilibrium (LTE).
The DM distribution function: $ F_d[p_c] $ freezes out at decoupling
becoming a function of the  comoving momentum $ p_c = $.
$ P_f(t) = p_c/a(t) = $ is the physical momentum. 
Basic physical quantities can be expressed in terms of the
distribution function as the velocity fluctuations,
$ \langle \vec{V}^2(t) \rangle = \langle \vec{P}^2_f(t) \rangle/m^2 $ 
and the DM energy density $ \rho_{DM}(t) $
where $ y = P_f(t)/T_d(t) = p_c/T_d $ is the integration variable and
$ g $  is the number of internal degrees of freedom of the DM 
particle; typically $ 1 \leq g \leq 4 $.

\medskip

{\bf Two} basic quantities characterize DM: its particle mass $ m $
and the temperature $ T_d $ at which DM decouples.  $ T_d $
is related by entropy conservation to the number of
ultrarelativistic degrees of freedom $ g_d $ at decoupling by
$ \quad T_d = \left(2/g_d \right)^\frac13 \; T_{cmb} \; ,
\; T_{cmb} = 0.2348 \; 10^{-3} \; $ eV.
One therefore needs {\bf two} constraints to determine the values of
$ m $ and $ T_d $ (or $ g_d $).

\medskip

One constraint is to reproduce the known cosmological DM density today.
$\rho_{DM}({\rm today})= 1.107 \; {\rm keV/cm}^3 $.

Two independent further constraints are considered in refs. [1-4].
First, the phase-space density $ Q=\rho/\sigma^3 $ [1-2] and second the
surface acceleration of gravity in DM dominated galaxies [3-4].
We therefore provide {\bf two} quantitative ways to derive the value $ m $ 
and $ g_d $ in refs. [1-4].

\medskip

The phase-space density $ Q $ is invariant under the
cosmological expansion and can {\bf only decrease} 
under self-gravity interactions 
(gravitational clustering). The value of $ Q $ today follows
observing dwarf spheroidal satellite galaxies of the Milky Way (dSphs):
$ Q_{today} = (0.18 \;  \mathrm{keV})^4 $ (Gilmore et al. 07 and 08).
We compute explicitly $ Q_{prim} $ (in the primordial universe) and it turns
to be proportional to $ m^4 $ [1-4].

\medskip

During structure formation $ Q $
{\bf decreases} by a factor that we call $ Z $. Namely, 
$ Q_{today} = Q_{prim}/Z $. The value of $ Z $ is galaxy-dependent.
The spherical model gives $ Z \simeq 41000 $
and $N$-body simulations indicate: $ 10000 >  Z > 1 $ (see [1]).
Combining the value of $ Q_{today} $ and $\rho_{DM}({\rm today}) $ with 
the theoretical analysis yields that $ m $ must be in the keV scale and 
$ T_d $ can be larger than 100 GeV. More explicitly, we get 
general formulas for $ m $ and $ g_d $ [1]:
$$ 
m = \frac{2^\frac14 \; \sqrt{\pi}}{3^\frac38 \; g^\frac14 } \; 
Q_{prim}^\frac14
\; I_4^{\frac38} \; I_2^{-\frac58} \; , \quad
g_d = \frac{2^\frac14 \; g^\frac34}{3^\frac38 \; 
\pi^\frac32 \; \Omega_{DM}} \; 
 \; \frac{T_{\gamma}^3}{\rho_c} \; Q_{prim}^\frac14 \; 
\left[I_2 \; I_4\right]^{\frac38}
$$
where $ I_{2 \, n} = \int_0^\infty y^{2 \, n} \; F_d(y) \; dy 
\quad , \quad n=1, 2 $ 
and $ Q_{prim}^\frac14 = Z^\frac14 \; \; 0.18 $ keV using the dSphs data,
$T_{\gamma} = 0.2348 \; {\rm meV } 
\; , \; \Omega_{DM} = 0.228 $ and $ \rho_c = (2.518 \; {\rm meV})^4$.
These formulas yield for relics decoupling UR at LTE:
$$ 
m = \left(\frac{Z}{g}\right)^\frac14 \; \mathrm{keV} \; 
\left\{\begin{array}{l}
         0.568 \\
              0.484      \end{array} \right. \; , \;
 g_d = g^\frac34 \; Z^\frac14 \; \left\{\begin{array}{l}
         155~~~\mathrm{Fermions} \\
              180~~~\mathrm{Bosons}      \end{array} \right. \; . 
$$
Since $ g = 1-4 $, we see that 
$ g_d \gtrsim 100 \Rightarrow  T_d \gtrsim 100 $ GeV.
Moreover, $ 1 < Z^\frac14 < 10 $ for $ 1 < Z < 10000 $.
For example for DM Majorana fermions $ (g=2) \; m \simeq 0.85 $ keV.

\medskip

We get results for $ m $ and $ g_d $ on the same scales for DM particles 
decoupling UR out of thermal equilibrium [1]. For a specific model of 
sterile neutrinos where decoupling is out of thermal equilibrium:
$$
0.56 \; \mathrm{keV} \lesssim m_{\nu} \;  
Z^{-\frac14} \lesssim 1.0 \; \mathrm{keV}
\quad ,  \quad 15 \lesssim g_d  \;  Z^{-\frac14}\lesssim 84
$$
For relics decoupling non-relativistic
we obtain similar results for the DM particle mass: keV 
$ \lesssim m \lesssim $ MeV [1].

\medskip

Notice that the dark matter particle mass $ m $ and decoupling temperature 
$ T_d $ are {\bf mildly} affected by the uncertainty in the factor $ Z $ through a 
power factor 
$ 1/4 $ of this uncertainty, namely, by a factor $ 10^{\frac14} \simeq 1.8 $

\medskip

The comoving free-streaming) wavelength, (Fig. 1), and the Jeans' mass are obtained in the range
$$
\frac{0.76}{\sqrt{1+z}} \; {\rm kpc} <\lambda_{fs}(z) <
\frac{16.3}{\sqrt{1+z}} \; {\rm kpc} \; , \; 0.45 \; 10^3 \; M_{\odot} 
< \frac{M_J(z)}{(1+z)^{+\frac32}} < 0.45 \; 10^7  \; 
\; M_{\odot} \; .
$$

These values at $ z = 0 $ are consistent with the $N$-body simulations 
and are of the order of the small dark matter structures observed today .
By the beginning of the matter dominated era $ z \sim 3200 $, the masses are of the 
order of galactic masses $ \sim 10^{12} \; M_{\odot} $ and the comoving free-streaming 
wavelength scale turns to be of the order of the galaxy sizes today 
$ \sim 100 \; {\rm kpc}$.

\medskip

 Lower and upper bounds for the dark matter annihilation cross-section $ \sigma_0 $ 
are derived: $ \sigma_0 > (0.239-0.956) \; 10^{-9} \; \mathrm{GeV}^{-2} $ and 
$ \sigma_0 < 3200 \; m \; \mathrm{GeV}^{-3} \; . $ There is at least five orders of 
magnitude between them, the dark matter non-gravitational self-interaction is 
therefore negligible (consistent with structure formation and observations, as 
well as by comparing X-ray, optical and lensing observations of the merging of 
galaxy clusters with $N$-body simulations).

\medskip

Typical "wimps" (weakly interacting massive particles) with mass $ m = 100 $ GeV 
and $ T_d = 5 $ GeV  would require a huge $ Z \sim 10^{23} $, well above
the upper bounds obtained and cannot reproduce the observed galaxy properties. 
They produce an extremely short free-streaming or Jeans length $ \lambda_{fs} $ today $ 
\lambda_{fs}(0) \sim 3.51 \; 10^{-4} \; {\rm pc} = 72.4  \; {\rm AU} \; $ that would
correspond to unobserved structures much smaller than the galaxy structure.
Wimps result strongly disfavoured. 

\bigskip

Galaxies are described by a variety of physical quantities:

(a) {\bf Non-universal} quantities: mass, size, luminosity, fraction of DM,
DM core radius $ r_0 $, central DM density $ \rho_0 $.

(b) {\bf Universal} quantities: surface density $ \mu_0 \equiv r_0 \; \rho_0 $
and DM density profiles. $M_{BH}/M_{halo}$ (or halo binding energy).
The galaxy variables are related by
{\bf universal} empirical relations. Only one variable remains free. 
That is, the galaxies are a one parameter family of objects.
The existence of such universal quantities may be explained by the
presence of attractors in the dynamical evolution. 
The quantities linked to the attractor always reach the same value
for a large variety of initial conditions. This is analogous
to the universal quantities linked to fixed points in critical
phenomena of phase transitions.The universal DM density profile in Galaxies has the scaling 
property: 
$$
\rho(r) = \rho_0 \; F\left(\displaystyle\frac{r}{r_0}\right) \quad , \quad  
F(0) = 1 \quad , \quad x \equiv  \displaystyle\frac{r}{r_0} \; , \quad (1)
$$
where $ r_0 $ is the DM core radius.
As empirical form of cored profiles one can take Burkert's form for $ F(x) $.
Cored profiles {\bf do reproduce} the astronomical observations
(see the contribution here by Salucci and the review by de Blok 2010).

\bigskip

The surface density for dark matter (DM) halos 
and for luminous matter galaxies is defined as: 
$ \mu_{0 D} \equiv r_0 \; \rho_0, $ 
$ r_0 = $ halo core radius, $ \rho_0 = $ central density for DM galaxies.
For luminous galaxies  $ \rho_0 = \rho(r_0) $
(Donato et al. 09, Gentile et al. 09).
Observations show an Universal value for $ \mu_{0  D} $: independent of 
the galaxy luminosity for a large number of galactic systems 
(spirals, dwarf irregular and spheroidals, elliptics) 
spanning over $14$ magnitudes in luminosity and of different Hubble types.
Observed values:
$$
\mu_{0  D} \simeq 120 \; \frac{M_{\odot}}{{\rm pc}^2} = 
5500 \; ({\rm MeV})^3 = (17.6 \; {\rm Mev})^3 \quad , \quad
5 {\rm kpc}  < r_0 <  100 {\rm kpc} \; .
$$
Similar values $ \mu_{0  D} \simeq 80  \; \frac{M_{\odot}}{{\rm pc}^2} $ are observed in 
interstellar molecular clouds of size $ r_0 $ of different type and composition over 
scales $ 0.001 \, {\rm pc} < r_0 < 100 $ pc (Larson laws, 1981).
Notice that the surface gravity acceleration is 
given by $\mu_{0  D}$ times Newton's constant.

{\vskip0.2cm} 

The scaling form eq.(1) of the density profiles
implies scaling properties for the energy and entropy.
The total energy becomes using the  virial theorem and the profile $F(x)$:
$$
E = \frac12 \; {\avg U} = - \frac14 \; G \; \int 
\frac{d^3r \; d^3r'}{|\rv-\rv'|} \; \avg{\rho(r) \; \rho(r')}=
- \frac14 \; G \; \rho_0^2 \; r_0^5 \int \frac{d^3x \; d^3x'}{|\vx-\vx'|} \; 
\avg{F(x) \; F(x')} \quad \Rightarrow \quad E \sim G \; \mu_{0  D}^2 \; r_0^3
$$ 
Therefore, the energy scales as the volume.

{\vskip0.1cm} 

The Boltzmann-Vlasov distribution function $ f(\pv,\rv) $ for consistency 
with the profile form eq.(1), must scale as
$$
f(\pv,\rv) = \frac1{m^4 \; r_0^3 \; G^{\frac32} \; \sqrt{\rho_0}} \; 
{\cal F} \left(\frac{\pv}{m \; r_0 \; \sqrt{G \; \rho_0}},
\frac{\rv}{r_0}\right)
$$
where $ m $ is the DM particle mass. Hence, the entropy scales as 
$$
S_{gal} = \int f(\pv,\rv) \; \log f(\pv,\rv) \; d^3 p \;  d^3 r
\sim r_0^3 \; \frac{\rho_0}{m} = r_0^2 \; \frac{\mu_{0 D}}{m}
$$
The {\bf entropy} scales as the {\bf surface} as it is the case for black-holes.
However, for black-holes of mass $ M $ and area $ A = 16 \, \pi \; G^2 \; M^2 $,
the entropy $ S_{BH} = A /(4 \; G) = 4 \, \pi \; G \; M^2 $.
That is, the proportionality coefficients $ c $ between entropy and area are
 very different:
$$
c_{gal}=\frac{S_{gal}}{r_0^2} \sim \frac{\mu_{0 D}}{m} \quad , \quad
c_{BH}= \frac{S_{BH}}{A} = \frac1{4 \; G} \quad {\rm which ~ implies} \quad
\frac{c_{BH}}{c_{gal}} \sim \frac{m}{\rm keV} \; 10^{36}
$$
showing that the entropy per unit area of the galaxy is much smaller 
than the entropy of a black-hole. In other words the Bekenstein 
bound for the entropy of physical is well satisfied here.

\medskip

In order to compute the surface density and the density profiles from
first principles we have evolved the linearized Boltzmann-Vlasov equation 
since the end of inflation till today [2-3].
We depict in fig. 10 the density profiles vs. $ x \equiv r/r_{lin} $
for fermions (green) and bosons (red) decoupling ultrarelativistic
 and for particles decoupling non-relativistically (blue).
These profiles turn to be universal with the same shape as, for example, the Burkert
profile.  $ r_{lin} \sim r_0 $ depends on the galaxy and is entirely determined
theoretically in terms of cosmological parameters and $ m $ [2-3].

\begin{figure}[ht]
\begin{turn}{-90}
\psfrag{"dengifd.dat"}{}
\psfrag{"dengibe.dat"}{}
\psfrag{"r2dengimb.dat"}{}
\includegraphics[height=9.cm,width=3.5cm]{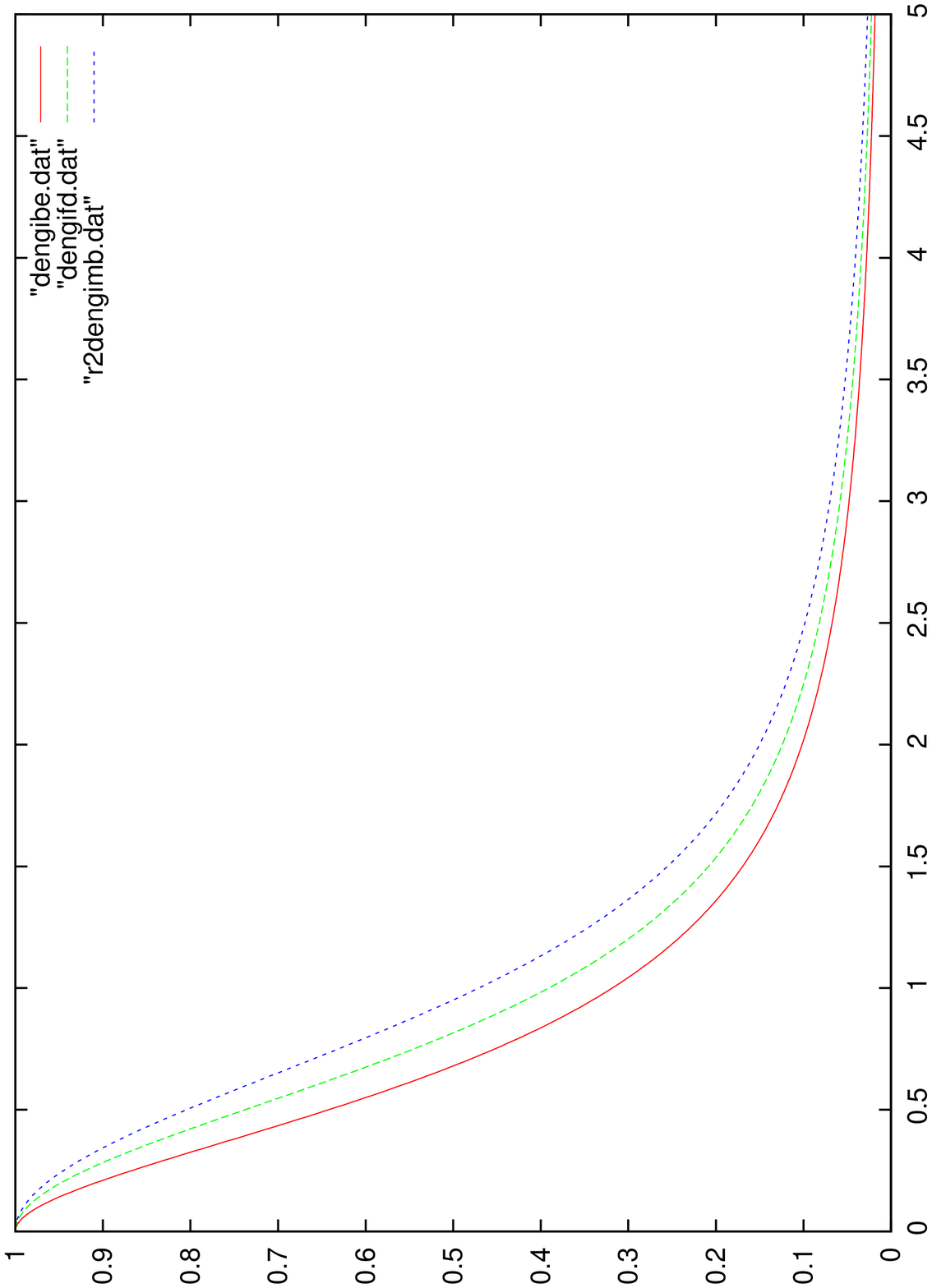}
\end{turn}
\caption{}
\label{perflin}
\end{figure}

\medskip

We obtain in refs. [3,4] for the galaxy surface density in the linear
approximation
$$ 
\mu_{0 \, lin} = 8261 \; \left[\frac{Q_{prim}}{({\rm keV})^4}\right]^{0.161}
\left[1+0.0489 \; \ln \frac{Q_{prim}}{({\rm keV})^4} \right] {\rm MeV}^3 
$$
where $ 0.161 = n_s/6 , \; n_s $ is the primordial spectral index
and fermions decoupling UR were considered.
Matching the {\bf observed values} from spiral galaxies $ \mu_{0 \, obs} $ 
with this $ \mu_{0 \, lin} $ gives the primordial phase-space density
$ Q_{prim}/({\rm keV})^4 $ and 
from it the mass of the DM particle. 
We obtain $ 1.6 < m < 2 $ keV for the dark matter particle mass [4].

\medskip

We obtain the linear density profiles for any DM particle mass $m$ [3-4].
At intermediate scales $ r \gtrsim r_{lin} $ we obtain [4],
$$
\rho_{lin}(r)\buildrel{r \gtrsim r_{lin}}\over= 
 \left(\frac{36.45 \; {\rm kpc}}{r}\right)^{1+n_s/2} \; 
\ln\left(\frac{7.932\; {\rm Mpc}}{r}\right) \times 
\left[ 1 + 0.2416 \; \ln \left(\frac{m}{\rm keV}\right)\right] \; 
10^{-26} \;  \frac{\rm g}{{\rm cm}^3} \quad , \quad 1+n_s/2 = 1.482
$$
The theoretical linear results {\bf agree} with the universal empirical behaviour 
$ r^{-1.6\pm 0.4} $: M. G. Walker et al.  (2009) (observations), 
I. M. Vass et al. (2009) (simulations).

At small scales $ r \ll r_{lin} \; ( \lesssim $ kpc) the linear 
density profiles turns to be {\bf cored} for keV scale DM particles, and {\bf cusped}
for wimps [3-4].

\medskip

We summarize in the Table the values for non-universal galaxy quantities 
from the observations and from the linear theory results.
The  larger and less denser are the galaxies, the better are the results from the linear
theory for non-universal quantities. The linear approximation turns to improve for 
larger galaxies (i. e. more diluted) [4]. Therefore, universal quantities as profiles 
and surface density are reproduced by the linear approximation.
The agreement between the linear theory and the observations is {\bf remarkable}.

\medskip

The last column of the Table corresponds to 100 GeV mass wimps.
The wimps values strongly disagree by several orders of magnitude with the observations.

\medskip

\begin{table*}
 \centering
\begin{tabular}{|c|c|c|c|} \hline  
      & Observed Values & Linear Theory & Wimps in linear theory \\
\hline 
  $ r_0 $ & $ 5 $ to $ 52 $ kpc &  $ 46 $ to $ 59 $ kpc & $ 0.045 $ pc \\
\hline 
  $ \rho_0 $ & $ 1.57  $ to $ 19.3 \times 10^{-25}  \; \frac{\rm g}{{\rm cm}^3} $  & 
$ 1.49  $ to $ 1.91  \times 10^{-25}  \; \frac{\rm g}{{\rm cm}^3} $  &
$ 0.73  \times  10^{-14}  \; \frac{\rm g}{{\rm cm}^3} $ \\ \hline 
  $ \sqrt{{\overline {v^2}}}_{halo} $ & $ 79.3 $ to $ 261 $ \; km/sec & 
$ 260 $ \; km/sec &  $ 0.243 $ \; km/sec \\
\hline   
\end{tabular}
\end{table*}

{\bf References}

\begin{description}

\item[1]  H. J. de Vega, N. G. Sanchez,  arXiv:0901.0922, 
Mon. Not. R. Astron. Soc. 404, 885 (2010).
\item[2] D. Boyanovsky, H. J. de Vega, N. G. Sanchez, 	
arXiv:0710.5180, Phys. Rev. {\bf D 77}, 043518 (2008).
\item[3] H. J. de Vega, N. G. Sanchez, arXiv:0907.0006.
\item[4] H. J. de Vega, P. Salucci, N. G. Sanchez, arXiv:1004.1908.

\end{description}

\newpage

\subsection{Anton V. Tikhonov}

\vskip -0.2cm

\begin{center}

Saint Petersburg State University, Astrophysics Inst., Saint Petersburg, RUSSIA\\
E-mail: avtikh@gmail.com

\end{center}

\bigskip

\centerline{{\bf Voids and Dwarf galaxies in the Local Volume: another 
$\Lambda$CDM-overabundance and possible solutions}}

\bigskip

At present, the reference cosmological model is a flat Friedmann universe whose 
mass-energy content is dominated by
a cosmological constant, a Cold Dark Matter (CDM) component and baryons. 
This $\Lambda$CDM model describes structure formation at large scales very well, 
however it fails on small scales: the standard model predicts
much more small scale structure than observed. 

\bigskip 

The problem of $\Lambda$CDM overabundance on small mass scales has its origin 
in the mismatch between the faint end of 
the observational luminosity function and the mass function of the DM halos predicted
by the $\Lambda$CDM model. 

\bigskip 

Numerically, the problem initially showed up as the "missing satellites
problem" - strong excess of substructures in simulated Milky Way sized DM halos with respect to
the observed number of dwarf galaxies in the Local Group (LG) (Klypin et al., 2009). 
More than 10 years of "solving" the problem has led to the statement that there should be many truly 
"dark" halos without stellar or gas components. But still none of the solutions became 
"essential" or "concordant".
For instance, in some recipes procedure of construction of LG satellites LF when accounting for SDSS
incompleteness assumes uniformity of missed dwarf galaxies population (Tollerud et al. 2009) 
which is a very poor approximation for known and newly discovered dwarf LG population. 

\bigskip 

Tikhonov \& Klypin (2009) have found that the $\Lambda$CDM model exhibits
the same problem concerning the problem of void dwarf galaxies. 
They found that  in the  $\Lambda$CDM model the minivoids are too small when
compared to observations. Our analysis led to the conclusion that the
$\Lambda$CDM model fits the observational
spectrum of minivoids defined by galaxies with $ M_B<-12 $ assuming 
that these galaxies can contain
only DM halos with circular velocities $ V_c>35 $ km/s (which roughly corresponds to 
$5 \cdot 10^9 \; M_{\odot} $).

\bigskip 

This is a quite interesting result from the theoretical side - according to some estimates, 
halos with $ V_c<35 $\,km/s
had essentially no gas infall and star formation inside them. The extrapolation of a 
`common mass scale' (Strigari et al., 2008) for LG's
dSphs points to a similar total mass of the smallest observable gravitationally bound 
systems in the Universe containing dark matter. 

\bigskip

The problem (for the $\Lambda$CDM model, not for Nature) is that about 100 of nearby dwarf galaxies
with $ M_B<-12 $ rotate slower than 35 km/s. There are some things special about 
dwarf isolated galaxies -- most of them are old, they
 rotationally support thin disks and they produce stars right now. 
We have a small but nonnegligible
collection of galaxies such as CamelopardalisB, CGCG269-049, DDO125 exhibiting regular 
rotation with maximum rotational velocities as low as 10-15\,km/s
with clear indications of modern star formation.

\bigskip 
  
$\Lambda$CDM-overabundance is quite spectacular when comparing the local Tully-Fisher relation for 
dwarf galaxies with model one in Tikhonov \& Klypin (2009)
(assuming  the simplest semi-analytical approach - abundance matching). 
   
\begin{figure}[ht]
  \includegraphics{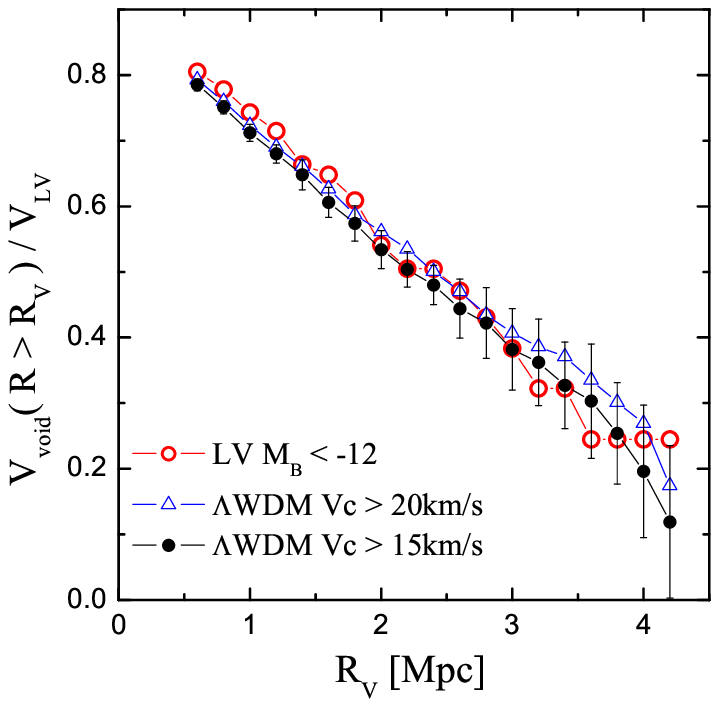}
   \caption{Volume fraction of the Local Volume (LV) occupied my mini-voids
   (VFF). The VFF of the observational sample with $ M_B<-12 $) (red
   circles) is compared with the mean VFF obtained from the 14 LVs in
   the $\Lambda$WDM simulation with haloes with circular velocity $ V_c > 20
   km/s$ (open blue triangles) and $ V_c > 15 \; km/s$ (filled black
   circles), for which the $ 1 \sigma $ scatter is also shown.}
   \label{fig:VFw}
\end{figure}

\bigskip

\begin{figure}[ht]
 \begin{center}
  \includegraphics[width=0.5\textwidth]{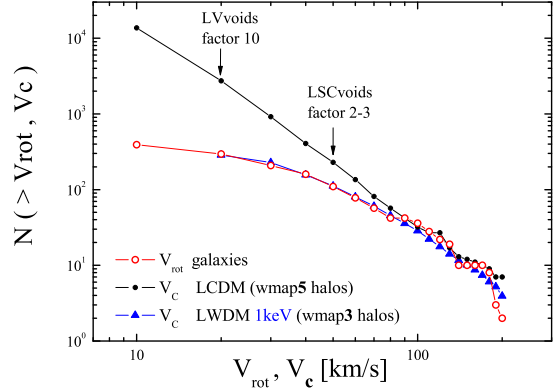}
   \end{center}
   \caption{Velocity Functions (VFs)  in 8Mpc-sphere. VF of the LV 8Mpc galaxies (red circles) 
is corrected for barion impact in $ V_{rot} $.
     Galaxy  $V_{rot}$ in LV sample is nearly complete down to 20 km/s. 
Error bar in VF on 20 km/s because of  galaxies with unknown $ V_{rot} $ is
   smaller than symbol size.
  VF of  $\Lambda$CDM CR CLUES simulation (black filled circles) and $\Lambda$WDM 1keV simulation 
(blue triangles) are corrected for adiabatic compression.   Arrows indicate 
$\Lambda$CDM-overabundance factors obtained with void-analysis.}  
 \label{fig:VF}
\end{figure}


Thus, with the reasonable assumption that halos with at least 
$ V_c>20 $\,km/s should contain galaxies we do 
have strong (about factor of 10) $\Lambda$CDM-overabundance. 
The problem is more severe than in the case of missing satellites
since much less physics may be involved to explain the discrepancy for isolated galaxies. 

\bigskip 

$\Lambda$CDM-overabundance is much more prominent in terms of observational Local Volume 
(LV -- Karachentsev et al., 2001) and model (high resolution wmap5 and wmap3 CLUES 
http://www.clues-project.org/simulations) cumulative Circular Velocity Functions -- VFs 
(with  corrections of DM halo peak circular velocity for adiabatic compression). 

\bigskip 

$\Lambda$CDM demonstrates evident discrepancy with observations which starts from 
rotational (circular) 
velocities $ V_c\sim60 $\,km/s and steeply increases towards smaller $ Vc $ ($ V_{rot} $). 
A factor of 3 $\Lambda$CDM overabundance is obtained with voids statistics on volume-limited (VL) 
Local Supercluster sample 
($ D < 25 $\,Mpc, $ M_K < -17.5 $, "Virgo" hemisphere) and it points in the direction of 
our previous results.
A factor of 10 $\Lambda$CDM overabundance on $M_B<-12$  LV galaxy sample 
(8 Mpc around MW) is obtained by a 
comparison of the observed spectrum of mini-voids in the Local Volume 
 with the spectrum of mini-voids determined from the $\Lambda$CDM simulations. 

\medskip

These two overabundance factors are nearly exact numerical points on real and model 
velocity function divergence if we use the LV Tully-Fisher dependence to correspond
 $ V_{rot} $ with $ M_B $. 
Different things coincide. The $\Lambda$CDM model predicts much more dwarf objects 
than we do see as dwarf galaxies in our very local neighbourhood.

\bigskip 

Still possibilities of quenching of star formation in small halos such as different kinds of 
feedback or (and) UV photoheating during
the epoch of reionization are proposed.   
 
\bigskip 
 
Observational possibility to solve the discrepancy relies on the large amount of LSB galaxies 
to discover. The population
of dSphs in the field is rather limited. There is only one confirmed nearby dSph in the field -- 
KKR25 (Karachentsev et al., 2001) and
there are indications that we cannot expect significant increase of new dSphs in the field. 
Other possible population should be of very low SB and thus such galaxies they should 
be quite extended. 
This possibility meets some difficulties. It is known that
blind HI surveys did not find HI clouds with masses larger than $\sim10^6 \; M_{\odot} $ 
which do not have optical counterpart.

\bigskip 

Another possibility -- dwarf galaxies locate in DM halos which are significantly more massive than we
may reasonably expect from galaxy dynamics i.e. by halo with circular velocity 
$ V_c \sim 1.5-2 V_{rot} $. Preliminary results indicate that they do not.

\bigskip 

Non-baryonic solution of the $\Lambda$CDM-overabundance with DM particles of keV mass scale 
($\Lambda$WDM - Warm Dark Matter model) is considered to be able to solve 
$\Lambda$CDM problems on small scales. 
WDM physics effectively acts as a truncation of $\Lambda$WDM power spectrum. 
$\Lambda$WDM CLUES simulation with 
1 keV particle gives much better answer than $\Lambda$CDM reproducing sizes of local minivoids. 
The velocity function of 1 keV WDM LV-counterpart reproduce observational  velocity function
remarkably well. 

\bigskip 

WDM model has its own difficulties such as uncertainties with star formation in dwarf 
WDM halos because of 
their late formation, problem of dwarf fake-halos appeared in WDM-like simulations by spurious 
fragmentation of filaments. By the way, it is still not clear how reliable are $L\alpha$-forest 
constraints which put a lower bound on probable mass of WDM particle (Boyanovsky, 2010).

\bigskip

{\bf Overall we may conclude that keV DM particles 
deserve experimental attempts of their detection.} 

\begin{figure}[ht]
 \begin{center}
  \includegraphics[width=0.5\textwidth]{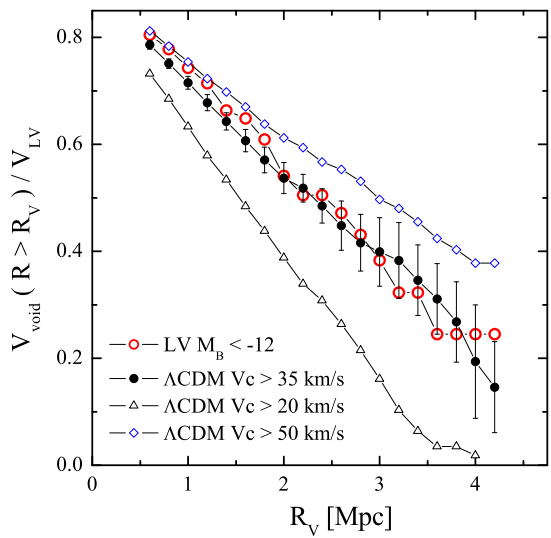}
   \end{center}
   \caption{Volume fraction of the LV occupied my mini-voids
   (VFF). The VFF of the observational sample with $M_B<-12$) (red
   circles) is compared with the mean VFF obtained from the 14 LVs in
   the $\Lambda$CDM simulation with haloes with circular velocity $ V_c > 20
   km/s$ (open triangles), $ V_c > 50 km/s $ (open diamonds), and $ V_c >
   35 \, km/s$ (filled black circles), for which the $ 1 \sigma $
   scatter is also shown.}  \label{fig:VFc}
\end{figure}

\newpage

\section{Summary and Conclusions of the Colloquium by
H. J. de Vega, M.C. Falvella and N. G. Sanchez}

\bigskip

About one hundred participants (from Europe, North and South America, Africa, Armenia, China,
India, Israel, Japan, Korea, New Zealand, Russia, South Africa, Taiwan, Ukraine) 
attended the Colloquium. 
Journalists, science editors and representatives of the directorates of several 
agencies were present in the Colloquium.

\bigskip

Discussions and lectures were outstanding. Inflection points in several current 
research lines emerged.  
New important issues and conclusions arised and between them, it worths to highlight:

\bigskip

Results and the current state of missions and ongoing projects
were reported by their teams: WMAP7, BICEP, QUAD, SPT, AMI, ACT, Planck, QUIJOTE, Herschel, SPIRE,
ATLAS and HerMES surveys.

\begin{itemize}

\bigskip

\item{The WMAP 7-year data and their cosmological interpretation 
were presented including the detection of primordial helium, significant improvements
in the high multipole temperature data and of polarization data at all multipoles, 
new limits on inflation and properties of neutrinos. The primordial spectral tilt
is $n_s=0.968\pm 0.012$~(68\%~CL), $n_s=1$ is excluded by 99.5\%~CL. The latest 95\% 
upper limit on the tensor-to-scalar ratio is $r<0.24$ (from WMAP+BAO+$H_0$)}.

\bigskip

\item{ Progress  with CMB secondary anisotropies continue rapidly:  
first `blank field' Sunyaev-Zeldovich (SZ) samples appearing and new constraints on high-$l$ 
CMB power spectrum. They were recently providing some `puzzles' as 
SZ amplitudes look sistematically smaller than theoretically expected.
The WMAP-7 SZ and the X-ray data are fine and agree but the existing 
{\it theoretical} models of the 
intracluster medium need revision since they overestimate
the amount of the gas pressure (and hence SZ) in clusters of galaxies. 
Distinction between relaxed and non-relaxed clusters becomes a crucial
issue here, WMAP-7 SZ analysis showed this distinction is significant}.

\bigskip

\item{ The QUIJOTE-CMB experiment will provide unique information on the polarization 
emission (synchrotron and anomalous) from our galaxy at low frequencies, which will be 
of value for future B-mode experiments. It will reach the sensitivity level to detect the 
B-mode signal due to primordial gravity waves if $r = 0.05$. 
Quijote will complement Planck at low frequencies, the combined
two experiments will allow improvement on  $ r $ information,
galactic magnetic field determination, primordial magnetic fields.}

\bigskip

\item{ Large scale CMB anisotropies provide information on the initial conditions of 
inflation. Early fast-roll inflation is generic, it mergers smoothly to slow roll
 and its inclusion provides a mechanism for lowest multipoles depression 
 in the standard cosmological model . 
Fast-roll depression of the quadrupole sets to about $64$ the total 
number of inflation e-folds.}

\bigskip

\item{The primordial CMB fluctuations are almost gaussian. The
effective theory of inflation \`a la Ginsburg-Landau predicts negligible primordial 
non-gaussianity, negligible running scalar index and the 
tensor to scalar ratio $ r $  in the range $0.021 < r <
 0.053$, with the best value $\sim 0.04-0.05$ at reach of the next CMB observations.
Forecasted  $r$-detection probability for Planck with 4 sky coverages is border line. 
Improved measurements on $n_s$ as well as on TE and EE modes will improve 
these constraints on $r$ even if a detection will be lacking. 
Results from Planck are eagerly expected.}

\bigskip

\item{First results from Herschel ATLAS and HerMES surveys were reported: number 
counts and surface density of sub-mm galaxies, clustering of the resolved sources, 
and the Herschel-SPIRE Legacy Survey (HSLS) programme: 2.5 to 3 million sources 
(at least 2000 bright lensed galaxies and 10,000
dusty galaxies at z larger than 5), cross-correlation
with CMB maps for the ISW and CMB lensing signal traced by dusty, starbursts at z of 1-3.}

\bigskip

\item{Most of the baryons that we expect to be associated with galaxies are missing. 
The cosmic baryon fraction is well quantified: ${f_b = 17 \pm 1\%}$.  
An inventory of the detected baryons in individual cosmic structures like 
galaxies and clusters of galaxies falls short of this universal value.  
On the largest scales of clusters, most but not all of the expected baryons are detected. 
The fraction of detected baryons decreases monotonically from the cosmic baryon 
fraction as a function of mass.  
In the smallest dwarf galaxies, fewer than 1\% of the expected baryons are detected.  
It is an observational challenge to identify the missing baryons, and
a theoretical one to understand the observed variation with the mass.}

\bigskip

\item{ A robust analysis of stellar kinematics in dSph galaxies in order to determine the true 
dark matter distribution have been presented: it is based on new studies of several thousand 
precision velocities in dSph galaxies, 
with sophisticated MCMC-based comparison of models and data. 
The stellar populations in dSph galaxies and in the Milky Way Halo are now well-established to be 
quite different.  The Milky Way is not formed from dSph-like systems. 
Very substantial progress is 
being made in quantifying feedback from the chemical distributions and star 
formation histories of 
the dSph: feedback does not operate as has been suggested, feedback was extremely 
mild in the lowest 
luminosity galaxies and does not substantially modify initial conditions.}

\bigskip

\item{An `Universal Rotation Curve' (URC) of spiral galaxies was discovered  from  
3200 individual observed Rotation Curves (RCs)
well reproduces out to the virial radius  the Rotation Curve of 
any spiral galaxy. This URC is the  observational counterpart to which the  circular  
velocity profile emerging in cosmological  simulations must comply.
Large N-body  numerical simulations performed in the $\Lambda$CDM
scenario lead to the well 
known NFW  halo cuspy density profile. However, a careful analysis 
from about 100 observed high quality rotation curves has now {\bf ruled out} the disk + NFW halo 
mass model, in favor of {\bf cored profiles}. \\
The observed galaxy surface density appears to be universal within $ \sim 10 \% $
with values around $ 100 \; M_{\odot}/{\rm pc}^2 $ 
for galactic systems spanning over 14  magnitudes, different Hubble Types, morphologies and
mass profiles determined by several independent methods.}

\bigskip

\item{ The features observed in the cosmic-ray spectrum by Auger, Pamela, Fermi, HESS, 
CREAM and others can be all quantitatively explained with the action of cosmic rays
 accelerated in the magnetic winds of very massive star explosions such as Wolf-Rayet stars, 
without any significant free parameter. All these observations of cosmic ray positrons and 
electrons and the like are due to normal astrophysical sources and processes, and do not require 
hypothetical decay or annihilation of heavy DM particles. 
The models of annihilation or decay of heavy 
dark matter become more and more tailored to explain these normal 
astrophysical processes 
and their ability to survive observations is more and more reduced.}

\bigskip

\item{ Sterile neutrinos with mass in the $\sim \,\mathrm{keV}$ 
range are suitable warm dark matter 
candidates that may help solve the small scale problems of the $\Lambda$CDM concordance model. 
These neutrinos can decay into an active-like neutrino and an X-ray photon. Abundance and phase 
space density of dwarf spheroidal galaxies constrain the mass to be in the 
$ \sim $ keV range.  
Small scale aspects of sterile neutrinos and different mechanisms of their production 
were presented: 
The transfer function and power spectra are obtained by solving the 
collisionless Boltzmann equation 
during the radiation and matter dominated eras: as a consequence,  
the power spectra features new WDM acoustic
oscillations on mass scales $ \sim 10^8-10^{9} \, M_{\odot} $.} 

\bigskip

\item{ A right-handed neutrino of a mass of a few keV appears as the most interesting candidate to 
constitute dark matter.  A consequence should be Lyman alpha emission and absorption at around a few 
microns; corresponding emission and absorption lines might be visible from molecular Hydrogen H$_2$  
and H$_3$  and their ions, in the far infrared and sub-mm wavelength range.  The detection at very 
high redshift of massive star formation, stellar evolution and the formation 
of the first super-massive 
black holes would constitute the most striking and testable prediction of this specific dark matter 
particle proposal.}

\bigskip

\item{CLUES numerical simulations with warm dark matter
of mass of $ m_{\rm WDM}=1 $ keV have been presented and its predicted 
galaxy distribution in the local universe for the $\Lambda$WDM cosmogony agrees well 
with the observed one in the ALFALFA survey. On the
contrary, the $\Lambda$CDM model predicts a steep rise in the velocity
function towards low velocities and thus forecasts much  more
sources both in Virgo-direction as well as in anti-Virgo-direction
than the ones observed by the ALFALFA survey. These results indicate 
problems for the cold dark matter paradigm, also the spectrum of mini-voids points to  
a problem of the $\Lambda$CDM model. The $\Lambda$WDM model
provides a natural solution to this problem.}

\bigskip

\item {The non-baryonic solution of the $\Lambda$CDM-overabundance with DM particles of 
keV mass scale 
($\Lambda$WDM - Warm Dark Matter) is considered to be able to solve $\Lambda$CDM 
problems on small scales. 
WDM physics effectively acts as a truncation of the $\Lambda$CDM power spectrum. $\Lambda$WDM 
CLUES simulation with 
1 keV particles gives much better answer than $\Lambda$CDM when reproducing sizes of local 
minivoids. 
The velocity function of 1 keV WDM Local Volume-counterpart reproduces the observational 
velocity function remarkably well. 
Overall, keV DM particles deserve dedicated experimental efforts of their detection.}

\bigskip

\item{Facts and status of DM: Astrophysical observations
points the existence of DM. Despite of that, proposals to
replace DM  by modifing  the laws of physics did appeared.
Notice that modifying gravity spoils the standard model of cosmology
and particle physics supported by CMB and LSS observations
not providing an alternative.}

\bigskip

\item{ After more than twenty active years the subject of DM is mature 
and  it appears divided in three sets:
(a) Particle physics DM model building beyond the standard
model of particle physics, dedicated laboratory experiments,
annhilating DM, all concentrated on wimps.
(b) Astrophysical DM: astronomical observations, astrophysical models.
(c) Numerical CDM (wimps) simulations. : The results of (a) and (b)
do not agree and (b) and (c) do not agree neither at small scales.
None of the small scale predictions of CDM wimps simulations 
have been observed: cusps, over abundance of substructures
by a huge factor. 
In addition, all direct dedicated searchs of wimps from more than twenty years 
gave null results. Something is going wrong in the DM research. 
What is going wrong and why?}

\bigskip

\item{Astronomical observations strongly indicate that
{\bf dark matter halos are cored till scales below 1 kpc}. 
More precisely, the measured cores {\bf are not} hidden cusps.
Numerical simulations with wimps (particles heavier than $ 1 $ GeV)
without {\bf and} with baryons yield cusped dark matter halos.
Adding baryons do not alleviate the problems of wimps simulations,
on the contrary adiabatic contraction increases the central density of cups
worsening the discrepancies with astronomical observations.}

\bigskip

\item{The results of numerical simulations must be confronted to observations.
The discrepancies of CDM wimps simulations with the astronomical
observations at small scales $ \lesssim 100 $ kpc 
 {\bf keep growing and growing}: 
satellite problem (for example, only 1/3 of satellites predicted by wimps
simulations around our galaxy are observed), voids problem, peculiar velocities
problem (the observations show larger velocities than wimp simulations),
size problem (wimp simulations produce too small galaxies).}

\bigskip

\item{The use of keV scale DM particles in the simulations alleviate all the
above problems. For the core-cusp problem, setting
the velocity dispersion of keV scale DM particles seems beyond
the present resolution of computer simulations. Analytic work in the
linear approximation produces cored profiles for keV scale DM particles
and cusped profiles for wimps. Model-independent analysis of DM from phase-space density
and surface density observational data plus theoretical analysis
points to a DM particle mass in the keV scale.
The dark matter particle candidates with high mass (100 GeV, "wimps") 
became strongly disfavored, while cored (non cusped) dark matter halos  and light (keV scale mass) 
dark matter are being increasingly favoured from theory and  astrophysical observations.}

\bigskip

\item{As a conclusion, the dark matter particle candidates with large mass 
($ \sim 100$ GeV, the so called `wimps') became strongly disfavored,
while light (keV scale mass) 
dark matter are being increasingly favoured both from theory, numerical 
simulations and a wide set of astrophysical observations.}

\bigskip

\item{Many researchers continue to work with $\Lambda$CDM at small scales
and to perform simulations with heavy DM candidates
(mass $ \gtrsim 1 $ GeV) despite the {\bf growing} evidence that these
DM particles do not reproduce the small scale astronomical observations
($ \lesssim 100 $ kpc). Why? [The keV scale DM particles naturally produce the
observed small scale structure].  
The answer to this strategic question is certainly not strictly scientific.}
\end{itemize}

It should be recalled that the connection between 
small scale structure features and the mass of the DM particle 
directly follows from the value of the free-streaming
length $ l_{fs} $ and is well known. Structures 
smaller than $ l_{fs} $ are erased by free-streaming.
DM particles with mass in the keV scale 
give $ l_{fs} \sim 100 $ kpc while 100 GeV DM particles produce an
extremely small $ l_{fs} \sim  0.1 $ pc. While $ l_{fs} \sim 100 $ kpc
is in nice agreement with the astronomical observations, a $ l_{fs} $ 
a million times smaller requires the existence of a host of
DM smaller scale structures till a distance of the size of the Oort's cloud
in the solar system. No structures of this type have ever been observed.

\bigskip

\begin{center}

 {\bf \em  `Examine the objects as they are and you will see their true nature;
look at them from your own ego and you will see only your feelings;
because nature is neutral, while your feelings are only prejudice and obscurity'}

\medskip

Shao Yong, 1011-1077 [quoted by Gerry Gilmore in his Lecture].

\end{center}

\bigskip

All the Colloquium lectures can be found at:

\begin{center}

{\bf http://www.chalonge.obspm.fr/colloque2010.html}

\end{center}

\bigskip
  
Best congratulations and acknowledgements to all lectures and participants which 
made the 14th Paris Cosmology Colloquium 2010 so fruitful and interesting, the 
Ecole d'Astrophysique Daniel Chalonge looks forward for you for the next Colloquium of this series.

\bigskip

\begin{figure}[ht]
\includegraphics[height=17.cm,width=17.cm]{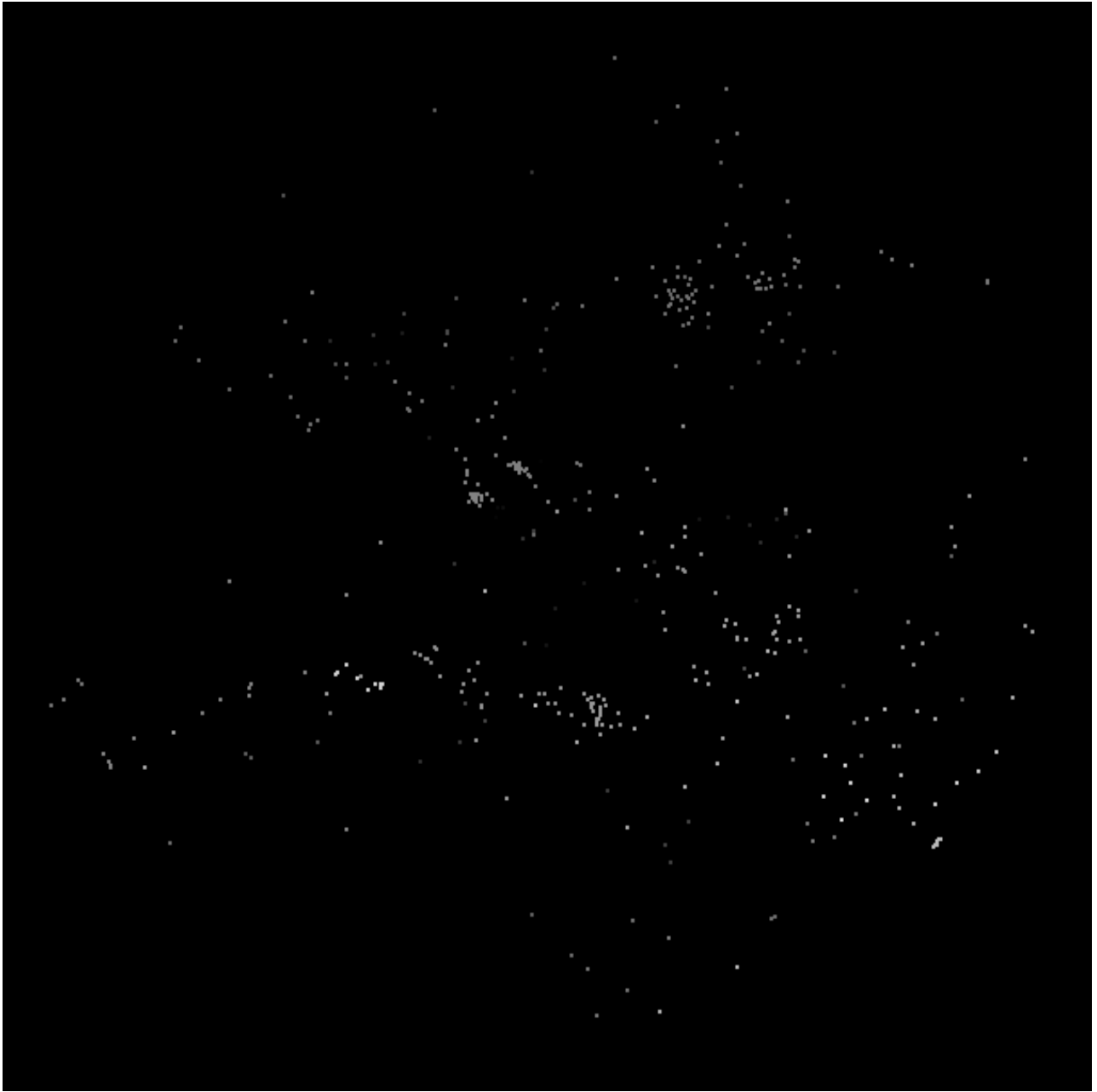}
\caption{Nearby galaxies observed in a 8 Mpc sphere around the Milky Way from data by 
Igor D. Karachentsev (SAO, RUS). More distant objects are dimer.} 
\label{obs}
\end{figure}

\begin{figure}[ht]
\includegraphics[height=17.cm,width=17.cm]{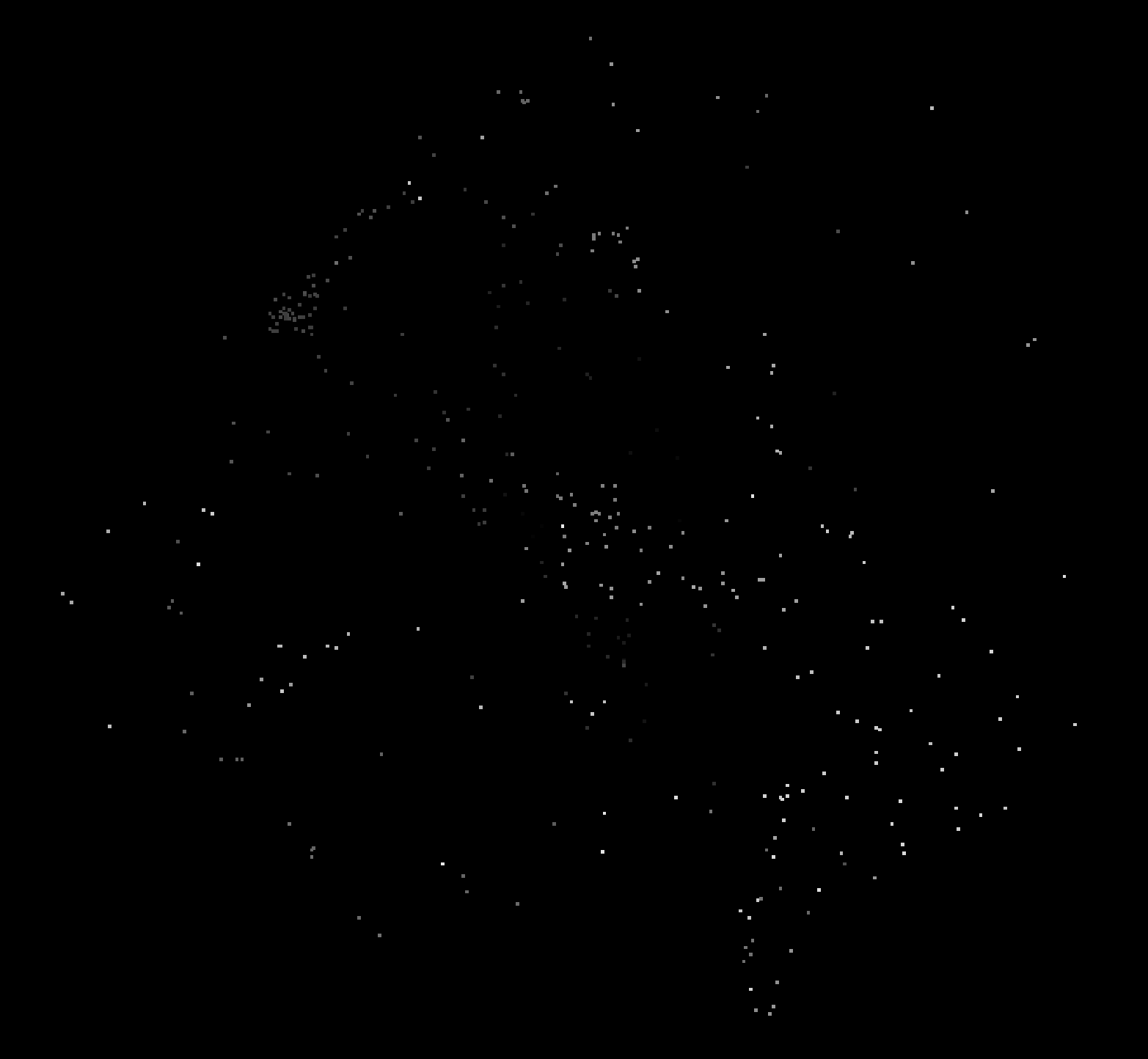}
\caption{Nearby galaxies in a 8 Mpc sphere from the $\Lambda$WDM 1 keV Clues wmap3
simulation with halos of circular velocity $ V_c > 20$ km/s from Anton Tikhonov. 
More distant objects are dimer.}
\label{keV}
\end{figure}

\begin{figure}[ht]
\includegraphics[height=17.cm,width=17.cm]{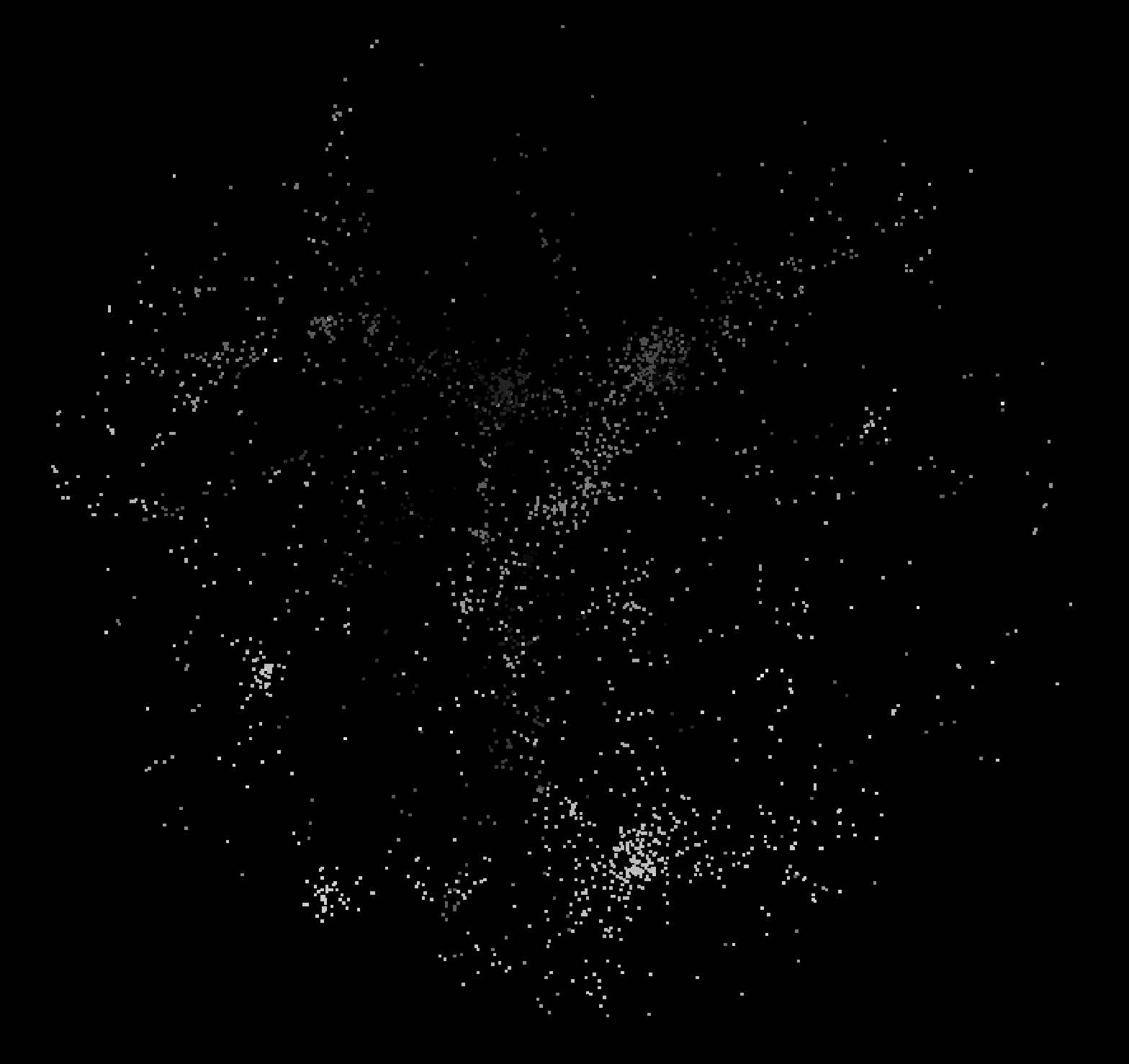}
\caption{Nearby galaxies in a 8 Mpc sphere from the $\Lambda$CDM wmap5 
simulation with $ V_c > 20$ km/s
from Anton Tikhonov. More distant objects are dimer.}
\label{wimp}
\end{figure}

We thank Anton Tikhonov for having kindly provided to us the three figures reproduced in 
the following pages (Nearby galaxy data by Igor D. Karachentsev fig. \ref{obs}; Clues
simulations with 1 keV DM particles fig. \ref{keV} 
and Clues simulations with wimps DM particles fig. \ref{wimp}).

\bigskip

The readers can compare the galaxy observations in fig. \ref{obs} with the
results of WDM 1 keV simulations in fig. \ref{keV} and with the wimps
CDM simulations fig. \ref{wimp} and see which simulations better agree with
the observations.

\bigskip

We recall that the Highlights and Conclusions of
the Chalonge 13th Paris Cosmology Colloquium 2010 \cite{cha13}
and the Chalonge CIAS Dark Matter Meudon Workshop 2010 are available
in the arXiv \cite{meu10}.

\newpage

\begin{center}

{\bf Galaxy Observations}

\end{center}

\newpage

\begin{center}

{\bf Simulations with 1 keV DM particles}

\end{center}

\newpage

\begin{center}

{\bf  Simulations with heavy WIMPS as DM particles}

\end{center}

\newpage

\section{Live Minutes of the Colloquium by Peter Biermann}

\def\simle{\lower 2pt \hbox {$\buildrel < \over {\scriptstyle \sim }$}}
\def\simge{\lower 2pt \hbox {$\buildrel > \over {\scriptstyle \sim }$}}
\def\intunits{{\rm s}^{-1}\,{\rm sr}^{-1} {\rm cm}^{-2}}


\begin{center}
{{\bf The nature of dark matter}}\\

\vskip0.5cm

P.L. Biermann$^{1,2,3,4,5}$ \\

\vskip0.5cm

$^{1}$ MPI for Radioastronomy, Bonn, Germany; 
$^{2}$ Dept. of Phys., Karlsruher Institut f{\"u}r Technologie KIT, Germany, 
$^{3}$ Dept. of Phys. \& Astr., Univ. of Alabama, Tuscaloosa, AL, USA; 
$^{4}$ Dept. of Phys., Univ. of Alabama at Huntsville, AL, USA; 
$^{5}$ Dept. of Phys. \& Astron., Univ. of Bonn, Germany ; 
\\
\end{center}

%


\subsection{CMB data}

	Anthony (Tony) N. {\bf Lasenby} (Cambridge):  The CMB in the standard model of the universe, a status report.  Big progress with secondary anisotropy experiments, new results on SZ samples and high-l CMB power spectrum;  SZ amplitudes look systematically smaller than expected; modern frontiers are polarization and high-wavenumber power spectrum; photon diffusion around recombination, Thomson scattering generates linear polarization with rms order 5 microK;  need to measure nanoK; gravity waves allow to approach physics at GUT scale; various experiments measure wavenumber around a few tens to a few thousands - many balloon ULD (ultra-long-duration) flights 2010 and future; premier European experiment is called QUIJOTE, 10 - 36 GHz, angular resolution about 1 degree; shows PLANCK one-year sky map; BICEP gives the best direct limit on polarization, tensor-to-scalar ratio;   BICEP2 test map shows dust polarization in Galactic plane at 1 - 3 percent level; polarization patterns of stacked hot and cold spots in WMAP7 03100273, 0405341 and 0710.5737 give special cold spot suggesting 3D texture; talks about biaxially symmetric BIANCHI IX models for the universe, a generalized FRW model; Dechant et al. PRD 79, 043524 (2009); structure on the largest scale could stem from an epoch, when the universe was oblate???; see Lasenby \& Doran 2005; Efstathiou (2003); dark flows Kashlinsky et al. 2008, 2009, 2010; Feldman, Hudson, Watkins 2008, 2009; Dechant, Lasenby \& Hobson 1007.1662; see Taub (1951 Ann.Math.Phys. 53, 472); AMI: eight 13m telescopes for SZ measurements in Cambridge;  for large clusters the SZ detection is independent of redshift of the cluster, a 32h observations should give 20 sigma detection;  candidate cluster detection in one of heir blank fields;  another bullet type cluster, A2146, Russell et al. 1004.1559; 13sigma detection in 9h with AMI; complex bulk flow motions; CBI and BIMA see excess emission at wave-numbers larger than 2000; Friedman et al. 0901.4334; QUAD disagrees with CBI; Luecker et al. 0912.4317; Vanderline et al. 1003.0003; Andersson et a. 1006.3068; Fowler et al. 1001.2934; Menateau et al. 1006.5126;  conflict with SZ expectations; CMB power spectrum goes flat and slowly increases with very high wavenumber;  upon questioning he states that indeed weak radio sources make up a large part of the signal; it does require modelling...
	
\vskip0.5cm

	Eiichiro {\bf Komatsu} (Austin):  WMAP7 results: cosmological interpretation; 9 years of data by August 2010; first detection of primordial helium; tilt less than unity; improved neutrino limits; confirmed polarization patterns around cold/hot spots; measurement of SZ: missing pressure? series of papers in 2010, all on arXiv; power-spectrum slightly better after 7 years than 5 years; $Y_p = 0.33 \pm 0.8$; more Helium implies fewer electrons, and so enhances damping;  ratio of first CMB power-spectrum peak to second gives baryon density, first to third gives matter density, including DM; sum mass of various neutrinos $< \; 0.6$ eV, using $H_0$; CMB polarization is generated by local temperature quadrupole anisotropy; on large angular scales $\Delta T / T$ Newton's grav pot/3; discusses temperature signal from potential disturbances, from velocity disturbances, gravitational waves ..;  here are 12387 hot spots, and 12628 cold spots - stack them; discusses E-mode and B-mode: E-mode detected, B-mode ``next holy grail"; WMAP data consistent with simple single-field inflation models; SZ effect (1969 + 1972) decrement $<$ 217 GHz, increment $>$ 217 GHz; significant difference between relaxed and non-relaxed clusters; SZ effect and CMB fluctuations can now be separated, example Coma cluster; using Arnaud et al profile outer part gives good convergence between many clusters: discrepancy Komatsu et al. 2010; key point is that cooling flow clusters are relaxed, all non-cooling flow clusters are non-relaxed clusters; Arnaud et al profile ruled out, reason Arnaud et al did not distinguish relaxed and non-relaxed clusters; cosmic ray and magnetic field pressure might be important, details highly uncertain; Fermi satellite  detects no gamma rays, so CR pressure in a certain energy range of the particles low; my comment that ISM already very complicated, determining for instance B and B$^2$ gives very discrepant results, so SZ discrepancies almost certainly imply that we learn physics of intra-cluster medium...

\vskip0.5cm

	Rafael {\bf Rebolo} (Tenerife):  CMB polarization: the QUIJOTE experiment:  Dodelson 2003 etc; scattering 
generates polarization; two orthogonal components, E-mode: tangential around any hot spot; B-mode gives vortex like 
pattern around a spot: Kamionkowske et al. 1997; Seljak \& Zaldarriga 1997; scalar perturbations give E-mode, no B-mode; 
gravitational waves gives E and B-mode - I wonder about vortices in the gas given by tidal forces (A. Lasenby answered 
this: non-linear effect); Kovac et al. 2002 Nature; Jarosik et al. 2010(WMAP7); Komatsu et al. 2010 (WMAP7); 
Chiang et al. 2010 (WMAP7); QUIJOTE science goals cosmological parameters from E-mode polarization measurements;  
built at Teide Observatory; Teide Obs. is run by IAC (Tenerife); first step goal five polarization maps in 
10 - 30 GHz to correct the 30 GHz from foreground emission and then determine the B-mode polarization; Tucci, M. 
et al. give the contribution of extragalactic radio sources (MN 386, 1729, 2008); discusses the contribution from 
spinning and magnetic dust (?); Lazarian \& Draine 2000 ApJ; status of experiment;  I expressed a bit of skepticism 
on our understanding of foreground emission, and its characteristics, such as spottiness and spectral index; I also 
asked about the influence of rotation of density perturbations and its effect on scattering - rotation is due to tidal
 forces most people think;...

\vskip0.5cm

\subsection{Cosmology with a keV DM particle}

	Hector J. {\bf de Vega} (Paris):  Inflation and keV DM in the standard model of the universe; out of equilibrium 
evolution of fast expansion, explosive particle production, see 0901.0549; inflation ends after finite number of e-folds, 
about 60; WMAP7 gives scale of inflation; it turns out that the scale of inflation is the GUT scale, $0.7 \; 10^{16}$ÊGeV, 
nice coincidence or key?; see 0703417, 1003.6108; double well potential favored; our present universe was built from the 
quantum fluctuations at the end of inflation;  DM distribution fct freezes out at decoupling; phase space density Q argument 
using Gilmore et al. 2007 and later, $Q = \rho/\sigma^3$; $Q_{today} = Q_{prim}/Z$; $Z$ between 1 and a few $10^4$;  
DM mass of particle $m = (Z/g)^{1/4}$, order keV; 0710.5180, 0901.0922; shows graph of $P(k)$ for WIMPS, keV DM 
particles, and 10 eV DM particles, P(k) cut for keV DM particles at $\simle \; 100$ kpc; for all larger scales irrelevant, 
no difference to WIMPS; argues for universal quantities, like $M_{BH}/M_{halo}$, surface density of DM, and DM density profiles; 
universal quantities may be attractors in the Poincare sense; energy scales as volume, but entropy scales as surface; satisfies 
Bekenstein bound (probably 1973 PRD paper); using then the surface densities gives about 2.6 keV (marginally higher for Fermions); 
see M.G. Walker et al. 2009 (observations), I.M. Vass et al. 2009 (simulations); his conclusion DM particle between 1.6 and 
2 keV; my comment: how about the argument on sub-thermal behaviour by Mikhail Shaposhnikov (2006/2007)?; in terms of scale 
$r_0$, central density, and halo velocity dispersion WIMPS contradict data by many orders of magnitude; argues that 
right-handed neutrinos most natural; see Galeazzi et al 2001 PRL 86, 1978; argues the empty slot of right-handed neutrinos 
can be filled with keV scale right-handed neutrinos; results: a) reproduces the phase space density, b) universal galaxy 
profiles HdV + NoSa 2009); c) universal surface density (Hoffman et al. 2007; HdV + NoSa 2010); d) alleviates satellite 
problem;  e) alleviates the void problem; also mentions our work, Blasi \& Serpico 2009, Kashlinsky et al. 2008, Watkins 
et al. 2009, Lee \& Komatsu 2010; HdV et al. 1004.1908, 0907.0006, 0901.0922; 
mentions a discrepancy in helioseismology (Asplund et al. 2009); in discussion NoSa and HdV state that $Z$ larger for 
spirals than for dSphs; big debate with Komatsu, incomprehensible; 
Tikhonov asks about Lyman alpha constraints on DM keV particles: Salucci answers that there are problems to give 
quantitative bounds from Lyman alpha since the data are not very reliable;
HdV says that these Lyman alpha  constraints only concern the Dodelson-Widrow model of sterile neutrinos.
I remember from another discussion that there is no serious problem here;...

\vskip0.5cm

	Daniel {\bf Boyanovsky} (Pittsburgh):  keV DM particle candidates, sterile neutrinos; mentions Strigari et al.; he mentions that rh neutrinos couple only to left-handed neutrinos via a seesaw mass matrix; rh neutrinos may also solve the other problems of voids, and massive halos; mentions Kusenko \& Loewenstein claim of the discovery of an emission line possibly from the decay of a rh neutrino, 0912.0552 now in ApJL; mixing angles $10^{-4}$ to $10^{-5}$ are necessary (see recent Kusenko review);  phase space density is conserved, so $n(t)/{<{p_f}^2>}^{3/2}$: phase space density diminished in violent relaxation mergers!; uses then Boltzmann-Einstein equation for DM, radiation, and baryonic density; distinguishes three stages, i) relativistic free streaming, radiation dominated, ii) non-relativistic free streaming, but still radiation dominated, and iii) matter dominated, and non-relativistic free streaming; decaying (Q) and growing (P) modes; production models:  A)  non-resonant Dodelson-Widrow production via sterile-active mixing, B) Boson decay: scalar and vector boson decay, all with thermal distribution fct  at 100 GeV; WDM acoustic oscillations at very small scales, in wavenumber $30 \, (Mpc)^{-1}$ one example; program from micro to macro...

\vskip0.5cm

	Stefan {\bf Gottloeber} (Potsdam):  Constrained Local UniversE Simulations (CLUES); nice pictures of simulations; Bolshoi simulations gives good integral mass fcts for galaxies, looks like $M_{gal}^{-0.5}$; require both very large scale and also very good mass resolution - impossible of present day computers, so use nested simulations; so ``constrained" local universe simulations, try to simulate the local environment of our galaxy; shows lots of nice 3D movies;  mass accretion histories of the DM halos; $ 10^{10.5} $ solar masses at redshift 4, and about $10^{12}$ today; satellites tend to enter our galaxy froma preferred directions; the matter stripped from these subhalos retains a memory of that direction; cosmology with WDM: shows parallel simulations, with CDM and WDM; he concludes, that $m_{WDM} \; = \; 1$ keV is a lower limit;  HI Arecibo survey gives LF, and CDM does not fit, WDM fits - lower LF, Zavals et al. 2009; spectrum of mini-voids (Tikhonov 2009; later talk); also suggests WDM models; critical mass $M_c$ of star formation: for halos with less mass no star formation: runs from $ 10^8 $ solar masses at redshift 6 to $10^{10}$ solar masses today;  make DM decay sky models, Antonio Cuesta et al.; some questions about this simulated sky map, contributions from Galactic halo, local universe; there must be an Olbers paradox for such models;...
	
\vskip0.5cm

	Claudio {\bf Destri} (Milano):  Fast-roll eras in the effective theory of inflation, low CMB multipoles and a MCMC analysis of the CMB + LSS data: claudio.destri@mib.infn.it; many eqs; 
horizon problem; entropy of the universe dominated by photons and neutrinos; tensor-scalar ratio in generic single-field inflation; paper with Claudio D.: Boyanovsky et al. 2009 IJMPA 24; 
using WMAP + other data tensor to scalar ratio $> $0.025, best fit 0.05; Destri-de Vega-Sanchez 0906.4102; pre-inflation stage $ \frac{1}{a H} \, \sim \,  a^2/\sqrt{a^6 + const.} $ .., 
slow-roll inflation stage 
$ \frac1{a H} \, \sim \, \frac1{a} $ .., radiation dominated stage $ \frac{1}{a H} \sim a $ .., matter dominated stage $\frac{1}{a H} \, \sim \,  \sqrt{a} $..; 
Destri-de Vega-Sanchez Phys. Rev. D81 (2010). Komatsu et al. ApJ 2010 CMB polarization; 
Ayaita et al. PRD 2010 on too few hot spots ; Bennett et al. 1001.4758; Destri + HdV + NoSa PRD 78, 023013 (2008); fast roll depression sets to about 64 the number of e-folds in inflation; 
1003.6108; additional explanations by NoSa;
	
\vskip0.5cm
	
\subsection{N-body Simulations}

	Carlos S. {\bf Frenk} (Durham):  Cosmology in our backyard: first detection possibilities for detecting of supersymmetric cold dark matter particles; a)  at LHC, b)  underground las, c)  indirectly via decay emission; Peebles 1982, Davis et al. 1985, Bardeen et al. 19..; Sanchez et al. 2006; at really small scales the nature of DM really plays a role; non-baryonic DM candidates, a) hot, neutrino, b) warm, sterile neutrino, c) cold, axion, neutralino; free-streaming cutoff, $10^{-6} \, M_{\odot}$ for 100 GeV WIMP; Vie et al. 2008, Boyarsky et al. 2009;$ M_{cut} \, \simeq \, $...; Davis et al. 1985 again; 30 eV neutrinos give too much clumping in simulations; warm DM gives less small structure, same large scale structure in simulations as CDM, Gao et al.; test CDM on a)  structure of dark matter halos, b)  number of satellites, c) remnants of merging, streams; shows movies; more massive halos and halos that form earlier have higher density, Navarro, Frenk, White 1997; there is no obvious density plateau at the center in these simulations; Springel et al. 2008; 6 different galaxy size halos simulated at varying resolution, with the highest at 15 pc softening scale, and $> \, 10^{9}$ particles; density profile of NFW goes down to very small radii; slight but significant deviations from similarity: simulations give cuspy profiles: spike at center; but how about nature?; halo likely to be modified by the galaxy forming in it; Vikhlinin in et al. 2006; central profiles of clusters are very well fitted by NFW; data from 2 to 0.01 of $r_{500}$; Carlos Frenk clearly says: CDM predicts cusps; Bradac et al. 2008; to study dwarf spheroidal galaxies with Jeans eq, terms stellar density profile, radial velocity dispersion and anisotropy; assumes anisotropy zero; Milky Way dwarfs test $\rho \, \simÊ\, x^{-a}$ in the limit of small radius $x$, all galaxies have either a = 0.5 or a = 1 (NFW); Strigari, Frenk, White 2010; conclusion: photometric and kinematic data for Milky Way satellites consistent with cuspy NFW profiles; he emphasizes that he is testing for cuspy profiles, he is NOT testing core-profiles; he is only testing data versus prediction, and his prediction was cusps, and finds that the data are fully consistent; NoSa contradicts him, and says CDM is more than his fits, of course also true; she emphasizes that a more complete theory fails for CDM; Carlos Frenk in turn emphasizes that he is not testing such things, he is just testing his CDM version via NFW-updates, and only to its limits, as done in the Aquarius runs; some arguments on the assumption of an-isotropy in phase space, which he does NOT use; next step in his argument sub-halos: simulations produce $> \, 10^{5}$ sub-halos, observations detect only a few tens;  argues about Strigari et al. 2008: WDM or astrophysics inside CDM halos?  normal conclusion is that there is a special scale in cosmology; he argues about astrophysics, see Gao \& Lovell 2011; questions: how many sub-halos make a visible galaxy; Millenium run compared with 2dF.. observations; Benson et al. 2003; SN feedback and photoionization; at high mass AGN feedback; with these two/three effects predicted mass factor comes down; Kauffmann et al. 1993, Bullock et al. 2001, Benson et al. 2002; reports about big argument with Scott Tremaine; 
Koposov et al. 2008; Cooper et al. 2009; Okamooto \& Frenk 2009; Okamoto et al.; Irwin; compares visible satellites and dark satellites; he has reionization at redshift 9; finds a critical velocity 
from reionization, and that defines the critical mass which is visible, not sterile neutrino; states strongly, if CDM is right, the dark sub-halos must be there; Dandan Hu + Aq 2009, 2010; 
Cooper et al. 2010; Bernard Sadoulet asks about the Cornell HI Arecibo project and what Carlos Frenk predicts; life is complicated is the answer; his main point is that CDM cannot so easily be ruled out; 
	
\vskip0.5cm

	Gerard (Gerry) F. {\bf Gilmore} (Cambridge): DM on small astrophysical scales: where are we with dwarf spheroidal galaxies?  a) sizes, b)  ages/chemical abundances/first stars, c)  kinematics, d)  masses; Sgr dSph (Ibata et al. 1995) proves that late minor merging occurs in MW, but not dominant in evolution of MW except ion outer halo $>$ 25 kpc; Oh et al., de Blok et al. 2008; states ``cusped DM is very hard to find"; Kuijken \& Gilmore 1989, 1991 remains the only experimental determination of DM density; satellite number problem, Moore et al 1999; one can limit feedback from chemical enrichment data; Belokurov et al. 2006a; Governato et al. 2010; updated field of streams; update from Gilmore et al. 2007; absolute magnitude versus half-light-radius show globular clusters versus small galaxies : show clear gap; Belokurov et al. 2009, 2010; chemistry?  big scatter from single SNe?  metallicity depletion and energy feedback connects clearly to gas loss; Wyse \& Gilmore 1993; shows diagram of O/Fe versus Fe/H to distinguish the various IMFs and SNe; Kobashi et al. 2006; Ruchti et all 2010; dSphs vs MW abundances, halo/thick disk is NOT dSph graveyard; Koch et al. 2008; Shetrone et al. 2008; Frebel et al. 2009; Aoki et al. 2009; ...; metallicity luminosity relation: Norris, Gilmore et al. 2010a, b, ..; dispersion in metallicity increases as luminosity decreases- consistent with stochastic inhomogeneous enrichment, gentle feedback; significant stripping ruled out; Wyse \& Gilmore 1995; Edvardsson et al. 1993; G-dwarf problem?  Nomoto, Komatsu et al stellar evolution of zero heavy element stars; finds Carbon rich extremely metal-poor stars! stars in dSphs are younger, and have different chemistry than the halo and thick disk stars; Strigari et al. 2008 Nature diagram extended to $10^{2.5} \, L_{\odot}$; shows Fe/H, M/L and mass within 300 pc across $10^{2.5}$  to $10^{7.5} \, L_{\odot}$; determines the true velocity dispersion of some systems to be of order 3 km/s; many velocity dispersions of order 5 km/s: clear implication, that this cannot be due to temperature effects, too cold; stellar density increases; Walker et al. 2006, 2009, Gilmore et al.; Koch, GG et al. 2007; cored and cusped halo profiles fit almost equally well!  cores slightly favored, but not conclusive; playing on anisotropy of phase space; reminds that the use of the Jeans equation assumes a given velocity dispersion profile - hat is the key mistake in using this eq (essentially hydrostatic equilibrium for a stellar distribution); now they use velocity distribution fcts; Wu \& Tremaine 2006, Wu 2007, Lukas 2002 + 2005, Wilkinson et al. 2002; dispersion profile as test - positive; Conclusions: a)  minimum physical scale for galaxies, half light radius $>$ 100 pc, b)  cored (?) profiles, with similar low mean mass densities $\sim \, 0.1 \, M_{\odot}/pc^{3}$ phase space density fairly constant , maximum for galaxies, are the the first halos?  c)  pre-galactic abundances;... long discussion between Carlos Frenk and Gerry Gilmore;... at tea/coffee had a discussion with GG: he confirmed that his key work was on the reionization and feedback argument for small galaxies, and his work demonstrates that a) the velocity dispersion for many is smaller than a temperature effect can reproduce, and b) that the required star formation history for the feedback argument fails; these small galaxies do show a special mass scale;... they show a mass distribution, and this distribution peaks around $5 \, 10^{7} \, M_{\odot}$;..

\vskip0.5cm

\subsection{Galaxy data}

	Paolo {\bf Salucci} (Trieste):  Universality properties in galaxies and cored density profiles (cores -$>$ no spike); 1996 MN, 2007 MN, 2009 MN; shows diagram central surface brightness vs magnitude; fraction of luminosity versus fraction of radial scale universal profile for spirals;  no DM in the innermost regions of galaxies; giant ellipticals: no DM at center, velocity dispersion not controlled by DM; the slope of ${d log V}/{d log R}$ is crucial: the slope measures the presence of DM; introduces the details of his ``Universal Rotation Curve (URC)"; Burkert profile provides excellent fit, better than NFW; halo mass fct $\sim M_h^{-1.84} \, d M_h$; tries out his parametrization on galaxy ESO 116-... and Burkert fits best; 50 objects investigates, NFW inconsistent; DDO47 also does not work with NFW; length scale of galaxy correlates with core radius, Donato et al. 2004; central baryonic surface density constant among galaxies (Gentile et al. 2009 Nature); shows some amazing colorful movies to illustrate things; Carlos Frenk defends the NFW work; Paolo S. points out that some of his work was done and published before NFW (1997);  long debate on core vs cusp (density spike), between Carkos Frenk and NoSa + HdV;...  Before start discussion with Paolo Salucci on the history of the arguments about rotation curves; suggested that many early papers misleading in the sense, that there was an argument only on invisible matter, not on any cosmological contribution.  Even Zwicky (1933) talked about baryonic missing matter, like Oort 1932.

\vskip0.5cm

	Stacy S. {\bf McGaugh} (Maryland):  Baryon content of cosmic structures and its relation to DM; $\Omega_b = 0.042$, and $\Omega_m = 0.24$, but structure? average ratio 0.17; McGaugh et al. 2010: clusters dominant baryons gas $\sim 10^{14} \, M_{\odot}$, groups stars? $10^{13} \, M_{\odot}$, ellipticals stars $10^{13} \, M_{\odot}$, spirals, gas rich late spirals, dSph satellites;   clusters may have less than the global average in baryonic fraction $f_b$; Giodini et al. 2009; $f_b$ for stars increase with lower mass clusters, while the dominant gas $f_b$ decreases with lower mass towards groups, where the two fractions overlap; spirals stars, atomic gas, molecular gas, Young \& Knezek 1989, McGaugh \& Blok 1997; molecular gas tricky, he uses scaling relation (if CO, bad);  NGC2403 Fraternali et al.; Lund et al. 2006; Xue et al. 2008; Sellwood \& McGaugh 2005; total mass of Galaxy  $1.2 \, 10^{12} \, M_{\odot}$; McGaugh 2004, 2005; baryonic Tully-Fisher relation; ellipticals faber-Jackson relation; he ignores gas in groups, says too little known (my comment: we detected hot gas in groups almost 30 years ago, and published in ApJL; in one case it was a lot of gas, NGC5846); Stark et al. 2009, Trachternach et al. 2009; Kuzio de Naray et al. 2006. 2007, etc); local dwarf data Walker et al. 2009, Mateo et al. 1998, Martin et al. 2008; $m_b \; \not \sim \; M_{500}$; various models 2010, Trujilo-Gomez et al., De Rossi, unreadable; detected baron fraction declines with decreasing total mass; stellar mass fraction peaks between $10^{12}$ and $10^{13} \, M_{\odot}$; galaxies suffer a baryon deficit problem; could be molecular, see Pfenniger \& Combes; Pfenniger \& Revaz 2005; Hoekstra et al. 2001; Pederson et al; Anderson \& Bregman 2010; McGaugh \& Wolf 2010; reionization does not seem to work; best bet ?? maybe gas (just as Jerry Ostriker says); 

\vskip0.5cm

	Asantha {\bf Cooray} (Irvine, he):  First large scale structure and cosmologiocal results from ATLAS and HerMES surveys with the Herschel Observatory (came out a few days ago): shows many maps with very high angular resolution in multiple-color, three long wavelength FIR; $ L(CO, Arp220) \simeq 10^{8} \, L_{\odot}$; achieved instrument noise in repeated 30 arcsec beam about 10 mJy/beam (Nguyen et al. 2010); SPIRE instrument noise = confusion noise in 2 repeat scans; at 2 micron and at 200 micron the same W m$^{-2}$ 
sr$^{-1}$, so same energy density; SPIRE source counts at 250 micron Oliver et al. 2010; Xu et al. 2003, Lagache et al. 2004, Negrello et al. 2007, Le Borgne et al. 2009, Pearson et al. 2009, Rowan-Rowinson 2009, Valiante et al. 2009, ..., number counts of bright galaxies (ULIRGS+) over-predicted by models; source counts reach 80 percent of BG at 250 micron, 80 percent at 350 percent, and 85 percent of 500 micron, using counts, $P(D)$ analysis, and stacking; colors generally spread redder than models predict, so colder dust or higher redshift (as we argued); the most likely explanation, he says, is actually high redshift; Gruppioni et al. 2010; Eales et al. 2010; starbursts dominate at high $L$ and high $z$; Elbaz et al. 2010; L. Shao 2010; argument that starburst and AGN luminosities coupled by mergers (sure..); Amblard et al. 2010;  Dowell et al. 2010; Zemcov et al. 2010; Cooray et al. 2010; bulk of far-IR bg produced by milky-way like halos at redshifts 1 to 3; special A\&A issue on Herschel out; ESA first results symposium talks online; Cooray et al. 2010; 

\vskip0.5cm

	Anton {\bf Tikhonov} (St. Petersburg):  Sizes of mini-voids and the Tully-Fisher 
relation in the Local Volume; another CDM overabundance problem and its possible solution;  
compares spectrum of void sizes in Local Volume sample with simulations; Karachentsev et al. 
2007, 2004; Tikhonov \& Klypin 2009; Tully et al. 2008 ApJ; defines cumulative void fct; 
Tikhonv \& Kararentsev 2007 as a fct of void size, about 1 at 1 Mpc, and about 0.2 at 4 Mpc, 
all in Local Volume (maybe volume fraction of voids above a certain size); the observed void 
fct disagrees strongly with LCDM simulations, theory a factor of 10 more than observed;  
Makarov \& Karantentsev 2010; Afanasiev \& Moiseev 2005; Begun et al. 2006; possible solutions: 
a)  hundreds of dSphs or LSBs still to find, b)  dwarf galaxies are hosted by significantly 
more massive halos, c)  dwarf formation was suppressed, d)  LWDM, truncation of scales, 
$ m_{DM} \, \simeq \,  1$ keV;... Chengalur \& Begum (GMRT);  
	
\vskip0.5cm

\subsection{Evidence for DM and Massive star explosions}

	{\bf PLB} (Bonn, Tuscaloosa):  Dark matter has been detected since 1933 (Zwicky) 
and basically behaves like a non-EM-interacting gravitational gas of particles.  
From particle physics Supersymmetry suggests with an elegant argument that there should be a 
lightest supersymmetric particle, which is a dark matter candidate, possibly visible via decay 
in odd properties of energetic particles and photons:  Observations have discovered i) an 
upturn in the CR-positron fraction (Pamela: Adriani et al. 2009 Nature), ii) an upturn in the 
CR-electron spectrum (ATIC: Chang et al. 2008 Nature; Fermi: Aharonian et al. 2009 AA), iii) a 
flat radio emission component near the Galactic Center (WMAP haze: Dobler \& Finkbeiner 2008 ApJ), 
iv) a corresponding IC component in gamma rays (Fermi haze: Dobler et al. 2010, Su et al. 2010 arXiv), 
v) the 511 keV annihilation line also near the Galactic Center (Integral: 
Weidenspointner et al. 2008 NewAR), and most recently, vi) an upturn in the CR-spectra of all 
elements from Helium (CREAM: Ahn et al. 2009 ApJ, 2010 ApJL; for H and He the upturn has been 
confirmed by Pamela, shown at the COSPAR meeting July 2010).  All these features can be 
quantitatively explained with the action of cosmic rays accelerated in the magnetic winds of 
very massive stars, when they explode (Biermann et al. 2009 PRL, 2010 ApJL), based on 
well-defined predictions from 1993 (Biermann 1993 AA, Biermann \& Cassinelli 1993 AA, 
Biermann \& Strom 1993 AA, Stanev et al 1993 AA).  While the leptonic part of these 
observations may be explainable with pulsars and their winds, the hadronic part clearly 
needs very massive stars, such as Wolf-Rayet stars, their winds and their explosions. 
What the cosmic ray work (Biermann et al., from 1993 through 2010) shows, that allowing 
for the magnetic field topology of Wolf Rayet star winds (see, e.g. Parker 1958 ApJ), 
both the leptonic and the hadronic part get readily and quantitatively explained, close 
to the predictions, without any significant free parameter, so by Occam's razor the 
Wolf-Rayet star wind proposal is much simpler.    This allows to go back to galaxy data 
to derive the key properties of the dark matter particle: Work by 
Hogan \& Dalcanton (2000 PRD, 2001 ApJ), Gilmore et al. (from 2006 MNRAS, 2007 ApJ, etc.), 
Strigari et al. (2008 Nature), Gentile et al. (2009 Nature); work by Boyanovsky et al. (2008 PRD), 
de Vega \& Sanchez (2010 MNRAS) clearly points to a keV particle.  
A right-handed neutrino is a candidate to be this particle (e.g. Kusenko \& Segre 1997 PLB; 
Fuller et al. 2003 PRD; Kusenko 2004 IJMP; for a review see Kusenko 2009 PhysRep; 
Biermann \& Kusenko 2006 PRL; Stasielak et al. 2007 ApJ; Loewenstein et al. 2009 ApJ; 
Loewenstein \& Kusenko 2010 ApJL): This particle has the advantage to allow star 
formation very early, near redshift 80, and so also allows the formation of supermassive 
black holes, possibly formed out of agglomerating massive stars in the gravitational 
potential of a dark matter clump; the stellar wind limit derived by Yungelson et al. 2008 AA 
does not apply for stars at near zero heavy elements, since such stars have weak winds.  
Black holes in turn also merge, but in this manner start their mergers at masses of a 
few million solar masses; the mass is given by the instability of stars at such a mass 
due to General Relativity and radiation effects.  This readily explains the supermassive 
black hole mass function as the result of mergers between black holes.  
The corresponding gravitational waves are not constrained by any existing limit, 
and could have given a substantial energy contribution at high redshift.  
Our conclusion is that a right-handed neutrino of a mass of a few keV is the most 
interesting candidate to constitute dark matter.  A consequence should be Lyman alpha 
emission and absorption at around a few microns; corresponding emission and 
absorption lines might be visible from molecular Hydrogen H$_2$ (Tegmark et al. 1997 ApJ) 
and H$_3$ (Goto et al. 2008 ApJ) and their ions, in the far infrared and sub-mm 
wavelength range.  The detection at very high redshift of massive star formation, 
stellar evolution and the formation of the first super-massive black holes would 
constitute the most striking and testable prediction of this specific dark matter 
particle proposal.  Questions about massive stars, and about the other right-handed 
neutrinos; I emphasized I) the arguments from CR physics, so debunking DM decay 
arguments, and II) the very early star formation, which should be testable by 
observation, and would be a phantastic discovery.

\vskip0.5cm

	Bernard {\bf Sadoulet} (Berkeley):  Direct detection of dark matter:  
Why WIMPS? future of direct detection?  dark matter could be due to new physics at the 
TeV scale; $ \Omega_m h^2 \, = \, 10^{-26.6} cm^2/\sigma v = 0.12 \; => \;
 \sigma = \alpha^2/M_{ZW}^2 $ 
points to WZ scale at point, when going non-relativistic; WIMPS; 
$ {\Lambda}$ CDM is still alive; halo WIMP scattering, annihilation products; 
LHC detection; elastic scattering in halo, energy deposition a few keV; 
signal nuclear recoil; bg electron recoil; estimated rate 1 event per kg per month; 
should show annual modulation (but would several thousand events); typical plot 
cross-section versus mass of WIMP in log-log; various experiments give exclusion regions, 
January 2009 compilation by Jeff Filippini, Savage et al.; DAMA claim April 2008 
still stands: if WIMPS exist, we expect a modulation in event rate - DAMA claims 
to have seen it; DAMA claims 3 keV peak cannot be fully explained by $^{40}$K escape peak; 
Bernard Sa states: ``not a WIMP, incompatible with other experiments"; 0907.1438; 
DAMA modulation is proportional to bg, very disturbing; other criticism 
"not blind analysis"; nobody finds the result plausible enough to repeat the experiment; 
conclusion: a different team has to repeat the experiment at the South Pole (e.g.); 
CDMS II Dec 2009:  ionization + a-thermal photons; CDMS bind analysis: 
timing discrimination to find two events; result in Science Feb 12, 2010; CDMS vs. 
EDELWEISS; EDELWEISS presented results July 18, last Sunday: one event of interest 
in nuclear recoil region; new results from ``Xenon 100" May 2010: exclusion limit 
similar to CDMS; Savage et al. 1006.0972;  Hooper et al. 1007.1005; new system using 
metastable detectors, so energy deposit creates bubble, or else; many various detection 
techniques, like liquid Argon;   there approaches: a) LHC, b)  laboratory, c) cosmos;

\vskip0.5cm

	Felix {\bf Mirabel} (Saclay - Buenos Aires):  Cosmic evolution of stellar BHs and the end of the dark ages; dark ages count from 400,000 years $< \, 10^{9}$ years; asking about reionization of the universe (Madau, Rees, Volonteri et al. 2004; Loeb et al. and many others); he proposes as a working hypothesis that BH high mass X-ray binaries may have played a complementary role, to that of their massive stellar progenitors in the process of the reionization of the universe (fits my ideas, only I use an even higher redshift perhaps); talks about stellar forensics, wonders about the implosion of massive stars without SNe; Heger et al. 2003, Georgy et al. A\&A 502, 611 - 622 (08/2009), Mapelli et al. 2010, Fryer 1999; the BHs in low mass galaxies have larger mass than those in our Galaxy; Zampieri \& Roberts 2009, Pakull et al. 2020, King et al. 2001; the occurrence rate of ULXs with $M_{BH} \, > \, 30 \, M_{\odot}$ per unit galaxy mass in starburst galaxies is a decreasing fct of the metallicity of the host galaxy; example ULXs in the Cartwheel galaxy (Gao et al. 2007); Rappaport et al. 2010 on Arp147; massive stars of high metallicity end as neutron stars rather than BHs; Mirabel et al. 1999, Vrba et al. 2000, Davies et al. 2009, Muno et al. 2006, Davies et al. 2009; progenitors of core-collapse SNe have masses $< \,  20 \, M_{\odot}$, Smartt et al ARAA 2009, Smith et al. 2010; Mirabel \& Irapuan Rodrigues 2001 - 2009; assumption: if BH binaries have no anomalous motion they must have been formed without energetic SN kick; for such an exercise it is necessary to determine the space motion of such systems: Mirabel et al. 1992, Mirabel \& Rodriguez 1994, Dhawan et al. 2007; shows the Galactic trip of Sco X-1 (Mirabel \& Rodrigues 2003); two run-away BHs Mirabel et al. 2001 Nature, Israelian et al. 1999 Nature; suggests that the $10 \, M_{\odot}$ BH in Cyg X-1 was born in the dark (Mirabel \& Rodrigues 2003 Science); gives the number of about $10^{8}$ stellar BHs in the Galaxy; GRS1915+105 and V404 Cyg have BHs with $> \, 10 M_{\odot}$ and have small velocities, especially small vertical velocity components; hosts of LLGRBs are small Irr galaxies, Fruchter et al. 2006 Nature, and Mirabel et al. 2003, Levesque et al. 2010;  suggests that the fraction of BHs should increase with redshift; Turk et al. 2009 Science, Krumholz et al. 2009 Science, Stacy et al. 2010 ApJ; Pop II stars were multiple systems dominated by binaries with $10 \; - \; 100 \, M_{\odot}$;  a GRB at z = 8.2 has similar properties to GRBs at lower redshift, Salvaterra et al. 2009 Nature; HST galaxies at z = 8 - 9 are photon starved to reionize the universe; unless there are small galaxies below detection, top heavy IMF (Lorenzoni et al., Bouwens et al. 2010 (my comment: that fits our ideas, expressed in Caramete \& Biermann 2010); estimates the number of ionizing photons from the microquasar to be about 3 times that of the progenitor star (Mirabel et al. work in progress); microquasars ionize multiple times due to the energetic photons (similar to our argument in Biermann \& Kusenko 2006 PRL, then for DM rh neutrino decay); conclusion: a large proportion of stellar BHs important for reionization;.. I comment with our BH paper (2010 arXiv), and suggest to go even further (agglomerating massive stars in a group inside a DM halo clump all the way to make BHs of a million $M_{\odot}$); 
	
\vskip0.5cm
	
\subsection{The near future: PLANCK}
	
	Paolo {\bf Natoli} (Rome), report for Reno Mandolesi (Bologna):  The Planck Satellite:  first fluctuation and GW generator, later fluctuation amplifier, but GW dissipator; 436 days since launch, 100 percent of sky covered; about half the sky has been covered twice; the currently approved mission operation will cover $>$ four sky surveys, until the end of the cold phase (Nov 2011); now end of cryo-lifetime expected to end of Jan 2012; lots of technical details; he suddenly mentions that PKS 1222+21 = 4C21.35 jumped up by a huge factor at TeV; they see every single flat spectrum radio source;... one wonders how many there are...(my comment: down to 0.25 Jy one every two degrees at 5 GHz even; S51803+78 is flat from 5 GHz to 60 microns);...
	
\vskip0.5cm

\subsection{Conclusion}

	Norma G. {\bf Sanchez} (Paris):  Predictions of the effective theory of inflation and keV dark matter in the 
standard model of the universe:  Baryonic matter 4.6 percent, DM 23.4 percent, 72 percent DE; matter-DE equality at 
$z \, \simeq \,  0.47$; end of inflation $z \, \simeq \,  10^{29}$; EW phase transition $z \, \simeq \, 10^{15}$; 
QCD phase ransitioon $z \, \simeq \, 10^{12}$; BBN $z \, \simeq \, 10^{9}$; DM outside standard model of particle physics; 
DE described by cosmological constant $\Lambda$;  during inflation the universe expands by $e^{62} \, \simeq \,  10^{27}$; 
inflation lasts $10^{-36}$ s and ends by $z \, \simeq \, 10^{29}$; energy scale, when inflation starts is about $10^{16}$ 
GeV, about GUT scale; Planck time $10^{-44}$ s; fast roll inflation $10^{-39}$ to $10^{-38}$ s, slow roll inflation $10^{-38}$ 
s to $10^{-36}$ s; fast roll inflation suppresses the CMB quadrupole;   Boyanovsky et al. 2006 PRD 74, 123006; 
$z \, \simeq \, 10^{56}$ at beginning of inflation; compares inflation with semi-classical quantum gravity a la Hawking;  
compares also BH evaporation with cosmic inflation, inverse; suggests that string temperature is a quantum gravity concept;  
Destri et al. 2008 PRD 77, 043509; r is tensor to scalar perturbation ratio, so a measurement of the primordial gravitons, 
lower bound $>$ 0.016 at 98 \% C.L., $>$ 0.049 at  \% C.L., 0.055 most probable level; 
MCMC stands for Monte-Carlo-Markoff-Chain = MCMC, and is a method of analysis; basically predictions for WMAP9 and PLANCK:  
Burigana et al. (incl. NoSa + HdV) 
1003.6108; $0.028 \, < \, r \, < \, 0.116$ at 95 \% C.L., best value at 0.04, 
tilt $n_s \, = \,  0.9608$. For dark matter distinguishes (a) particle physics, (b) astrophysics, (c) 
numerical simulations: and b as well as b and c do not agree; answers in their papers: 
2008 to 2010, PRD, MNRAS and astro-ph; her expression: 
$0.45 \, 10^{3} \, M_{\odot} \; < \; {M_J}(z) \; (1 + z)^{-3/2} \, < \,  0.45 \, 10^{7} \, M_{\odot}$: based on Biermann \& Kusenko 2006 
I conclude for z = 100 this gives $10^{5.65} \, M_{\odot} \; < \;  M_J \; < \;  10^{9.65} \, M_{\odot}$; 
interesting for massive star clumps; if I take the geometric average, 
I get $10^{7.76} \, M_{\odot}$, then $10^{-1}$ for baryonic matter, so $10^{6.76} \, M_{\odot}$, so just 
what you need to make a massive star clump; good to make the first SMS BHs, 
necessary near $10^{6} \, M_{\odot}$;...

\vskip0.5cm

My conclusion: Things are beginning to hang together, and we can now make quite specific 
predictions as a consequence of the keV DM model.  
If the right-handed neutrino were this particle, star formation and the first super-massive 
black holes could be formed quite early, possibly earlier than redshift 50.  
A confirmation would be spectacular.

\bigskip

{\bf Acknowledgements}

\medskip

PLB would like to thank G. Bisnovatyi-Kogan, J. Bl{\"u}mer, R. Engel, T.K. Gaisser, 
L. Gergely, G. Gilmore, A. Heger, G.P. Isar, P. Joshi, K.H. Kampert, Gopal-Krishna, 
A. Kusenko, N. Langer, M. Loewenstein, I.C. Mari\c{s}, S. Moiseenko, B. Nath, 
G. Pavalas, E. Salpeter, N. Sanchez, R. Sina, J. Stasielak, V. de Souza, H. de Vega, 
P. Wiita, and many others for discussion of these topics.  


\bigskip

\begin{itemize}

\item  Adriani, O., et al. (Pamela Coll.), \Nature {\bf 458}, 607 - 609 (2009); arXiv 0810.4995

\item  	Aharonian, F., et al., (H.E.S.S.-Coll.), \AA {\bf 508}, 561 - 564 (2009); arXiv:0905.0105

\item  Ahn, H.S. et al. (CREAM-Coll.), \ApJ  {\bf 707}, 593 - 603 (2009); arXiv:0911.1889

\item  Ahn, H.S. et al. (CREAM-Coll.), \ApJL  {\bf 714}, L89 - L93 (2010); arXiv:1004.1123

\item  Biermann, P.L., \AA {\bf 271}, 649 (1993) - paper CR-I; astro-ph/9301008

\item  Biermann, P.L., \& Cassinelli, J.P., \AA {\bf 277}, 691 (1993) - paper CR-II; astro-ph/9305003
   
\item  Biermann, P.L., \& Strom, R.G., \AA {\bf 275}, 659 (1993) - paper CR-III; astro-ph/9303013
   
\item  Biermann, P.L., 23rd ICRC, in Proc. ``Invited, Rapporteur and Highlight papers"; Eds. D. A. Leahy et al., World Scientific, Singapore, p. 45 (1994)
 
\item  Biermann, P. L., Becker, J. K., Meli, A., Rhode, W., Seo, E.-S., \& Stanev, T., \PRL {\bf 103}, 061101 (2009); arXiv:0903.4048
	
\item  Biermann, P.L., Becker, J.K., Caceres, G., Meli, A., Seo, E.-S., \& Stanev, T., \ApJL {\bf 710}, L53 - L57 (2010); arXiv:0910.1197

\item Bisnovatyi-Kogan, G. S., {\it Astron. Zh.} {\bf 47}, 813 (1970)

\item Bisnovatyi-Kogan, G. S., Moiseenko, S. G., {\it Chinese J. of Astron. \& Astroph. Suppl.} {\bf 8}, 330 - 340 (2008)

\item Chang, J., et al. \Nature {\bf 456}, 362 (2008)

\item  Dobler, G., Finkbeiner, D.P., \ApJ  {\bf 680}, 1222 - 1234 (2008); arXiv:0712.1038
	
\item Dobler, G., Finkbeiner, D. P., Cholis, I., Slatyer, T. R., Weiner, N., eprint arXiv:0910.4583 (2009)

\item  Gopal-Krishna, Peter L. Biermann, Vitor de Souza, Paul J. Wiita, 
in press \ApJL (2010); arXiv:

\item  Schlickeiser, R., Ruppel, J., \NJPh {\bf 12}, 033044 (2010); arXiv:0908.2183  

\item  Stanev, T., Biermann, P.L. \& Gaisser, T.K.,   \AA {\bf 274}, 902 (1993) - paper CR-IV; astro-ph/9303006

\item  Stawarz, L., Petrosian, V., \& Blandford, R.D., \ApJ  {\bf 710}, 236 - 247 (2010); arXiv:0908.1094

\item  Weidenspointner, G., et al. \NewAR  {\bf 52}, 454 - 456 (2008)

\end{itemize}

\newpage

\section{Photos of the Colloquium}

\bigskip

Photos of the Colloquium are available at:

\bigskip

http://www.chalonge.obspm.fr/colloque2010.html

\bigskip

\bigskip

\begin{figure}[ht]
\includegraphics[scale=.9]{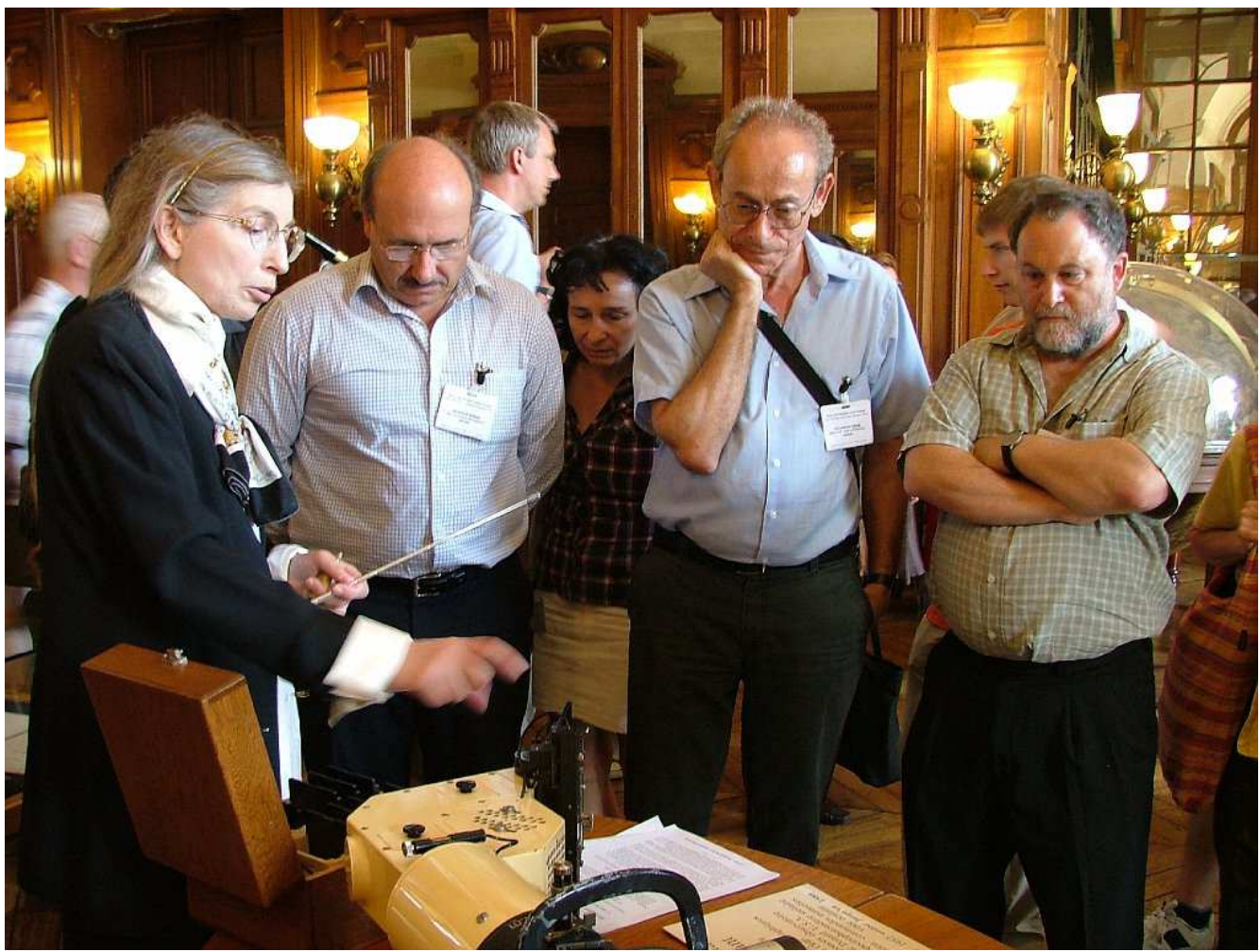}
\caption{The original Chalonge spectrograph at the Grande Gallerie}
\end{figure}

\begin{figure}[ht]
\includegraphics[scale=.9]{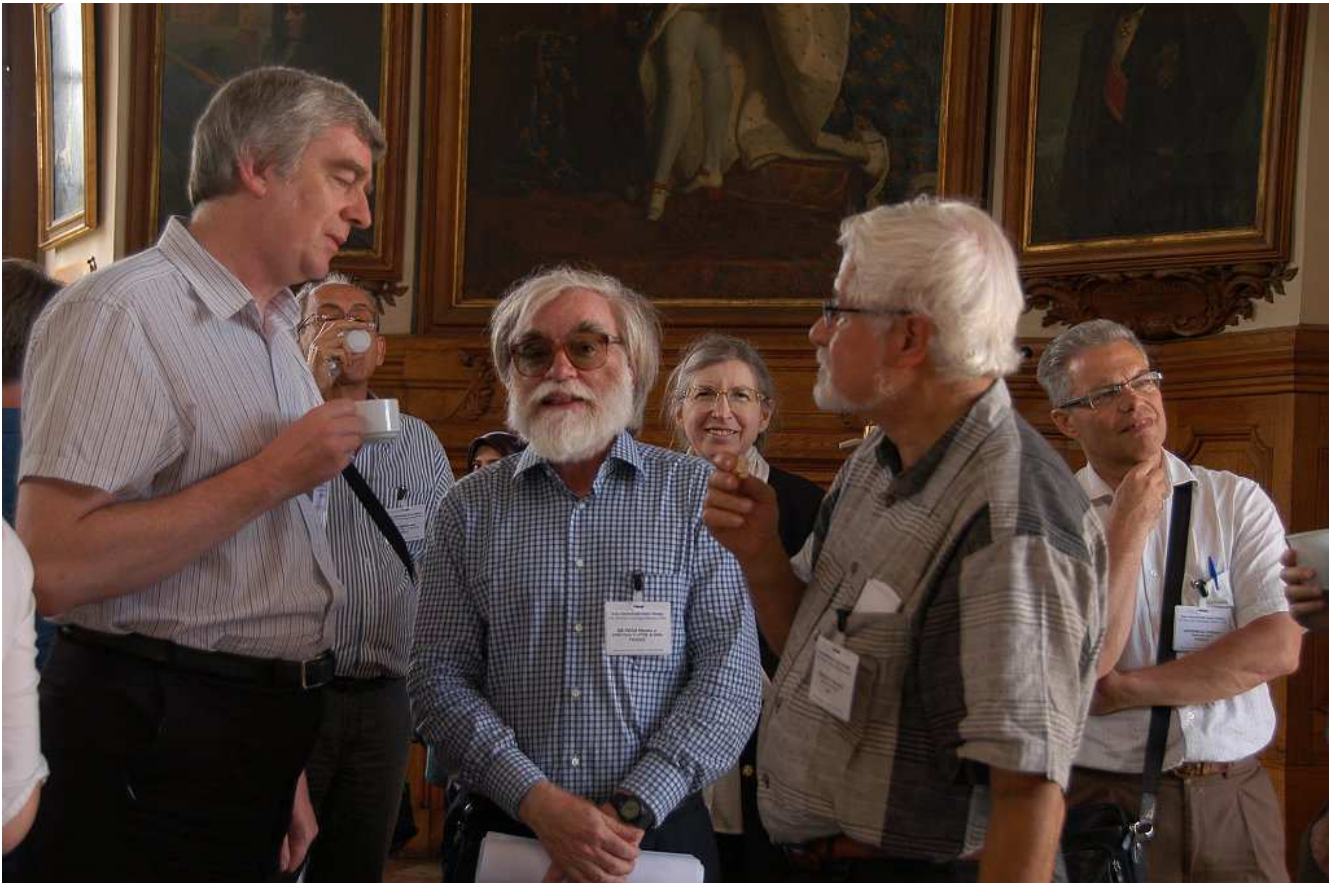}
\caption{At the Salle du Conseil}
\end{figure}

\begin{figure}[ht]
\includegraphics[scale=.9]{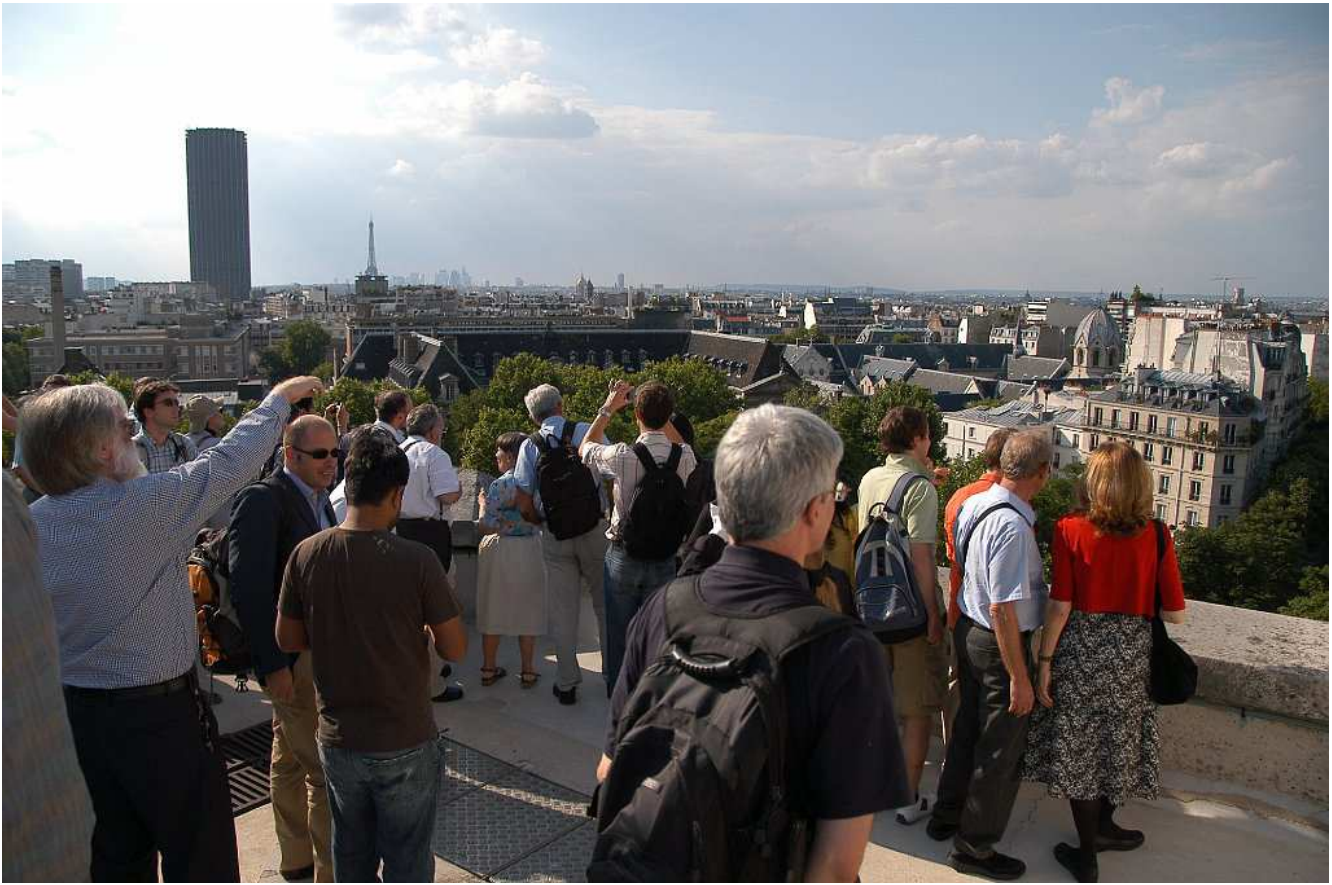}
\caption{On the Terrasse of the Observatoire}
\end{figure}

\begin{figure}[ht]
\includegraphics[scale=.9]{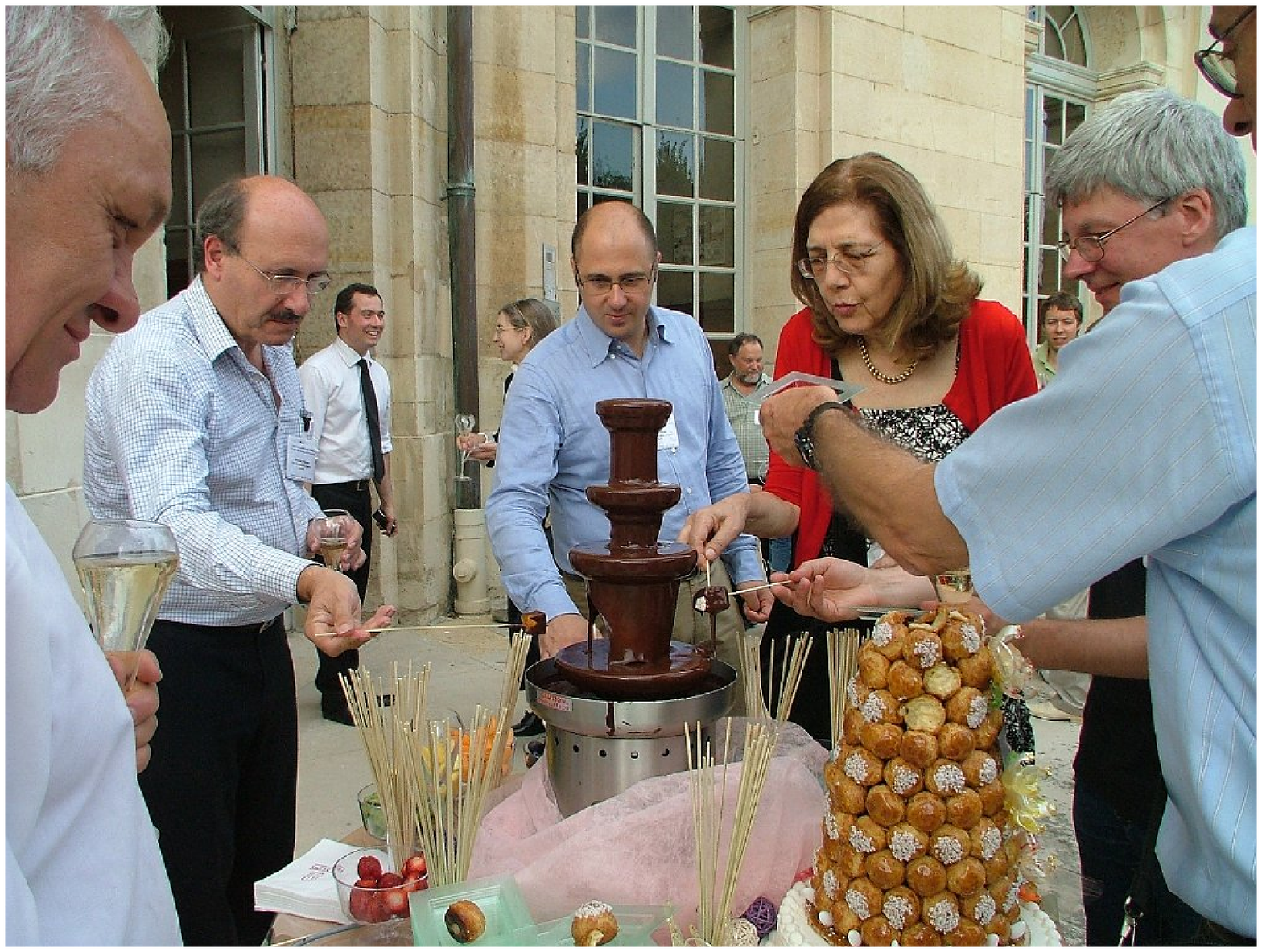}
\caption{The chocolate fountain}
\end{figure}

\newpage

\section{List of Participants}

\bigskip

\noindent
ARKHIPOVA Natalia, Astro Space Center of Lebedev Physical Institute, Moscow RUSSIA \medskip\\
\medskip
ASLANYAN	Petros,	Joint Institute For Nuclear Research, LHEP	Dubna	RUSSIA\\
\medskip
BHADRA	Arunava,	University of North Bengal	Siliguri	INDIA\\
\medskip
BIERMANN	Peter L.,	MPIfR Bonn Germany and UA Tuscaloosa, AL, USA	\\
\medskip
BIESIADA	Marek,	Institute of Physics, University of Silesia,	Katowice	POLAND\\
\medskip
BLAU	Steve,	American  Institute of  Physics, College Park, Maryland	USA\\
\medskip
BLUHM 	Robert,	Colby College, Waterville 	USA\\
\medskip
BONIFACIO	Paolo,	University of Aberdeen,	Aberdeen	UK\\
\medskip
BOYANOVSKY	Daniel,	University of Pittsburgh, Dept of Physics \& Astronomy,	Pittsburgh, PA	USA\\
\medskip
BRADU	Pascal,	Ecole Polytechnique, Palaiseau	FRANCE\\
\medskip
BUCHARD	Albert,	IBENS (ENS),	Paris	FRANCE\\
\medskip
BURDE	Georgy,	Ben-Gurion University of the Negev,	Sede-Boker Camp	ISRAEL\\
\medskip
CALABRESE	Erminia,	University of Rome "La Sapienza",	Rome	ITALY\\
\medskip
CEA	Paolo,	Dipartimento di Fisica di Bari \& INFN di Bari,	Bari	ITALY\\
\medskip
CHARGUI	Yassine,	Faculty of Sciences of Tunis,	Tunis	TUNISIA\\
\medskip
CHAUVINEAU	Bertrand,	Observatoire de la C\^ote d'Azur,	Grasse	FRANCE\\
\medskip
CHEN	Xuelei,	National Astronomical Observatories, Chinese Academy of Sciences,	
Beijing	CHINA\\
\medskip
CHO	Inyong,	Seoul National University of Technology,	 Seoul	SOUTH KOREA\\
\medskip
CIOBANU 	Nellu  Institute of Applied Physics, Academy of Sciences,  Chisinau MOLDOVA\\     
\medskip
CNUDDE Sylvain,	 Observatoire de Paris LESIA Meudon,	FRANCE\\    
\medskip
COLLINS	Hael,	Niels Bohr International Academy,	Copenhagen	DENMARK\\
\medskip
COMIS	Barbara,	"Sapienza", University of Rome,	Roma	ITALY\\
\medskip
COORAY	Asantha,	University of California, Irvine	Irvine	USA\\
\medskip
COSMAI	Leonardo, 	Università di Bari \& INFN Sezione di Bari, Bari ITALY\\
\medskip
DE VEGA 	Hector J.,	UPMC Paris VI LPTHE Jussieu \&  CNRS,	Paris	FRANCE\\
\medskip
DECHELETTE	Typhaine,	IAP,	Paris	FRANCE\\
\medskip
DESTRI 	Claudio,	Univ Milano-Bicocca-INFN Dipt di Fisica G Occhialini,	Milano	ITALY\\
\medskip
DOKUCHAEV Vyacheslav,	Institute for Nuclear Research of the Russian Acad, Moscow	RUSSIA\\
\medskip
DUMIN	Yurii	IZMIRAN, Russian Academy of Sciences	Troitsk, Moscow	RUSSIA\\
\medskip
EASSON	Damien,	Arizona State U. and IPMU, U. of Tokyo,	Tokyo	JAPAN\\
\medskip
ECHAURREN	Juan,	Codelco Chile, North Division, Calama	CHILE\\
\medskip
FABRETTI	Alexandre,	University Paris VI,	Paris	FRANCE\\
\medskip
FALVELLA	Maria Cristina,	Italian Space Agency \& MIUR-Direzione Generale Ricerca,	Rome	ITALY\\
\medskip
FLIN	Piotr	Jan Kochanowski, University, Institute of Physics,	Kielce	POLAND\\
\medskip
FREITAS DINIZ	Edgard,	National Institute for Space Research - INPE,	São José 	BRAZIL\\
\medskip
FRENK	Carlos S.,	Center for Computational Cosmology- Univ of Durham,	Durham	UNITED KINGDOM\\
\medskip
GHODSI	Hoda,	University of Glasgow,	Glasgow	UNITED KINGDOM\\
\medskip
GHOSH	Shubhrangshu,	IIA Academia Sinica, 	Taipei	TAIWAN\\
\medskip
GILMORE	Gerard F.,	Institute of Astronomy, University of Cambridge,	Cambridge	UNITED KINGDOM\\
\medskip
GIOCOLI	Carlo	ZAH/ITA University of Heidelberg,	Heidelberg	GERMANY\\
\medskip
GOLDMAN	Itzhak,	Afeka College , Department of Exact Sciences, 	Tel Aviv	ISRAEL\\
\medskip
GOTTL\"OBER 	Stefan, 	Astrophysikalisches Institut Potsdam,	Potsdam, 	GERMANY\\
\medskip
HANZEVACK	Emil,	College of William \& Mary,	Williamsburg 	USA\\
\medskip
HASHIM	Norsiah,	University of Malaya,	Kuala Lumpur	MALAYSIA\\
\medskip
IVANOV	Mikhail,	Sternberg Astronomical Institute, M.V.Lomonosov, 	Moscow	RUSSIA\\
\medskip
JOURNEAU	Philippe,	Discinnet Labs,	Puteaux	FRANCE\\
\medskip
KAHNIASHVILI	Tina,	Carnegie Mellon University,	Pittsburgh, PA	USA\\
\medskip
KAMIYA	Noriaki,	University of Aizuwakamatsu,	Aizuwakamatsu	JAPAN\\
\medskip
KARCZEWSKA	Danuta,	University of Silesia,	Katowice	POLAND\\
\medskip
KHADEKAR	Goverdhan,	RTM Ngapur University, Nagpur	Nagpur,	INDIA\\
\medskip
KOMATSU	Eiichiro,	University of Texas at Austin, Dept of Astronomy,	Austin	USA\\
\medskip
KOSTRO	LUDWIK,	University of  Gdansk, 	  Gdansk	POLAND\\
\medskip
KRAWIEC 	Adam,	Jagiellonian University,	Krakow	POLAND\\
\medskip
LASENBY	Anthony, Cavendish Laboratory, Astrophysics Group, Univ. Cambridge, UNITED KINGDOM\\
\medskip
LETOURNEUR	Nicole,	Observatoire de Paris LESIA Meudon, 	Meudon	FRANCE\\
\medskip
LI	Nan,	National Astronomical Observatories, Chinese Acad. Sciences,	Beijing	CHINA\\
\medskip
LIMA NETO	Gastao,	IAG - Universidade de São Paulo,	São Paulo	BRAZIL\\
\medskip
LIN	Hai,	University of Santiago de Compostela,	Santiago de Compostela	SPAIN\\
\medskip
Mc GAUGH Stacy, University of Maryland, College Park Maryland USA\\
\medskip
MEHTA	Kushal,	University of Arizona, Steward Observatory,	Tucson	USA\\
\medskip
MIRABEL	F\'elix, CEA-Saclay, France \& IAFE-Buenos Aires, ARGENTINA and	Gif-sur-Yvette	FRANCE\\
\medskip
MOSKALIUK	Stepan,	Bogoliubov Institute for Theoretical Physics of NA,	Kiev	UKRAINE\\
\medskip
MUSAKHANYAN	Viktor,	Gavar and HayBusak Universities,	Yerevan	ARMENIA\\
\medskip
MUSSA	Atifah,	University College London  UCL,	London	UK\\
\medskip	
NATOLI Paolo, Universit\`a Roma 2 Tor Vergata and ASI Science Data Center, 
Frascati ITALY \\
\medskip		
NOH	Hyerim,	Korea Astrronomy and Space Science Institute,	Taejon	KOREA\\
\medskip
ORANI	Stefano,	Imperial College London,	London	UK\\
\medskip
PAGANO	Luca,	University of Rome "Sapienza"	,Roma	ITALY\\
\medskip
PANDOLFI	Stefania,	University of Rome "La Sapienza",	Rome	ITALY\\
\medskip
PANDYA	Aalok,	Department of Physics, University of Rajasthan,	Jaipur	INDIA\\
\medskip
PANKAJ	Kumar,	IGIDR, Mumbai	Mumbai	INDIA\\
\medskip
PARISI	Maria Florencia	,Facultad de Matematica, Astronomia y Fisica, Univ C\'ordoba	ARGENTINA\\
\medskip
PEÑA SUAREZ	Vladimir Jearim,	Universidad Industrial de Santander,	Bucaramanga	COLOMBIA\\
\medskip
PILO	Luigi,	University of L'Aquila and INFN,	L'Aquila	ITALY\\
\medskip
RAMON MEDRANO	Marina,	Universidad Complutense Dept Fisica Teorica,	Madrid	SPAIN\\
\medskip
REALDI	Matteo,	Department of Physics, University of Padua,	Padua	ITALY\\
\medskip
REBOLO	Rafael,	Instituto Astrofisico de Canarias, Tenerife	Tenerife	SPAIN\\
\medskip
RICOTTI	Massimo,	University of Maryland,	College Park	USA\\
\medskip
ROBLES	Sandra,	Universidad Autonoma de Madrid,	Madrid	SPAIN\\
\medskip
ROCHUS Pierre, Université de Liège, Centre Spatial de Liège, Liège, BELGIUM\\
\medskip
RODRIGUEZ	Ivan,	Cinvestav	Mexico, D.F.	MEXICO\\
\medskip
ROMANO	Antonio Enea,	Yukawa Institute Theoretical Physics,	Kyoto	JAPAN\\
\medskip
RUIZ	Andres Nicolas,	Instituo de Astronomia Teorica y Experimental (IATE),	Cordoba	ARGENTINA\\
\medskip
SABBATINI	Lucia,	University of Roma Tre, Dept. of Physics,	Roma	ITALY\\
\medskip
SADOULET	Bernard	,Particle Cosmology Group, University of California,	Berkeley	USA\\
\medskip
SALUCCI 	Paolo, 	SISSA-Trieste- Astrophysics Group,	Trieste	ITALY\\
\medskip
SANCHEZ	Norma G.,	Observatoire de Paris LERMA and CNRS,	Paris	FRANCE\\
\medskip
SEVELLEC	Aurelie,	Observatoire de Paris LESIA Meudon, 	Meudon	FRANCE\\
\medskip
SMOOT	George,	LNBL-Univ California and Univ Paris Denis Diderot,	Berkeley and Paris 	USA\\
\medskip
SOARES	Ivano Damiao, 	Centro Brasileiro de Pesquisas Fisicas - CBPF/MCT,	Rio de Janeiro	BRAZIL\\
\medskip
SORENSEN	Peter,	LLNL,	Livermore	USA\\
\medskip
STARIKOVA	Svetlana,	Department of Astronomy, University of Padova,	Padua	ITALY\\
\medskip
STIVOLI	Federico,	INRIA,	Paris	FRANCE\\
\medskip
SUCIU	Oana Elena,	Faculty pf Physics and Faculty of Mathematics, Bab	Cluj-Napoca	ROMANIA\\
\medskip
SZYDLOWSKI	Marek,	Jagiellonian University,	Krakow	POLAND\\
\medskip
TABATABAEI	Seyed, Alireza	Queen Mary, University of London,	London	UK\\
\medskip
TARTAGLIA	Angelo,	Politecnico di Torino,	Torino	ITALY\\
\medskip
TEDESCO	Luigi,	Dipartimento di Fisica di Bari and INFN di Bari,	Bari	ITALY\\
\medskip
THOMAS	Daniel,	Imperial College,	London	UK\\
\medskip
TIKHONOV	Anton,	Saint-Petersburg State University, Astronomical Institute,	Saint-Petersburg	RUSSIA\\
\medskip
TING	Yuan Sen,	Ecole Polytechnique Paris,	Palaiseau	FRANCE\\
\medskip
TRINIDAD	Rodrigo,	Universiad de San Carlos de Guatemala,	Guatemala	GUATEMALA\\
\medskip
TURZYNSKI Krzysztof, Institute of Theoretical Physics, University of Warsaw,	Warsaw	POLAND\\
\medskip
UNGKU	Ferwani Salwa,	University of Malaya,	Kuala Lumpur	MALAYSIA\\
\medskip
URTADO	Olivier,	Orsay M1, Astrophysique,	Versailles	FRANCE\\
\medskip
VAZQUEZ-MATA	Jose Antonio,	University of Sussex,	Brighton	UK\\
\medskip
VERMA	Murli Manohar,	Department of Physics, Lucknow University,	Lucknow	INDIA\\
\medskip
VAN ELEWYCK	Veronique,	APC - Université Paris 7,	Paris	FRANCE\\
\medskip
VISHWAKARMA	Ram Gopal,	University of Zacatecas,	Zacatecas	MEXICO\\
\medskip
VON KNOP Jan Heinrich Heine, University D\"usseldorf, D\"usseldorf	GERMANY\\
\medskip
WAGSTAFF	Jacques,	Lancaster University,	Lancaster	ENGLAND\\
\medskip
WANDELT	Benjamin,	Institut d'Astrophysique de Paris,	Paris 	FRANCE\\
\medskip
WANG Xiang-Yu,  Department of Astronomy, Nanjing University, Nanjing CHINA\\
\medskip
XU Xiaoying,  Steward Observatory, University of Arizona, Tucson USA\\
\medskip
ZANINI Alba,  INFN Sezione di Torino,  Turin ITALY \\
\medskip
ZIAEEPOUR Houri,  Max-Plank Institute fur Extraterrestrisch 
Physik,  Garching b. Munchen,  GERMANY\\
\medskip
ZIDANI Djilali,  Observatoire de Paris - CNRS, Paris FRANCE \\
\medskip
ZOLOTKHIN Ivan, Observatoire de Paris and Sternberg Astr. Inst.,  Paris 
FRANCE\\

\end{document}